\documentclass[journal]{IEEEtran}

\ifCLASSINFOpdf
\else
\fi
\usepackage{amsthm}
\usepackage{multicol,multirow}
\usepackage{graphicx}
\usepackage{amssymb, amsmath}
\usepackage{xcolor}
\usepackage{cite}
\usepackage{hyperref}
\usepackage{enumerate}
\usepackage{esint}
\usepackage{balance}
\usepackage{textcomp}
\usepackage{booktabs}
\usepackage{bbold}
\usepackage{booktabs}% http://ctan.org/pkg/booktabs

\DeclareMathOperator*{\argmin}{argmin}

\begin{document}
%
% paper title
% Titles are generally capitalized except for words such as a, an, and, as,
% at, but, by, for, in, nor, of, on, or, the, to and up, which are usually
% not capitalized unless they are the first or last word of the title.
% Linebreaks \\ can be used within to get better formatting as desired.
% Do not put math or special symbols in the title.
\title{Localization as a key enabler of 6G wireless systems: A comprehensive survey and an outlook}
%
%
% author names and IEEE memberships
% note positions of commas and nonbreaking spaces ( ~ ) LaTeX will not break
% a structure at a ~ so this keeps an author's name from being broken across
% two lines.
% use \thanks{} to gain access to the first footnote area
% a separate \thanks must be used for each paragraph as LaTeX2e's \thanks
% was not built to handle multiple paragraphs
%

\author{Stylianos E. Trevlakis,~\IEEEmembership{Member,~IEEE,}
        Alexandros-Apostolos A. Boulogeorgos,~\IEEEmembership{Senior Member,~IEEE,} 
        Dimitrios Pliatsios,~\IEEEmembership{Member,~IEEE,}
        Konstantinos Ntontin,~\IEEEmembership{Member,~IEEE,}
        Panagiotis Sarigiannidis,~\IEEEmembership{Member,~IEEE,}
        Symeon Chatzinotas,~\IEEEmembership{Fellow,~IEEE,} and
        Marco Di Renzo,~\IEEEmembership{Fellow,~IEEE,}
        % <-this % stops a space
\thanks{Manuscript received January~XX, 2023; revised~XX; accepted~XX}% <-this % stops a space
\thanks{S. E. Trevlakis is with the Department of Research and Development, InnoCube P.C., 17th Noemvriou 79, 55535 Thessaloniki, Greece (email: trevlakis@innocube.org)}% <-this % stops a space
\thanks{A.-A. A. Boulogeorgos, D. Pliatsios, P. Sarigiannidis are with the Department of Electrical and Computer Engineering, University of Western Macedonia, ZEP Area, 50100 Kozani, Greece (e-mail: aboulogeorgos@ieee.org; dpliatsios@uowm.gr; psarigiannidis@uowm.gr)}% <-this % stops a space
\thanks{K. Ntontin and S. Chatzinotas are with the SnT, University of Luxembourg, Luxembourg. E-mails: kostantinos.ntontin@uni.lu,  symeon.chatzinotas@uni.lu.}% <-this % stops a space
\thanks{M. Di Renzo is with Universit\'e Paris-Saclay, CNRS, CentraleSup\'elec, Laboratoire des Signaux et Syst\`emes, 3 Rue Joliot-Curie, 91192 Gif-sur-Yvette, France. (marco.di-renzo@universite-paris-saclay.fr)}% <-this % stops a space
\thanks{This work has received funding from the European Unions Horizon-CL4-2021 research and innovation programme under grant agreement No. 101070181 (TALON) as well as the the European Unions HORIZON-JU-SNS-2022 research and innovation programme under grant agreement No. 101096456 (NANCY).}}

% note the % following the last \IEEEmembership and also \thanks - 
% these prevent an unwanted space from occurring between the last author name

% The paper headers
\markboth{IEEE Communications Surveys \& Tutorials,~Vol.~XX, No.~XX, First~quarter~2023}%
{Trevlakis \MakeLowercase{\textit{et al.}}: Localization: A key enabler or an outcome of \\ 6G wireless systems?}
% The only time the second header will appear is for the odd numbered pages
% after the title page when using the twoside option.
% 
% *** Note that you probably will NOT want to include the author's ***
% *** name in the headers of peer review papers.                   ***
% You can use \ifCLASSOPTIONpeerreview for conditional compilation here if
% you desire.

% make the title area
\maketitle

% As a general rule, do not put math, special symbols or citations
% in the abstract or keywords.
\begin{abstract}
When fully implemented, sixth generation (6G) wireless systems will constitute intelligent wireless networks that enable not only ubiquitous communication but also high-accuracy localization services. They will be the driving force behind this transformation by introducing a new set of characteristics and service capabilities in which location will coexist with communication while sharing available resources. To that purpose, this survey investigates the envisioned applications and use cases of localization in future 6G wireless systems, while analyzing the impact of the major technology enablers. Afterwards, system models for millimeter wave, terahertz and visible light positioning that take into account both line-of-sight (LOS) and non-LOS channels are presented, while localization key performance indicators are revisited alongside mathematical definitions. Moreover, a detailed review of the state of the art conventional and learning-based localization techniques is conducted. Furthermore, the localization problem is formulated, the wireless system design is considered and the optimization of both is investigated. Finally, insights that arise from the presented analysis are summarized and used to highlight the most important future directions for localization in 6G wireless systems. 
\end{abstract}

% Note that keywords are not normally used for peerreview papers.

% For peer review papers, you can put extra information on the cover
% page as needed:
% \ifCLASSOPTIONpeerreview
% \begin{center} \bfseries EDICS Category: 3-BBND \end{center}
% \fi
%
% For peerreview papers, this IEEEtran command inserts a page break and
% creates the second title. It will be ignored for other modes.
\IEEEpeerreviewmaketitle

\section{Introduction}\label{S:Introduction}
To satisfy the proliferating demands of next generation wireless applications, such as multisensory extended reality (XR)~\cite{Saad2020,Martin2020,Hu2021}, connected robots~\cite{Weiss2021,Savazzi2021,Sankhe2021}, wireless brain computer interactions~\cite{Perdikis2020}, digital twins~\cite{Khan2022b}, industrial internet-of-things (IoT)~\cite{Zhou2021}, tactile IoT~\cite{Aijaz2017a,Saeed2020,Yang2019b}, internet of underwater things~\cite{Abdulkadir2020,Filho2020},  self-driving ground and air vehicles~\cite{Pretto2021, Liu2022f,Zhong2022f}, and others, while dealing with the spectrum scarcity of the radio and microwave bands~\cite{Boulogeorgos2017,Boulogeorgos2018a,PhD:Boulogeorgos,Matthaiou2021},  the research, innovation and industrial communities turned its attention to communications in the millimeter wave (mmWave)~\cite{Xi2019,Jie2020,Ntontin2021,Ntontin2021b,Le2022}, terahertz (THz)~\cite{Boulogeorgos2021a,Boulogeorgos2020b,Boulogeorgos2021e,Tsiftsis2022}, and optical bands~\cite{Trevlakis2020,Boulogeorgos2021d,Abdulkadir2022,Anas2022,Naser2022}. Despite the unprecedented bandwidth that high-frequency systems offer, in order to achieve the promised performance excellence in terms of throughput, latency, and reliability, in acceptable transmission distances, both the transmitter (TX) and the receiver (RX) require knowledge of each others relative position and orientation~\cite{Azari2022,Pliatsios2022,Boulogeorgos2022f,Boulogeorgos2022g,Boulogeorgos2022h}. As a consequence, localization is expected to become a vital component of high-frequency wireless systems.

Scanning the technical literature, several surveys and tutorials for localization can be identified~\cite{Bresson2017,Ferreira2017,Kuutti2018,Laoudias2018,Peral2018,Keskin2018,Shit2019,Zafari2019,Saeed2019,Wen2019,Burghal2020,Zhu2020,Motroni2021,DeLima2021,Kanhere2021a,Xiao2022,Ding2022,Laconte2022,Liu2022,Chen2022}. In more detail, in~\cite{Bresson2017}, the authors presented an overview of the simultaneously localization and mapping approaches in self-driving vehicles. In~\cite{Ferreira2017}, the authors reported the requirements, technology enablers, techniques and approaches, as well as existing system models and their performance in indoor localization for emergency response scenarios. In~\cite{Kuutti2018}, a state-of-the-art (SotA) vehicle localization approaches comparison and assessment was conducted and their role to vehicle-to-everything scenarios was discussed. A list of the enabling localization technologies that were developed for cellular systems and wireless local area networks until 2018 was documented in~\cite{Laoudias2018}. In~\cite{Peral2018}, the authors presented an overview of the standardized localization methods from the first to the fifth generation cellular systems. In~\cite{Keskin2018}, the authors revisited and discussed visible light positioning (VLP) approaches.

The authors of~\cite{Shit2019} conducted an overview of the device-free localization approaches and presented the 2019 research and development trends. In~\cite{Zafari2019}, the authors surveyed the 2019 indoor localization systems and methods accounting for different radio technologies. In~\cite{Saeed2019}, the authors revisited and explained multidimensional scaling (MDS)-based localization techniques. In~\cite{Wen2019}, a review of the SotA of localization techniques for massive multiple-input multiple-output (MIMO) systems was conducted. Moreover, the authors of~\cite{Burghal2020} reviewed machine learning (ML)-based localization methodologies that employ radio signals. In~\cite{Zhu2020}, a summation of ML-enabled indoor localization approaches was reported.

The authors of~\cite{Motroni2021} presented a survey of radio frequency identifier (RFID) based localization techniques. In~\cite{DeLima2021}, the authors focused on the identification of key enabling technologies and applications of localization in the sixth generation (6G) networks. In~\cite{Kanhere2021a}, the vision of cm-level localization was presented and map-based approaches was identified as possible enablers. In~\cite{Xiao2022}, a survey on localization fundamental approaches accompanied by SotA results and future directions was reported. The authors of~\cite{Ding2022} focused on the recent developments and applications of simultaneous localization and mapping (SLAM) with emphasis in complex and unstructured agricultural environments. In~\cite{Laconte2022}, the authors overviewed a number of localization methods for autonomous vehicles. The authors of~\cite{Liu2022} conducted a survey on the fundamental limits of integrated communication and localization systems. Finally, in~\cite{Chen2022}, the authors provided a survey concerning THz empowered localization techniques. 

\begin{table*}
\centering
\caption{Summary of recent localization surveys and tutorials (2017-2019)}
\label{tab:Survey1719}
\begin{tabular}{|c|c| p{2.3cm} | p{1.6cm} | p{1.5cm}| p{1.8cm}| p{2.5cm} | p{4.0cm} |}
    \hline
    \textbf{Year} & \textbf{Ref.} & \textbf{Applications} & \textbf{Use-cases} & \textbf{Technologies} & \textbf{System models} & \centering \textbf{KPIs} & \textbf{Algorithms}  \\ 
    \hline \hline
    2017 & \cite{Bresson2017} & Autonomous vehicle & Outdoor & Sensors & \centering -- & Accuracy, scalability, availability, recovery, updatability, dynamicity & SLAM \\ 
    \hline 
    2017 & \cite{Ferreira2017} & Emergency response & Indoor \par Outdoor  & Radio signals, Inertial measurement units (IMUs) & \centering -- & Accuracy,
    information~accessibility, adaptability, scalability, physical robustness, assembling complexity, equipment's size and weight, energy efficiency, cost & Proximity, Triangulation, Lateration/ trilateration/ multilateration, Maximum likelihood estimation, Dead reckoning, Kalman filters, Centroid, Particle filters, Fingerprinting, Visual analysis, Map matching \\
    \hline
    2018 & \cite{Kuutti2018} & Autonomous vehicles & Outdoor & GPS/IMU, Sensors, Radio and microwave signals, VLP & \centering -- &  Accuracy, Dead reckoning, Position error, Odometry, Packet loss, Robustness, Time-to-first-fix & Differential GPS, Assisted GPS, Real time kinematic based techniques, SLAM, Random sample consensus, Frequency modulation continuous wave (FMCW)-based solutions \\
    \hline
    2018 & \cite{Laoudias2018} & Network planing and optimization, Emergency response & Indoor \par Outdoor & Cellular, WLANs, Sensors, mmWave & Cellular & Accuracy & Probabilistic radio signal strength (RSS) fingerprint matching, Semi-supervised and unsupervised learning, Cell-identifier and GPS position sequence matching, Hidden Markov model, Bayesian learning, Assisted GPS, Advanced forward link trilateration, RF pattern matching, Intelligent probability hierarchy based solutions, SLAM \\
    \hline 
    2018 & \cite{Peral2018} & \centering -- & Indoor \par Outdoor & Radio signals, WLAN/ Bluetooth, Sensors & \centering --  & Accuracy & Cell-ID, RF pattern matching, Time of arrival (ToA) and time difference of arrival (TDoA) based methods, Advanced forward link trilateration, Stand-alone, Differential, and Assisted GNSS \\
    \hline 
    2018 & \cite{Keskin2018} & \centering-- & Indoor & VLP & VLP & Cram\'er-Rao lower bound (CRLB), Root mean square error, Power efficiency & Direct positioning, proximity, geometric- and statistical-based approaches, fingerprinting, mapping\\
    \hline
    2019 & \cite{Shit2019} & Smart agriculture, Smart grid, Smart media, Smart cities, Social computing, Social IoT, e-Business, Affective computing, Cyber physical systems, Wearable applications & Indoor \par Outdoor & Radio signals, Sensors & \centering-- & Static and dynamic backscattering model & Directional and ambient radio imaging, channel diversity, Shadowing, Comprehensive sensing, Radio grid, Diffraction theory, Extreme learning machine, Markov model, Probabilistic classification, Gradient fingerprinting, Support vector machine (SVM), Deep learning, Dictionary learning  \\
    \hline 
    2019 & \cite{Wen2019}  & \centering -- & Indoor \par Outdoor & mmWave & mmWave MIMO & Accuracy, Complexity & Indirect localization, Direct localization, Fingerprinting \\
    \hline 
    2019 & \cite{Zafari2019} & Personalized context-aware networks, e-Health, Disaster management and recovery, Security, Asset management and tracking & Indoor & WiFi, Bluetooth, Zigbee, RFID, VLP, Acoustic, Ultrasound & \centering -- & Availability, Cost, Energy efficiency, Reception range, Accuracy, Latency, Scalability & Fingerprinting/scene analysis, RSS, Channel state information (CSI), Angel of arrival (AoA), Time of flight (ToF), TDoA, Return time of flight (RFoF), and Phase of arrival-based approaches \\
    \hline
    2019 & \cite{Saeed2019} & Disaster management, Security, IoT, Underwater exploration & Indoor \par Outdoor \par NTN \par Underwater & Cellular, RFID, Sensors, Acoustic, VLP & \centering-- & Accuracy, Complexity & MDS \\
    \hline
\end{tabular}
\end{table*}

\begin{table*}
\centering
\caption{Summary of recent localization surveys and tutorials (2020-2022)}
\label{tab:Survey2022}
\begin{tabular}{|c|c| p{2.3cm} | p{1.6cm} | p{1.5cm}| p{1.8cm}| p{2.5cm} | p{4.0cm} |}
    \hline
    \textbf{Year} & \textbf{Ref.} & \textbf{Applications} & \textbf{Use-cases} & \textbf{Technologies} & \textbf{System models} & \centering \textbf{KPIs} & \textbf{Algorithms}  \\ 
    \hline \hline
    2020 & \cite{Burghal2020} & Autonomous vehicles, Mission-critical applications, IoT, Beam alignment in massive MIMO systems & Indoor \par Outdoor  & WiFi, Cellular, Bluetooth, Sensors & \centering -- & Accuracy & K-nearest neighbors (KNN), Gaussian processes, Ensemble methods, Neural networks (NNs), Convolution NNs (CNNs), Recurrent NNs (RNNs), Autoencoders, Generative and statistical models, Deep belief networks, MDS, Tranfer learning \\
    \hline 
    2020 & \cite{Zhu2020} & Positioning and navigation of indoor environment, Nursing people and tracking, People management, Fire rescue and other safety needs & Indoor & WiFi, Bluetooth, VLP, Magnetic field & \centering -- & Accuracy, latency, complexity, coverage, robustness & Fingerprint \\
    \hline 
    2020 & \cite{Nguyen2020a, Nguyen2020b} & Social distancing & Indoor \par Outdoor \par NTN & WiFi, GNSS, cellular, bluetooth, UWB, RFID & \centering -- & Accuracy, latency, coverage & Trilateration, Kalman filters, RSS-based, TDoA-based, Assisted-GNSS, Enhanced Cell-ID \\
    \hline 
    2021 &  \cite{Motroni2021} & Vehicle localization, navigation, tracking & Indoor & RFID, Sensors & \centering -- & Accuracy, cost, energy efficiency, complexity  &  -- \\
    \hline
    2021 & \cite{DeLima2021} & Environment mapping, Robot localization, Tracking, Localization and sensing for eHealth, Context awareness, Radar-based applications & Indoor \par Outdoor & mmWave, THz, Reconfigurable intelligent surfaces (RISs) & \centering -- & \centering -- & Supervised, unsupervised, reinforcement learning \\
    \hline 
    2021 & \cite{Kanhere2021a} & People, objects, and vehicles monitoring & Indoor \par Outdoor & mmWave, Cellular & mmWave & Accuracy & Fingerprinting, Kalman filter, Extended Kalman filter \\
    \hline 
    2021 & \cite{Lemic2021} & Nanonetworks & Nanoscale & THz & \centering -- & Accuraxy, complexity & -- \\
    \hline 
    2022 & \cite{Xiao2022} & Mutisensory XR, Tele-presence, Smart transportation, Connected robotics and autonomous systems, UAVs & Indoor \par Outdoor \par NTN & mmWave, THz, Sensors & \centering -- & Accuracy, latency, precision, energy efficiency & Tracking, SLAM, NNs Extednde Kalman filter \\
    \hline
    2022 & \cite{Ding2022} & Precision agriculture, mapping, navigation & Outdoor &  Sensors & \centering -- & Accuracy & Extended Kalman filter SLAM, Monte Carlo localization, Visual SLAM, Sensor fusion SLAM \\
    \hline 
    2022 & \cite{Laconte2022} & Autonomous vehicles & Outdoor & Sensors & \centering --  & Accuracy, F1 Score & Extended Kalman filters, NNs \\
    \hline 
    2022 & \cite{Liu2022} & IoT connected home, Internet of vehicles, UAV communications and navigation, Target detection & Indoor \par Outdoor \par NTN & WiFi, mmWave & Cooperative and non-cooperative localization with MIMO and virtual MIMO & Accuracy, coverage, complexity, stability & -- \\
    \hline 
    2022 & \cite{Chen2022} &\centering --  & Indoor \par Outdoor & mmWave, THz & THz with and without RISs & Accuracy, Coverage, Latency, Update rate, Stability, Scalability, Mobility, Complexity & ToA-based, TDoA-based, AoA-based, AoD-based, Expected maximization, NNs, Kalman filter, Particle filter  \\
    \hline
    \multicolumn{2}{|c|}{This work} & Autonomous~supply chain, Smart cities, Manufacturing,~XR, Earth monitoring, Network~expansion, Mapping, Sensor infrastructure~web, Context-aware networks, Precision healthcare, Security, Gesture/motion recognition, Robots, Nanoscale, e-Health, Agriculture & Indoor \par Outdoor \par NTN \par Underwater \par Nanoscale & mmWave, THz, hardware, beamforming, RIS, AI, channel charting, radars, sensors, VLP & mmWave/THz, VLP & Accuracy, precision, latency, coverage, complexity, stability & Triangulation, Kalman filters, Compressive sensing, MDS, Direct localization, Swarm intelligence, Fingerprinting, SLAM, kNN, SVM, Decision trees, Gaussian processes, NNs, Autoencoders, CNNs, RNNs, Unsupervised learning, Semi-supervised learning, clustering, Dimensionality reduction, Federated learning, Reinforcement learning (RL), Deep RL, Transfer learning, Manifold learning, Cooperative approaches\\ 
    \hline 
\end{tabular}
\end{table*}

Tables~\ref{tab:Survey1719} and \ref{tab:Survey2022} summarize the surveys existing in literature that investigate localization applications and use cases as well as the relevant technologies, system models, key performance indicators (KPIs) and approaches. However, the focus and aim of each survey is different both in terms of use cases, technologies and methods enlisted.  For instance, almost every survey includes indoor and outdoor terrestrial use cases~\cite{Bresson2017, Ferreira2017, Kuutti2018, Laoudias2018, Peral2018, Keskin2018, Shit2019, Wen2019, Zafari2019, Saeed2019, Burghal2020, Zhu2020, Motroni2021, DeLima2021, Kanhere2021a, Xiao2022, Ding2022, Laconte2022, Liu2022, Chen2022}, while only~\cite{Saeed2019, Xiao2022, Liu2022} take into account non-terrestrial networks (NTN) and only~\cite{Saeed2019} investigates underwater use cases. As far as technology enablers are concerned, each contribution focuses on specific subsets of them. Specifically, the majority of the surveys before $2020$ take into account sensors, radars, and cellular systems~\cite{Bresson2017, Ferreira2017, Kuutti2018, Peral2018, Keskin2018, Shit2019, Zafari2019, Saeed2019}, while the contrary is valid for contributions between $2020$ and $2022$ that mainly investigate mmWave/THz technologies~\cite{DeLima2021, Kanhere2021a, Xiao2022, Liu2022, Chen2022}. In addition, enablers like visible light positioning (VLP), beamforming, channel charting and reconfigurable intelligent surfaces (RISs) are taken into consideration in a very small number of surveys. From a system model point of view, only selected contributions provide mathematical modelling for the discussed technologies, such as~\cite{Laoudias2018, Keskin2018, Wen2019, Kanhere2021a, Liu2022, Chen2022}. From these tables, it becomes evident that a complete survey of the applications, use cases, technology enablers, system models, KPIs, and methods with focus on the future 6G wireless networks is missing from the bibliography. To cover this gap and with respect to the aforementioned contributions, the overall objective of this survey is to deliver a timely and comprehensive review of the localization applications, use cases, system models, enabling technologies, methods and KPIs in the 6G era as well as to identify research gaps and highlight possible research directions. In particular, this survey investigates the following:
\begin{figure}[t]
    \centering\includegraphics[width=1\linewidth]{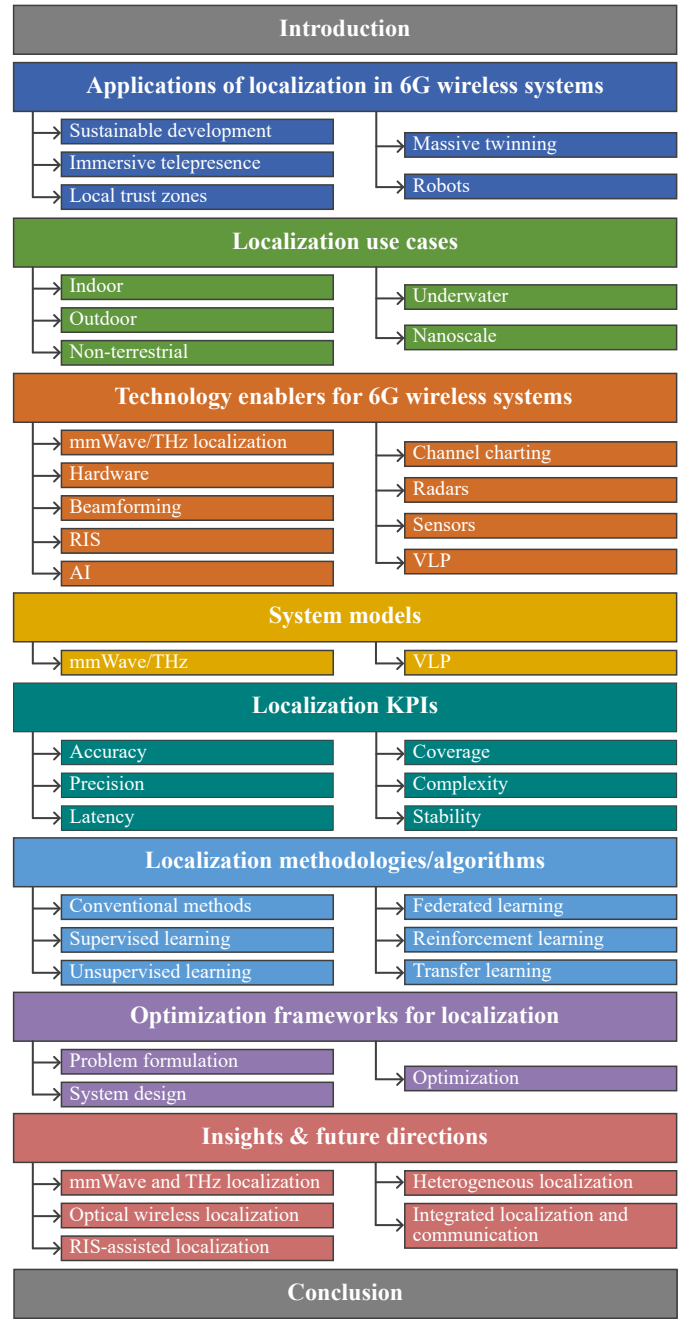}
    \caption{The overall structure of this survey.}
    \label{fig:Structure}
\end{figure}
\begin{itemize}
    \item The main localization-specific applications and use cases of future 6G wireless networks are presented.
    \item The localization-enabling technologies are investigated with regard to their contribution towards advancing the SotA of wireless systems.  
    \item System models for line-of-sight (LOS) and non-LOS (NLOS) channels in mmWave/THz and VLP localization wireless systems are analyzed taking into account both classic reflectors and RISs.
    \item The major KPIs used by localization methods to measure the performance of the system are discussed and accompanied by mathematical expressions.  
    \item The SotA of conventional and learning-based localization methodologies/algorithms are investigated and conceptual figures are provided to aid the reader in understanding their basic principles.
    \item The formulation, system design and optimization of the localization problem is described.
    \item Directions and insights for the design and improvement of future 6G wireless systems are discussed. 
\end{itemize}

The reminder of the survey article is organized as follows: Section~\ref{S:6G_aplications} describes the applications of 6G wireless systems, while Section~\ref{S:Use cases} delves into the use cases and their relations with the various 6G applications. Next, the technology enablers of localization are presented in~Section~\ref{S:Technology_enablers}. Section~\ref{S:System_model} focuses on the system models for mmWave/THz localization and VLP, while the various KPIs of the localization approaches are presented in~Section~\ref{S:KPIs}. Section~\ref{S:Methodologies} describes the SotA conventional and learning-based localization methodologies and algorithms that are envisioned to be applied in future 6G wireless networks. Section~\ref{S:Theoretical_framework} highlights the major aspects of the localization problem formulation, systems design as well as its optimization. Finally, in~Section~\ref{S:Future_directions}, the findings of this contribution are translated into insights and future directions, while concluding remarks are provided in~Section~\ref{S:Conclusion}. The  structure of this survey in a glance is reported in~Figure~\ref{fig:Structure}.

\textit{Abbreviations and notations}: The abbreviations that can be found throughout this survey are presented in~Table~\ref{tab:Abbreviations}, given at the top of the next page. Unless stated otherwise, {bold} low and capital letters respectively denote vectors and matrices, $(\cdot)^T$ stands for the transpose matrix, $|\cdot|$ denotes absolute value, $e^x$ represents the exponential function, $(\cdot)^{-1}$ is the inverse matrix, while $a^b$ is $a$ in the power of $b$. In addition, $\cos(\cdot)$, $\sin(\cdot)$, $\arctan2(\cdot)$, and $\arcsin(\cdot)$ represent the cosine, sine, 2-argument arctangent, and the arcsine functions, respectively. Also, $\log_{2}(\cdot)$ stands for the binary logarithm, $\mathrm{Hilb}[\cdot]$ represents the Hilbert function, and $\mathbb{E}[\cdot]$ denotes the expected value. Moreover, $tr(\cdot)$ is the trace of a matrix and $I^{-1}(\cdot)$ denotes the inverse Fisher matrix. $\argmin(\cdot)$ represents the argument of minimum, $\mathrm{diag}(\cdot)$ is the diagonal matrix, and $\mathrm{rank}(\cdot)$ denotes the rank of a matrix. Finally, $\sqrt{\cdot}$ represents the root and $\sum^N_{i=1}x_i$ is the sum of $x_i$ with $i$ in the range of $[1,N]$. 

\begin{table*}
\centering
    \caption{Abbreviations}
    \label{tab:Abbreviations}
\begin{center}
    \begin{tabular}{ll|ll}
         1D & One dimensional                           & MR & Mixture reality \\ 
         2D & Two dimensional                           & MSE & Mean square error \\ 
         3D & Three dimensional                         & NASA & National aeronautics and space administration \\ 
         3G & Third generation                          & NG-RAT & Next generation radio access technology \\
         4D & Four dimensional                          & NLOS & Non line of sight \\  
         4G & Forth generation                          & NN & Neural network \\ 
         5G & Fifth generation                          & OFDM & Orthogonal frequency division multiplexing \\
         6G & Sixth generation                          & PARP & Peak to average power ratio \\
         AI & Artificial intelligence                   & PCA & Principal component analysis\\ 
         AoA & Angle of arrival                         & PD & Photo-detector \\
         AoD & Angle of departure                       & QoE & Quality-of-experience \\ 
         AR & Augmented reality                         & QoPE & Quality-of-physical-experience \\
         BSE & Beam split effect                        & QoS & Quality-of-service\\ 
         CDF & Cumulative distribution function         & RaF & Random forest \\  
         CMOS & Complementary metal oxide semiconductor & RBF & Radial basis function \\ 
         CNN & Convolutional neural network             & RF & Radio frequency \\    
         COA & Curvature of arrival                     & RGB & Red green blue \\  
         CS & Compresive sensing                        & RIS & Reflective intelligent surfaces \\    
         CSI & Channel state information                & RL & Reinforcement learning \\   
         D2D & Device to device                         & RMS & Root mean square \\   
         DL & Deep learning                             & RRN & Recurrent neural network \\   
         DoA & Direction of arrival                     & RSS & Received signal strength  \\    
         DoD & Direction of departure                   & RTT & Round trip time \\    
         DOF & Degree of freedom                        & RX & Receiver \\    
         DT & Digital twin                              & SDG & Sustainable development goals \\    
         EMND & Enhanced MND                            & SDN & Software defined network \\  
         EKF & Extended Kalman filter                   & SINR & Signal to interference and noise ratio\\ 
         FDM & Frequency division multiplexing          & SLAM & Simultaneous localization and mapping \\ 
         FIM & Fisher information matrix                & SNR & Signal to noise ratio \\  
         FL & Federated learning                        & SotA & State of the art\\     
         FMCW & Frequency modulated continuous wave     & SVM & Support vector machine \\       
         FOV & Field of view                            & TDoA & Time-difference of arrival \\     
         GALILEO & European GNSS                        & TDM & Time division multiplexing \\      
         GDoP & Geometrical dilution of precision       & THz & Terahertz \\      
         GLONASS & Russian GNSS                         & TL & Transfer learning \\    
         GNSS & Global navigation satellite system      & ToA & Time of arrival \\    
         GP & Gaussian process                          & ToF & Time of flight \\    
         GPS & Global positioning system                & TX & Transmitter \\  
         GRU & Gated recurrent units                    & UAV & Unmanned aerial vehicle \\            
         IoT & Internet of things                       & UFK & Unscented Kalman filter \\   
         IVM & Import vector machine                    & UGV & Unmanned ground vehicle \\  
         KF & Kalman filter                             & UN & United Nations \\   
         kNN & k-nearest neighbors                      & URLLC & Ultra-reliable low latency communications \\     
         KPI & Key performance indicator                & UWB & Ultra wide band \\       
         LED & Light-emitting diode                     & VLC & Visible light communications \\ 
         LiDaR & Light detection and ranging            & VLP & Visible light positioning \\ 
         LOS & Line of sight                            & VNF & Virtual network function \\
         LSTM & Long short-term memory                  & WiFi & Wireless fidelity \\ 
         MDS & Multidimensional scaling                 & WSN & Wireless sensor network \\
         MEM & Micro-electro-mechanical                 & XLM & Extended learning machine \\  
         MIMO & Multiple input multiple output          & XR & Extended reality \\ 
         ML & Machine learning                          & & \\                   
         mmWave & Milimeter wave                        & & \\                 
         MND & Malicious node detection                 & & \\ 
    \end{tabular}
\end{center}
\end{table*}

\section{Applications of localization in 6G wireless systems} \label{S:6G_aplications}
\begin{figure*}[!ht]
    \centering\includegraphics[width=1\textwidth]{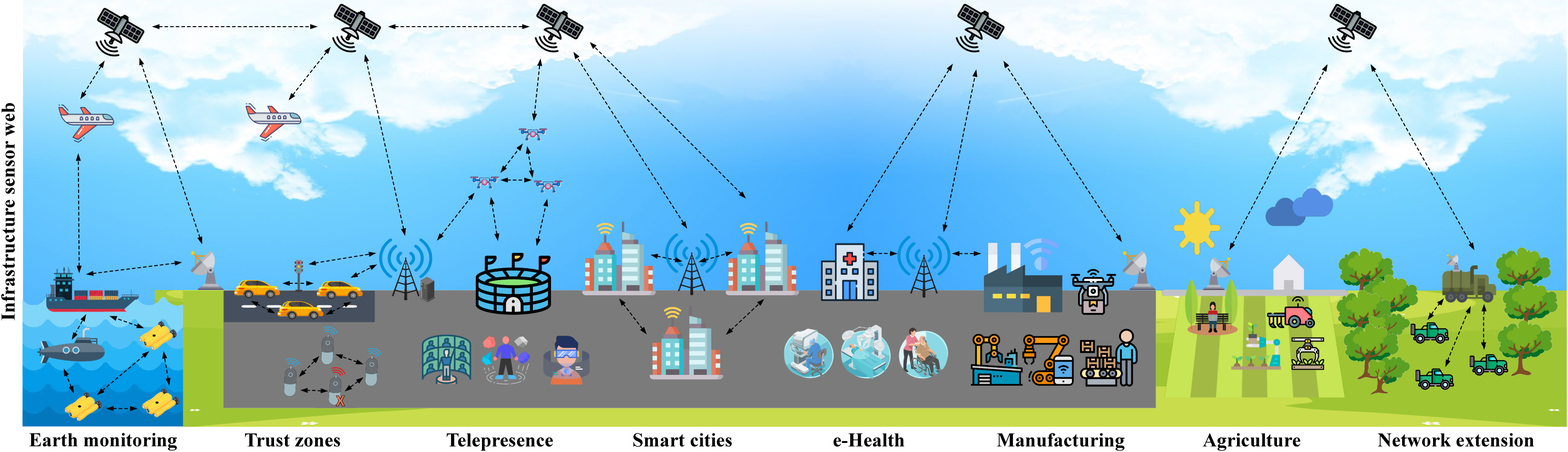}
    \caption{Localization-assisted applications}
    \label{fig:applications}
\end{figure*}
Localization has been completely transformed by the advancement of multiple technologies, such as radars, sensors, AI, mmWave, THz, VLC, and RISs. As illustrated in~Fig.~\ref{fig:applications}. It is expected to play a dominant role in the next generation of communication networks through its incorporation to a variety of applications. Some indicative examples include pinpointing a device's outdoor location with cm- to mm-level precision, accurate indoor localization with reliable interpretation of 3D data in addition to the 2D data that are presently available, or the incorporation of roll, pitch, and yaw into the localization process. To achieve these, 6G promises to use a combination of radio and VLC localization to achieve cm- to mm-level accuracy whether the user, object, or device is inside and/or~outside.

A  detailed analysis of specific applications of localization in 6G wireless systems is presented in the following sections. In more detail, the first two sections distinguish between the various applications of localization based on the nature of the utilized hardware. In~Section~\ref{Ss:Sustainable_development}, the applications that fall under the sustainable development umbrella are analyzed, while in~Section~\ref{Ss:Immersive_telepresence}, the ones that will lead to achieving immersive telepresence are detailed. Section~\ref{Ss:Trust_zones} highlights promising 6G applications that involve setting up local trust zones for human and machine localization. Finally, Sections~\ref{Ss:Massive_twinning} and~\ref{Ss:Robots} describe the applications of localization in next-generation massive twinning and robotics settings, respectively.

\subsection{Sustainable development}\label{Ss:Sustainable_development}
In order to achieve the United Nations' (UN) sustainable development goals (SDGs) and lessen the environmental impact, the next generation of radio access technology (NG-RAT) promises to incorporate measures to protect the environment and empower people to improve their own lives and the world around them. NG-RAT is planned to serve clusters of use cases such as autonomous supply chain, earth monitoring, e-health, and personalized context-aware networks in order to meet the major UN SDG. All these use cases have a common need for localization, which in turn have a set of criteria on major KPIs that must be satisfied.

\subsubsection{Autonomous supply chain}\label{Sss:Autonomous_supply_chain}
With the ability to monitor products from the time they are created all the way through their final stages of use and disposal, an autonomous supply chain may guarantee material and energy savings as well as increased efficiency. Specifically, 6G-enabled micro-tags may be used to (i) promote a flexible and adaptable supply chain; (ii) track and record the whereabouts and condition of all products; (iii) allow a decentralised, worldwide asset monitoring and management system; as well as (iv) link all products to the internet~\cite{Nemati2022, Viswanathan2020}. For example, consequences for the supply chain sector may be seen in the novel approaches to final-mile delivery made possible by the widespread use of computing and mobile communication technologies. Up to half of total distribution expenses are attributable to the last mile of the supply chain's journey. Packages spend much of their time in transit dead-zone locations, and very little real-time information is available, as they make their way from a site of storage, like a warehouse to a point of sale or straight to the premises of ultimate costumers. Lastly, a conventional supply chain is unable to meet the demand for instantaneous delivery for online purchases. Therefore, the management of supply chains has placed a premium on innovative technical solutions. Thus, the ongoing maturation of technologies based on 6G wireless systems and fog computing will unquestionably provide a plethora of chances for the next wave of technological innovation~\cite{Baccarelli2019}. The latter may provide a wide variety of services due to the use of VNFs, SDN, AI, and resource orchestration that works with both physical and virtual resources. This will pave the way for an ideal answer in terms of logistics management, protocol compatibility, and network architecture for the next-generation delivery systems, which rely on UAVs and robot delivery workers.

\subsubsection{Earth monitoring}\label{Sss:Earth_monitoring}
The UN SDGs include global real-time monitoring. In order to achieve this, a large number of low-cost biodegradable devices are expected to be deployed to gauge climate, collect meteorological data, and track species distributions around the globe. Autonomous and robust placement is required to map the gathered measurements from the massive deployment of sensors to the site where the data will be used. The so-called ``Destination earth'' allows for near real-time observation and monitoring of environmentally important factors including solute concentrations, biodiversity, and weather~\cite{DestinationEarth}. Environmentally friendly and pervasive sensors linked together in terrestrial wide area networks and wireless mesh networks make it possible to conduct such monitoring tasks. This encourages data collecting even in underdeveloped regions with limited access to modern technology. Cases include conserving endangered ecosystems through human activities like preventing the spread of disease via adjusting climate models and/or monitoring environmental conditions~\cite{Hofmann2022}. As a result, the objectives of future 6G network should include the development of earth monitoring systems. The network is strengthened by the incorporation of tiny sensors that work together to build a ``sensor infrastructure web'', which incorporates a worldwide sensor network for tracking the state of the planet's ecosystem~\cite{Torres2003}. The sustainability of human existence on earth is specifically threatened by climate change. To this end, the climate deal aims for net positive benefits on the climate budget. Due to the exponential growth of data, future networks will need to significantly lower their energy footprint. To this end, green localization approaches are perfectly suited for energy-saving usage in this context.

\subsubsection{e-Health}\label{Sss:e-Health}
By using a low-cost 6G networks, it is possible to increase access to healthcare for a huge fraction of the world's population. Remote healthcare is widely used all over the world to serve people who live in remote areas or who are unable to travel for medical treatment. The use of telemedicine and other forms of remote health monitoring make this service possible~\cite{Pramanik2019}. Telemedicine is the practise of treating, diagnosing, and evaluating patients remotely. Using AR/VR/XR, a patient and doctor may have a conversation that is almost identical to the one that would take place during a face-to-face visit. Telemedicine now makes use of high-quality video teleconferencing; but, with the fast development of holographic technology, is anticipated to be employed in the near future. Wireless body area networks and wearable devices are only two examples of the Internet-of-Medical-Things technologies that have begun to see widespread deployment for remote health monitoring~\cite{Wei2020}. The same way that real-time monitoring systems benefit society as a whole by tracking emergency situations, future 6G-enabled e-Health systems will be able to provide immediate treatment. However, there is a significant barrier to this aim caused by the slow rate of acceptance of new technologies. Also, those who require constant monitoring, like diabetics, may rest easy es that their dosages can be modified with the help of digital medication. Finally, sample collection for testing may be facilitated by deploying drones to out-of-the-way locations. Strict requirements on localization precision in the range of $0.1-0.3$ m in the vertical and $0.1-0.5$ m in the horizontal plane need to be satisfied to guarantee a valid sample collection by deploying drones to improve remote health care. For tracking purposes, especially during landing and sample collection/delivery operations, a high level of availability of up to 99.99\% is required.

\subsubsection{Personalized context-aware networks}\label{Sss:Context_aware}
6G is envisioned to allow contextual interpretations in both communication and localization protocols. Humans are able to comprehend and understand their activities because of the organizational, cultural, and physical circumstances in which they participate. This is accomplished by a combined comprehension of time and activity components of an action along with the quantitative characterisation of entities, like things, places, or people. The ability to recognize human and technological contexts is equally crucial for the realization of 6G networks~\cite{DeLima2021}.

Contextual understanding will be useful in several 6G applications. By communicating only when it is most efficient applications may significantly reduce their overall energy footprint. Insightful prediction of data transmission based on context paves the way for increased throughput on demand. To achieve hyper-personalization, context-awareness may be utilized to intelligently relocate personalization algorithms and sensor data to portions of the network, where storage and computing are quick and practical. For instance, in healthcare contexts, it is crucial to fuse sensor data in order to discover patterns and deviations, and this is made possible by intelligent storage and distributed processing capabilities~\cite{Dey2001,Dourish2001}. Additionally, multi-modal localization is made possible by context-awareness, allowing mobile devices to transition between channels and communication technologies based on their present location. With multi-modality, devices may choose between national and private technologies, reducing power consumption and improving service quality. This diversity in communication styles necessitates the use of context-aware localization approaches~\cite{Nust2019}.

The following developments in awareness of context is expected to benefit from 6G networks. Context awareness techniques will be able to keep track of history and interpret temporal data, allowing them to spot patterns and outliers, which, in turn, can improve anomaly detection protocols. For instance, this is essential in hyper-personalization, where instantaneous changes in a user's status are tracked. In addition, there are unique privacy risks with sophisticated temporal context detection algorithms since they integrate private information, i.e., human physiological data, with publicly available information, such a as environmental variables. Intelligent security solutions are needed to store and analyze this data in various portions of the infrastructure. Finally, a lot of standardization of context parameters is needed for distributed context awareness. It is hard to construct networks that can transform the context into assets for applications without defined interpretation rules for context parameters. Standardizing context operating rules and parameters to achieve a fair data economy that permits the combinations and flow of multiple data sources, which are regulated by a diverse ecosystem, poses a significant challenge to the implementation and adoption of such context-aware features.

\subsection{Immersive telepresence}\label{Ss:Immersive_telepresence}
Immersive telepresence enables users to interact with a distant environment, including devices, people, and/or objects, utilising all of their senses. The foundations of a fully-cyberphysical world would be strengthened by its capacity to provide unobstructed near-real-time holographic experience and mixed reality. This enhanced capacity is expected to be useful for mixed reality work, as well as fully immersive entertainment, such as sports, gaming, live events, etc. All of the use cases under the telepresence umbrella have one thing in common: the need for localization in order to enable features such as gesture and motion recognition and XR.

\subsubsection{Gesture and motion recognition}\label{Sss:Gesture_recognition}
6G's higher frequency range provides improved precision and resolution, allowing for the capturing of finer gestures and actions. With the advent of advanced computer technology and pervasive AI methods~\cite{Behravan2022}, a new age of gesture and activity recognition has begun. Localization and sensing in the near future will not be confined to a single space, but rather a massive and complex setting. This allows for widespread gesture recording and activity detection without the need for any additional devices. Each base station is envisioned to serve as a sensor that works together with user devices to get a more complete picture of their environment, leading to a significant increase in performance. Moreover, the base station enables a notably increased range that was not attainable through only user devices, as well as the dissemination of the gathered data to the cloud or to neighbouring buildings.

A smart environment could use high classification accuracy to include features like behaviour detection, gesture and motion recognition, intruder detection, and more~\cite{Tan2021b}. Likewise, patients in a smart hospital may benefit from automated patient monitoring thanks to the facility's medical rehabilitation system. In such environments, improper gestures or actions will trigger notifications, greatly enhancing security as well as the provided services (i.e., rehabilitation). More accurate localization capabilities and high classification precision would be necessary for certain superior contactless activities like playing virtual piano. Assuming that the black keys have an average width of $10-11$ mm, the expected accuracy for determining the finger's vertical distance from the board and its horizontal location in relation to the virtual keys is $3$ mm. In addition, the recognition probability should be higher than $99\%$ for a pleasant tune.

\subsubsection{XR}\label{Sss:XR}
Across the whole AR/VR spectrum, XR will provide several applications for 6G. Due to their limitation to provide extremely low latencies for data rate-heavy applications, upcoming 5G systems still fall short of offering a fully immersive XR experience that captures all sensory inputs. Creating an immersive AR/VR/XR experience calls for a collaborative design that takes into account not just technical but also perceptual considerations. Engineering methods have to account for both the maximum and minimum boundaries of human perception. To achieve this goal, traditional QoS, i.e., rate and latency, and QoE, like mean opinion score, inputs need to be combined with physical elements from the users themselves, necessitating a new notion of quality-of-physical-experience (QoPE) metric. QoPE may be influenced by a number of actions, such as non-verbal cues, physical processes, and mental processing. In~\cite{Kasgari2019}, for instance, we demonstrate that under the URLLC regime, the human brain may not be able to differentiate between various latency metrics. Meanwhile, as shown in~\cite{Park2019} the sensory input, particularly touch and sight, is critical for optimising use of available resources. Furthermore, with XR, it is possible to test out ideas before committing to a final design, and it allows corporations to provide customers services that are tailored to their own needs and circumstances. To offer an XR solution that is both appropriate and realistic for the user's current location, 6G localization of the device providing the service is essential. Thus, an XR user would require accurate localization in order for the device to receive the correct data. A shopper, for instance, may employ XR-enhanced advertising while strolling through a retail centre. As another use case, XR can superimpose a computer-generated image or simulation onto the real world. However, much greater precision is needed in this particular setting.

\subsection{Local trust zones}\label{Ss:Trust_zones}
Local trust zones related use cases highlight the need of a framework that calls for the NG-RAT to enable adaptable architectures that let enterprises and services to keep their most sensitive data inside their own networks. Some examples that fall within this broad category are: (i) infrastructure-less network extensions; (ii) sensor infrastructure web; (iii) precision healthcare; and (iv) security and privacy. Localization is required in all the aforementioned use cases clusters.

\subsubsection{Infrastructure-less network extensions}\label{Sss:Network_extensions}
This use case requires extending the network coverage temporarily, particularly if the number of devices to be served are located near the network's edge. Use cases are almost ubiquitous. Normal instances take place in rural locations, where coverage is restricted and performance levels are inconsistent. In production settings, bigger populations of the same vendor's modules, computers, and vehicles may be linked together via infrastructure-free networks that serve as an underlay for the infrastructure network and require the adaptive and fast solution of geometric consensus problems~\cite{Mostaani2021}. For example, autonomous vehicle platoons have been used for agricultural harvesting. Each vehicle should keep its distance from the others to prevent accidents and work together to find their way around the operational area. It is also important to guarantee the vehicles' safe operation even when on the borders of the coverage area. The application will need an update rate proportional to the current pace of the cars' movement. Cooperating localization is a term that is occasionally used to describe these kinds of applications. Finally, most of these use cases include either a population of nodes that make up the infrastructure-free network and may move independently to enable direct device-to-device connection, or a collection of sensor nodes travelling along strongly linked trajectories.

\subsubsection{Sensor infrastructure web}\label{Sss:Sensor_infrastructure_web}
The sensor infrastructure web was introduced by NASA~\cite{Delin2001}. It consists of smart infrastructure that enables coordinated localization and sensing. It is made up of disparate nodes equipped with sensors that not only gather information, but also communicate with each other to modify their behaviour~\cite{Xiao2018c}. Tens of millions of mobile, airborne, in-situ, and space-borne sensors have been deployed in diverse domains and for varied purposes recently due to the growth of IoT, smart cities, and the Global Earth Observation System of Systems (GEOSS). Earth's urban areas have never been detected so dynamically and in-depth in the past. To accommodate this scenario, 6G is expected to provide localization data to devices with almost zero sensing capabilities, such as guiding a vehicle through an unfamiliar area that has inadequate on-board sensors to comprehend its surroundings. Accurate device localization is required to convey relevant sensing information to devices lacking sensing capabilities, which is necessary to enable such use cases. For example, in disaster management application, the sensor infrastructure web is able to use a combination of in-situ and space sensors in order to collect data for volcano hazard monitoring, or use orbital space sensors, like satellites, to locate flooding indicators or wildfire hot-spots~\cite{Zhang2018c}.

\subsubsection{Precision healthcare}\label{Sss:Precision_healthcare}
Precision medicine has emerged as a mean to treat and prevent illness in an individualised fashion. Dispensing medication with the help of wireless nano-scale robots that move within the human body's soft tissues is fundamental to the individualised therapy, which differs from the standard one-size-fits-all method. To maintain tabs on a patient's health, 6G connection will be used to gather sensor-based data. Utilizing the massive coverage area provided by 6G, patient monitoring and tracking may be utilised to ensure the safety of patients regardless of the nature of their illness or their current health status. The anticipated 6G localization capacity benefits all these distinct applications by providing enhanced precision healthcare, such as medical equipment placement on the body, small-scale robots, patient monitoring and tracking, as well as telesurgery.

From the aforementioned applications, telesurgery is a unique use case of cutting-edge precision healthcare that employs a surgeon and an expert operator stationed in a different location to operate one or more robotic arms. Data for receiving feedback and controlling the surgical tools must be reliably sent between distant locations. High-data rate URLLCs that are expected to be made possible by 6G technology will allow robotic arms to mirror the surgeon's natural hand motions and offer haptic input, boosting the surgeon's senses and dexterity. More advanced technologies, known as cooperative surgical systems, let doctors and machines work together to achieve better results during procedures. The tissues and surgical equipment both need to be precisely localised for successful remote surgery.

\subsubsection{Nanoscale localization}
Recent breakthroughs in nanotechnology are bringing light to nanometer-size devices that will allow a multitude of innovative applications. Among the many fascinating fields where nanotechnology is predicted to reap benefits is in-body treatment. Antibiotic resistance might be detected at the molecular level, drugs could be delivered with pinpoint accuracy, and neurosurgery could be performed on specific areas of the brain. In order to implement such applications, a network of nanodevices will need to be created within the body, which will then travel through the circulatory system, respond in response to orders at specific areas, and report their findings to a more robust body area network (BAN)~\cite{Dressler2015}. Nanodevices in this context are assumed to have dimensions similar to those of red blood cells in order to prevent clotting as a result of their entry into the circulation. Because of their little size, these nanonodes are likely to rely only on energy harvesting for their operation~\cite{Piro2015}. These nanonodes are expected to be passively flowing, i.e. without the potential of mechanical guiding toward the specified point, due to their limited energy and small form factors.

Based on the aforementioned, localizing the nanonodes is of great importance for regulating their operation. To issue control commands, for example, two way communication between the nanonodes and the outside world is essential. THz communications constitutes a viable technology enabler in such scenarios as the primary prerequisite for in-body nanonodes, namely small transceiver form-factors, is only possible with transmission at these frequencies. The THz band, however, is entangled with various limitations, such as high scattering and spreading losses that limit the propagation distance of THz waves. The main challenges of THz-enabled nanoscale localization include:(i) maintaining low complexity and energy demands of nanonodes; (ii) overcoming the substantial attenuation of in-body THz propagation; and (iii) allowing network scalability~\cite{Lemic2021}.

Research on nanoscale networks and their localization approaches is currently limited and constrained in tackling a subset of these challenges. Specifically, a THz-enabled network architecture was proposed in~\cite{Lemic2022}, which enables fine-grained localization of the energy-harvesting in-body nanonodes, as well as their two-way communication with the outside world. This approach elisted software-defined metamaterials and location-aware and wake-up radio-based wireless nanocommunication paradigms to enable the novel energy-harvesting capabilities for in-body nanonetworks. The authors suggested that the proposed design can take advantage of the large number of nanonodes, surpassing the limitations of limited range of THz in-body propagation and severely confined nanonodes. Moreover, in~\cite{Gomez2022a, Gomez2022b}, the authors explore the potential for flowing nanosensors in the blood flow to identify and localize and report anomalies in the human body. This work focuses on the identification of quorum sensing molecules and evaluate their performance. The authors use a Markov chain model to simulate the nanosensors' motion through the blood arteries, and apply ML models to predict their trajectory. The results validate the detection and localization capabilities of the investigated method across a variety of body areas, demonstrating their utility in identifying vascular anomalies.

\subsubsection{Security and privacy}\label{Sss:Security}
Addressing the security elements of localization is a growing area of study alongside enhancing localization precision, sampling efficiency, and resilience against intermittent connection loss. Jamming and spoofing are two versions of intentional interference that degrades or prevents the calculation of a GNSS signal~\cite{Elango2022}. In this context, protecting personal privacy entails blocking the functioning of GNSS receivers that may track the user's whereabouts and communicate such information to other parties. In case jamming signals are not correctly identified and their consequences are not mitigated, the employment of jammers leads in reduced positioning accuracy or complete loss of GNSS signals, which may result in catastrophic damage. Intentional intervention is also done by political activists, cybercriminals, and foreign nations to disrupt the networks of others. Spoofing, on the other hand, refers to the malicious broadcast of false GNSS-like signals that trick the receiver. The hardest part of spoofing is its identification. Fortunately, the attacker must be able to produce authentic GNSS signals, including data, modulation, and timing, maintain temporal synchronisation near to authentic GNSS time, and modify the signal power levels to correspond to those of the authentic signals in order to be able to conduct a realistic spoofing assault.

There are a plethora of localization approaches that take security into account. SotA security-aware techniques frequently have large computing needs, prioritize structure-based assaults, assume static networks and even make bold assumptions that may not hold true under real-world circumstances. A common problem is that when harmful conduct from anchor or unknown nodes is expected, either no explicit attack model is supplied or detection of such hostile nodes is all that is defined, with no further treatment of the discovered nodes being specified. For instance, exploiting the noise characteristics induced by external distance assaults, the authors of~\cite{Xie2021} suggested a lightweight secure ToA-based localization technique that enables protection against impersonation attacks in WSNs. 

Clustering and consistency evaluation-based malicious node detection (MND) and its improved and secure variant, enhanced MND (EMND), constitute a solution to identify the anomalous node clusters by applying spatial clustering based on density~\cite{Liu2019}. Afterwards, they utilize a sequential probability ratio test to find the bad actors in the networks. Simulations reveal that the proposed algorithms are more accurate and efficient than current SotA approaches. One downside of the proposed algorithms is that there is no way for maliciously labeled anchor nodes to redeem themselves and reclaim their trusted status in the network. In addition, in~\cite{Bochem2022}, the authors present a extremely computationally-lightweight while yet retaining the benefits of range-free techniques including low cost and complexity of deployment as well as increased resilience against assaults of secure localization systems. Due to the distributed nature of this approach, it relies on no external resources or single points of failure, thus being able to identify and filter out malicious or malfunctioning anchor nodes, improving localization precision. 

Finally, intruders might exploiting security flaws to harass the users, break into their homes, steal their identities, and more. Privacy concerns about the impending proliferation of THz communications on smartphones and wearables have been voiced several time in the past~\cite{Lohan2017,Sarieddeen2020}. THz-based remote sensing and see-through imaging may be used by an attacker or malicious equipment, which might compromise users' privacy. Also, radar-like localization systems are being developed using 4G and 5G signals and developments in full-duplex communications~\cite{Barneto2019,Barneto2019b}, with the potential for expansion beyond 5G. Radio frequency fingerprinting~\cite{Balakrishnan2019,Soltanieh2020} may identify user devices even without providing a device identifier, highlighting the potential privacy risks of combining high-resolution photography with machine learning algorithms.

\subsection{Massive twinning}\label{Ss:Massive_twinning}
Future RAN aims to bridge the gap between the physical and digital worlds, making it possible to create a digital counterpart of anything in the actual world~\cite{Yuan2022}. The ability to create a trustworthy, efficient, and effective digital duplicate of any physical object will open up hitherto inconceivable avenues of exploration, such as (i) more effective use of resources has a positive effect on the long-term viability of agriculture; (ii) to facilitate a fully smart city by developing 4D spatio-temporal interactive maps; and (iii) tracking the operations at every corner of a smart factory at once by constructing a digital map of the space.

The development of a digital twin (DT) may fundamentally benefit from data about events occurring in the real world at a certain time and place. 6G-connected tags may be affixed to real-world items to generate a digital duplicate, with the added benefit of gathering sensory data that can be paired with time instances to show exactly when the data was gathered~\cite{Yuan2022}. There are, however, difficulties in correctly localising these devices, particularly when they are installed in GPS-denied locations. 6G WANs are being constructed to handle the communication demands of DTs; these networks might be used for localization as well, solving this issue~\cite{Uusitalo2021}. In addition, localization and detection of items that are not linked to the digital world via 6G tags or other 6G-enabled devices is required in certain use cases. Radio-signal-based sensing may be used in this context to gather data about passive items, such as their position and the time they were present. Adding data on inactive things is a useful supplement to the DT, which typically only includes digital copies of network-accessible physical items.

\subsubsection{Agriculture}\label{Sss:Agriculture}
Numerous advancements have been made in the study and use of robotic solutions for the agricultural industry, and fresh contributions are anticipated in the near future~\cite{Skvortsov2018}. Farmers are becoming more aware of its influence on agriculture, which has increased the need for autonomous machinery in this field. A wide range of duties, including application of fertilisers, watering, harvesting, planting, and more, are being carried out by robots~\cite{Roldan2018}. The design and development of technological concepts that enable robots to travel safely across various settings is crucial in this situation. The fundamental necessity imposed by these advancements is to localise the robots in various agricultural settings. Using Global Navigation Satellite System (GNSS) is the most typical technology used so far~\cite{Guo2018c}. The use of GNSS is questionable in many agricultural areas due to signal blocking and multi-reflection of satellite signals~\cite{Santos2020}. Therefore, it is crucial to investigate and create intelligent systems that calculate the robot's localization using a variety of sensor modalities and data sources. The most advanced method for doing this is simultaneous localization and mapping (SLAM)~\cite{Cadena2016}. This method involves mapping the immediate area while also predicting the status of a robot. The robot model often includes the robot's position as well as, in certain situations, its velocity, calibration factors, and sensor offsets. The map is a 3D depiction of the agents that the robot's onboard sensors have detected and are used as a guide throughout the localization process. Creating a map is often crucial to provide details about the surroundings. Additionally, maps may be updated and reused by the robots each time they go across the environment. Finally, visual odometry is one of the most popular alternatives to SLAM, which is capable of predicting the motion of the on-board camera using just image data as input~\cite{Zhan2020}.

\subsubsection{Smart cities}\label{Sss:Smart_cities} 
To improve the quality of life in a city, a comprehensive smart city must give a wealth of data, such as a digital reproduction of the city with real-time data on control utilities, pollution maps, traffic, etc~\cite{Ulusar2020}. An interactive 4D map is a key component of an immersive smart city, since it can be used to better organise and design essential infrastructure and services, like transportation, waste management, plumbing, electrical wiring, and more~\cite{Brincat2019}. Sensors installed to track, monitor, and update status of these services must be localised for this kind of dynamic map to be possible. Moreover, units may be stationed by the side of the road to aid in communication, while simultaneously operating as traffic monitors in various sections of the city.

Radio signals received by a user at a certain position are impacted by the features of their immediate physical surroundings in a predictable manner~\cite{Gope2018}. For instance, because of the existence of urban infrastructure, such as buildings, a user's device in the street level may pick up a signal that has travelled by way of several reflections. The presence of trees near the target causes a reflection of radio signals. By inferring the sort of topography at the user's position, we can better map the terrain around the cell towers. Finally, due to the virtual depiction of each piece of infrastructure, huge digital twining may be of great use in the smart buildings of the future. To do this, you must determine the precise location of every heater, lamp, light switch, and other controlled fixture in the facility. Not only will it be put to use in the daily running of the building, but its intelligence and automation will also allow for a significant decrease in the cost of the building's commissioning.

\subsubsection{Manufacturing}\label{Sss:Manufacturing}
As we go towards Industry 5.0, machines and robots will become more self-sufficient, with only little human oversight~\cite{Chaccour2022}. As a result, cutting-edge control mechanisms are essential for high-precision production processes. In particular, connection density of $10^7\text{/km}^2$, latency on the scale of hundreds of microseconds, and data speeds on the order of Tbps are required for the fast development industrial systems as well as the automation of their operations~\cite{Giordani2020}. THz networks are a natural solution for achieving such high data speeds. However, there are several obstacles that need to be addressed when using THz networks for Industry 5.0 applications, such as precision-driven control mechanisms, dense coverage, and zero-latency. Moreover, the positioning capabilities of 6G networks have the greatest potential to advance industrial use cases~\cite{Mogyorosi2022}. Many different industries, from logistics and manufacturing to mining and transportation, will benefit from using position data to improve and automate their operations~\cite{Lu2018}. Both the (mobile) terminals and robots as well as the whole network will benefit greatly from location data in factory automation and industrial control. The former can better allocate and regulate resources and boost processing efficiency, while the latter can better direct the movement of the terminals. 

\subsection{Robot coordination and interaction}\label{Ss:Robots} 
Local pose tracking and global localization are two basic types of the localization challenge for mobile robots. Once the robot's starting stance has been determined, the local pose tracking problem may focus on maintaining that pose over time. In contrast, the global localization issue requires the robot to independently localise itself and minimise the uncertainties in its posture predictions. A mobile robot's first posture estimate at startup is required for relocating the robot in the event of a pose tracking failure, autonomously navigating, and more. Moreover, numerous localization strategies have been investigated over the last several decades, but probabilistic methods using Bayes filters, Monte Carlo particle filters, Markov particle filter, and extended Kalman filter are now the most popular and well-proven options~\cite{Zhang2019f}. There are two main uses for localization with respect to robots: (i) interaction with humans in order to accomplish goals in the everyday life and industrial settings, and (ii) localization and mapping of the environment, where robots may determine where they are located in space and how they are oriented in relation to one another as well as their surroundings.

\subsubsection{Interactions}\label{Sss:Interactions}
It is anticipated that 6G wireless systems will offer the means through which a technological basis for human-robot collaboration may be laid. As a result, collaborative robots (cobots) will be used extensively in both the domestic and commercial settings. Moreover, cobots will become more important in the home and consumer sectors, evolving beyond the robotic household helpers that they are today. Cobots will also be used in the manufacturing sector to facilitate flexible production of a wide variety of items, including those with a high level of personalization and customization. Accurate 3D localization of cobots that supports high update rates is essential for their efficient operation, as are capabilities to perceive the environment, with the former in particular allowing for synchronised, uninterrupted, and smooth collaboration. Cobots are expected to use this data as input to plan and conduct actions, either alone or in swarms in order to achieve their goal. Additionally, the cobots use a symbiotic autonomy technique, where the robots are self-aware of their perceptual, motor, and cognitive limits and actively seek human assistance for tasks like object manipulation. The cobots' success in the 1,000-km challenge demonstrates their localization system's stability and accuracy during long-term deployments~\cite{Biswas2016}, despite the existence of environmental fluctuations. To achieve this, a monitoring script running on the robots would follow the status of the tasks' execution and contact researchers in the occasions when the robots required help. As a result, over ninety percent of deployments of robots did not need human involvement to reset the robot localization.

\subsubsection{Mapping}\label{Sss:Mapping}
When the GPS signal is poor or unavailable, a mobile robot may use SLAM to create a consistent map of its surroundings and simultaneously calculate its position~\cite{Durrant2006}. In the realm of autonomous navigation and localization, SLAM is widely recognized as a crucial challenge~\cite{Ding2022}. The autonomous capability of mobile robots in unfamiliar surroundings is greatly enhanced by the solution of the robust long-term and real-time SLAM issue, which focuses on two primary components: localization and mapping~\cite{Chong2015,Fuentes2015}. Through accurate and reliable localization, the robot's present location in space is determined. Simultaneous mapping without prior knowledge of the robot's position unifies the fragmented understanding of the surroundings into a coherent whole~\cite{Khairuddin2015}. Autonomous mobile robots equipped with powerful SLAM algorithms have played an important role in a wide variety of application scenes, including exploration in hostile environments such as aerial space, underground mining, rough terrain, and underwater surveillance~\cite{Erfani2019}.

Since its introduction in the 1980s, SLAM has evolved significantly over the course of more than three decades~\cite{Smith1988}. There has been a shift from filter-based to optimization-based algorithms and from single to multi thread methods. Sonar has been replaced by 2D/3D light detection and ranging (Lidar) systems and monocular, stereo, RGB-D, time of flight (ToF), and other cameras. There are now three primary SLAM methods in use: Visual SLAM, multi-Sensor Fusion SLAM, and Lidar SLAM. Visual SLAM estimates its posture with the use of prior knowledge of multi-frame pictures and multi-view geometry, and then utilizes this estimate together with depth information derived from the accumulation of pose changes. Most notably, RGBD cameras can get depth data without any further processing. To complete localization, the SLAM algorithm in Lidar SLAM matches and compares Point Cloud at various periods to determine the distance traveled by the robot as it moves relative to the surroundings and as its attitude changes. Due to the limitations of individual 2D/3D Lidar or cameras, multi-Sensor Fusion SLAM has gained popularity in recent years. The most common sensors for fusion algorithms include Lidar, cameras, IMUs, wheel odometers, and Global Navigation Satellite Systems (GNSS).

SLAM technology is in high demand for use in a wide variety of applications, including but not limited to unmanned aerial vehicle (UAVs)~\cite{Demim2016}, unmanned ground vehicle (UGVs)~\cite{Gupte2012}, indoor autonomous mobile robots~\cite{Faragher2012}, and VR/AR hardware~\cite{Newcombe2011}. The vast majority of AR systems need pre-existing knowledge about their surroundings in order to function. In contrast, the advancement of SLAM algorithms and hardware has allowed sensors to rebuild free-form interior scenes without the need for real-time map initialization. Since GNSS signals cannot completely penetrate buildings, it is more challenging to place robots inside, and conventional localization relies on community landmarks and signage. With SLAM and inertia measurement units working together, we can solve this issue~\cite{Faragher2012}. For UGVs and UAVs, SLAM is also a crucial tool. Specifically, Sensor Fusion SLAM is quickly replacing other methods for building the map concurrently, positioning unmanned vehicles, and implementing autonomous navigation in an uncharted area~\cite{Demim2016,Gupte2012}. All things considered, SLAM has been heavily used across many different disciplines and has promising future applications.

\section{Localization use cases}\label{S:Use cases}
\begin{figure*}[!ht]
    \centering\includegraphics[width=1\textwidth]{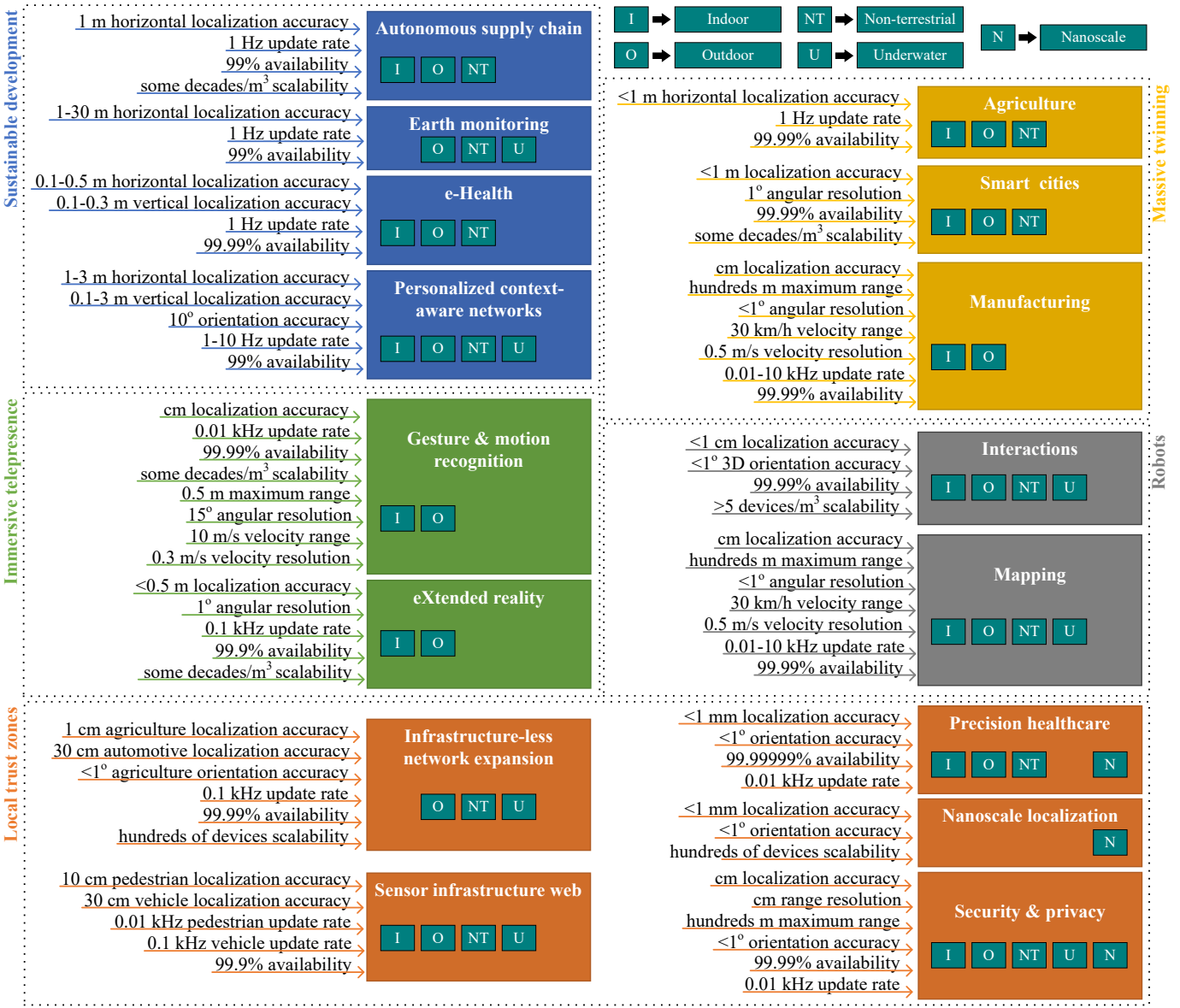}
    \caption{Localization use cases}
    \label{fig:use_cases}
\end{figure*}
In this section, the use cases of future wireless systems are described based on the  localization requirements and applications mentioned in Section~\ref{S:6G_aplications}. To this end, Fig.~\ref{fig:use_cases}, which is given at the top of the next page, illustrates each application alongside its corresponding KPIs and related use cases. 

\subsection{Indoor}\label{Ss:Indoor}
Services such as indoor localization have emerged as a consequence of the rapid use of wireless devices in recent years. Obtaining the position of a person or device within a building is referred to as ``indoor localization''. Extensive studies on indoor device localization have been conducted over the last decade, mostly in industrial settings and for wireless sensor networks and robotics. Since the widespread adoption of smartphones and wearable devices with wireless communication capabilities, the localization and tracking of such devices has become synonymous with the localization and tracking of the corresponding users, enabling a vast ecosystem of related applications and services. The ability to precisely pinpoint the location of a user or device has several practical applications in the fields of autonomous supply chain, e-Health, precision healthcare, personalized context-aware networks, XR, gesture and motion recognition, sensor infrastructure web, security and privacy, agriculture, smart cities, manufacturing, as well as robotic interactions and mapping.

\subsection{Outdoor}\label{Ss:Outdoor}
Accurate positioning data is essential for many outdoor applications, such as navigating in uncharted territory and monitoring the location of company vehicles. However, in urban canyons with high-rise buildings or subterranean parking lots, where building walls obstruct the GNSS reference signal, GNSS coverage is limited. Current mmWave cellular networks and upcoming sub-THz cellular networks offer good prospects for localization where GNSS fails to fulfill the application's KPIs. At mmWave frequencies, adjustable channel bandwidths of up to $400$ MHz are possible in 5G and phased arrays with hundreds of antenna components and narrow half power beamwidths are commercially available. Future 6G wireless systems will migrate to frequencies greater than 100 GHz, when channel allocations lasting several gigahertz are possible. Strategies that take the estimated condition of the wireless propagation channel into account have been the focus of research in wireless systems' localization. Channel impulse response, propagation delay, AoA, and RSS are only some of the propagation parameters that are measured or estimated by networks at one or more locations, usually base stations and user equipment. Moreover, user-location fingerprinting has also been interpreted geometrically, compared to existing databases, or used to locate the user on maps. Finally, the outdoor localization use case includes all applications discussed in the present contribution.

\subsection{Non-terrestrial}\label{Ss:Non_terrestrial}
Non-terrestrial localization has become ubiquitous as a result of the fast advancement of satellite localization technology, making formerly inconvenient tasks easier for the general public. When it comes to non-terrestrial localization, GNSS is the most applied solution and is made up of constellations of artificial satellites in geosynchronous orbit, transmitting real-time position and timing information. During the time of writing, four major not-terrestrial systems are in use: (i) Europe's Galileo, (ii) the United States' GPS, (iii) China's Beidou, and (iv) Russia's GLONASS. These systems provide of $1-10$ m accuracy and low device autonomy. GPS, which was originally designed for use by the United States Navy, is a radio navigation and positioning system that contains location and time information, while GLONASS, a second-generation military satellite navigation system is technologically superior to GPS in its capacity to withstand interference, yet it lacks GPS's pinpoint single-point precision. Moreover, BDS is a Chinese-created satellite navigation system with worldwide coverage and GALILEO, which uses two ground control centres and 30 satellites, is more widespread in Europe. Based on these, non-terrestrial localization applications of future wireless systems include autonomous supply chain, earth monitoring, e-Health, personalized context-aware networks, agriculture, smart cities, robotics-enabled mapping, infrastructure-less network expansion, sensor infrastructure web, precision healthcare, as well as security and privacy.

\subsection{Underwater}\label{Ss:Underwated}
In recent years, underwater wireless systems have applied localization techniques for many different purposes. Each implementation is crucial in its own field, while some of them can advance ocean exploration to accommodate a range of underwater applications, such as underwater surveillance, natural disasters alert systems, oceanographic data collection, assisted navigation, etc. For instance, sensors may evaluate specific characteristics, such as base intensity and mooring tension, for offshore engineering applications to monitor the structural condition of the mooring environment. Therefore, the basic applications of underwater localization wireless systems include earth monitoring, sensor infrastructure web, personalized context-aware networks, mapping, security and privacy, as well as infrastructure-less network expansion. 

\subsection{Nanoscale}\label{Ss:Nanoscale}
THz nanonetworks are expected to enable a wide variety of applications, many of which will need localization or even tracking of the nanonodes. As a result of the nanoscale nature of the nodes, it is essential that localization and tracking capabilities work under very tight energy budgets while yet maintaining a high degree of precision due to the small sizes of the nanonodes. Furthermore, because of the limited range of THz band nano-communication, multi-hopping may be required for localization, which has the disadvantage of increasing localization error with each hop. Few efforts have been made towards localization of THz-operating nanonodes, with the most notable ones using range and hop counting to make educated guesses about where all nanonodes are positioned in a given region. In the former, the locations of the two examined nanonodes are calculated by counting the number of hops between them, while the latter assumes that all nanonodes are clustered together to cut down on overhead and energy dissipation. One limitation of these techniques is that, as the number of hops rises, their accuracy will inevitably decrease due to the spread of localization errors. The main KPIs of nanoscale localization are low energy consumption and high precision, while the related open challenges include modeling of such systems in non-free space propagation, hardware imperfections, and frequency- and angle-dependent response of the nanonodes. 

\section{Technology enablers for 6G wireless systems} \label{S:Technology_enablers}
This section aims to shine light into the key technology enablers of localization in future 6G networks. Specifically, Section~\ref{Ss:mmWave_THz_localization} presents the current SotA of mmWave and THz localization techniques. Next, in Section~\ref{Ss:Hardware}, the main hardware advancements related to localization approaches are investigated, while Sections~\ref{Ss:Beamforming} and~\ref{Ss:RIS} illustrates the role of beamforming and RIS in the localization paradigm of future 6G networks. The impact of AI and channel charting approaches are presented in Sections~\ref{Ss:AI} and~\ref{Ss:Charting}, respectively. Section~\ref{Ss:Radars} discusses radar-based technologies that enable localization applications, while Section~\ref{Ss:Sensors} sensor-based ones. Finally, visible light-based localization methods are analyzed in~Section~\ref{Ss:visible_light_positioning}. 

\subsection{mmWave/THz localization}\label{Ss:mmWave_THz_localization}
With the explosion of cellular services in recent years comes a surge in demand for higher data transfer speeds~\cite{Stratidakis2019}. Using the vast amount of unstandardized bandwidth available in the mmWave and THz range, mmWave and THz wireless systems hold great potential for meeting the data rate need. However, substantial channel attenuations hinder communications at these frequencies~\cite{Kokkoniemi2018}. mmWave and THz systems counteract these higher losses by using pencil-beams made from massive antenna arrays at both the transmitter and the receiver side~\cite{Boulogeorgos2018,Boulogeorgos2018b}. As a result, mmWave and THz links are highly directional. In order to ensure high-reliability and avoid defects, the base station must know and monitor the orientation of the user equipment (UE). Reduced beamwidth necessitates a much higher cost for current localization and tracking methods. Because of this, they are not suitable for use in mmWave and THz systems. Because of this, it is necessary to work on developing efficient localization methods for THz systems.

Due to the complementarity between mmWave and THz systems, the need arises to compare the benefits and drawbacks of these techniques in various diverse localization scenarios especially since THz are likely to be utilized as an extension of mmWave systems~\cite{Kouzayha2021}. We anticipate bigger array sizes, smaller footprints, wider bandwidths, and higher frequencies as we move towards 6G networks. Thus, new possibilities are enabled both in the sense of larger arrays with the same dimensions as well as antennas with constraint sizes. Moreover, increased availability of bandwidth in conjunction with higher frequencies result in robust delay estimation, lower path loss, and decreased multipath.

The aforementioned changes will impact a plethora of features of future networks. For instance, new synchronization problems emerge and hardware needs to keep up with novel mmWave and THz-enabled system designs. Another already reported problem is the requirement for increased peak to average power ratio (PAPR) compared to 5G systems, which may render the classic orthogonal frequency division multiplexing (OFDM) inappropriate~\cite{Tarboush2022}. Another important aspect that needs to be taken into account in the development of next generation localization algorithms is the impact of near-field circumstances along with the beam-split effect (BSE) that can greatly influence the geometry of MIMO channel localization models. Finally, the high path loss plagued mmWave and THz channels can greatly influence the energy-efficiency of the overall system and thus, they must be accounted during the design of future networks. All in all, the increased performance anticipated by mmWave and THz networks comes hand in hand with important difficulties in complexity, overhead, area coverage, and hardware architectures that this affects. 

THz localization is already the subject of active study, with researchers delving into simulation environments, localization methods, and system architectures. For instance, a high level of estimated accuracy and reduced deafness in 2D settings are sought for with the proposal of cooperation-aided localization procedures~\cite{Stratidakis2019}. Also, Kalman filtering-based indoor localization approaches have been proven to offer great advantages for time-variant channel modeling~\cite{Nie2017}. Another example is the angle of arrival (AoA) estimation by a forward-backward algorithm that offers improved localization and human motion characterization~\cite{Peng2016}, while a near-field model with massive antenna arrays has been examined for the purpose of using the curvature of arrival (COA) as a sixth degree of freedom (DOF) in determining the location of the source~\cite{Guerra2021}. Furthermore, by adopting a beam zooming mechanism combined with delay-phase precoding for mmWave and THz beam tracking, it is possible to monitor numerous users and blockers using a single RF chain with significantly reduced beam training overhead~\cite{Tan2021}. Finally, in addition to geometry-based strategies, DL-based strategies show great potential for 3D mmWave and THz indoor localization by enhancing the localization accuracy by approximately 60\%~\cite{Fan2020}. Each of the aforementioned localization efforts addresses a unique issue associated with mmWave and THz localization, namely near-field effects, tracking, improper alignment, and BSE. Despite these advances, important gaps in our understanding still need to be filled. It is yet unclear how signals in the mmWave- and THz-band might boost localization accuracy in both active and passive approaches~\cite{Chen2022}.

\subsection{Infrastructure}\label{Ss:Hardware}
Most of the work being done to close the gap between current and potential applications of THz technology is still at the device level. Several competing technologies have recently illustrated great potential for realizing THz components. The most established technology is electronic silicon-based devices that have already been used in mmWave systems~\cite{Hillger2020}. For instance, complementary metal oxide semiconductor (CMOS) and its variants are characterized by manufacturing compatibility and small size~\cite{Nikpaik2017}. However, such approaches are plagued by limited maximum available power gain frequency and power handling capacity. To overcome the power constraints of CMOS and increase coverage area, MIMO-based systems with high array gains combined with beamforming techniques provide a promising alternative. 

In contrast to electronic devices, photonic THz systems, such as uni-traveling carrier photodiodes, quantum cascade lasers, photoconductive antennas, and optical downconversion systems, provide both data speeds and larger carrier frequencies (over 300 GHz)~\cite{Nagatsuma2016}. The downside of such approaches is the larger form factor that poses power and integration constraints. To this end, fully integrated hybrid electronic-photonic systems have been proposed. Such systems use photonic transmitters and electronic receivers and come with increased requiremetns for synchronization between transmitters and receivers~\cite{Sengupta2018}. Furthermore, photonics can provide extensive network coverage in the base station level, but the aforementioned constraints of photonic devices make it difficult to implement MIMO approaches in the user device level.

Graphene and other novel plasmonic materials are being explored for manufacturing THz devices due to their capacity to provide solutions with great reconfigurability~\cite{Hafez2018}. In plasmonics, antenna arrays may be made considerably smaller and more versatile since they are characterized by substantially lower wavelengths~\cite{Singh2020}. Plasmonic solutions are advantageous for adaptable system designs and on-site reconfiguration due to their small size and frequency-interleaving features. Unlike conventional transceivers, plasmonic ones do not need upconversion or downconversion to function at THz frequencies, where the creation of energy-efficient brief pulses is especially advantageous~\cite{Zakrajsek2017}. The downside of such devices is that due to the power output constraint, they have a short range making them impractical for localization.

\subsection{Beamforming}\label{Ss:Beamforming}
Data transmission towards a specific direction in order to increase the SNR and consequently the throughput (beamforming) is a fundamental requirement in high-frequency wireless systems~\cite{DeLima2021}. For mmWave and microwaves, 3D beamforming is a useful tool for mitigating the detrimental effects of high pathloss and for improving SNR. In order to localize and sense objects or people, the gathered channel response can be analyzed for any geographical information about the relationship with them. The channel estimation, which includes the angular and delay domain characteristics that are necessary for localization algorithms, is a prerequisite for beamspace processing~\cite{Ahmed2018}. Especially in challenging NLOS and high-mobility conditions, channel estimation is indespensible~\cite{Miao2019}. Beams are dynamically controlled based on the channel estimate of the AoA and angle of departure (AoD) to find and track mobile users in a dynamic environment~\cite{Huang2019}.

Accurate and rapid AoA or AoD estimate of the primary multipath can suffice forin scenarios of mobile users who connect with the BS through a LOS path or strong NLOS route. Uplink AoA estimates from LOS connections may be used to directly infer the position of users, whereas AoA estimates from NLOS links can be used for the scatterer's location. NLOS links originate from multipaths of the signal emitted by an active user and subsequently arrive at the BS. For the latter, NLOS cases, accurate estimates of both AoD and AoA are required to pinpoint the user's position.

Beamspace processing for locating mobile users in NLOS and LOS settings is a significant challenge; ``device-free'' localization~\footnote{Passive localization methods are capable of localizing targets without requiring them to carry a device or tag.} presents an even greater one~\cite{Denis2019}. Here, additional environment indicators are required so that the target(s)' spatial characterisation may be differentiated from that of the background objects. This is made more challenging as the number of targets increase~\cite{Miao2019} and the complexity and computational requirements for identifying said targets increase exponentially~\cite{Hong2016}. Very fine spatial resolution may be achieved by using a pencil-shaped beam operating in the mmWave and THz spectrum. Beamspace channels' combined angular-delay characteristics are used to detect and locate passive objects. In this example, the instantaneous beamspace may be compared with a reference to detect a passive target, and with a prior sample (on-line or real-time) to track a moving target. Target identification relies on learning algorithms that can tell the difference between the varying angular-delay profiles of various targets. High-precision localization and detection of scattering objects may be enabled by the use of specialized beamspace processing signals and the strategic placement of monostatic/multistatic MAS.

\subsection{RIS}\label{Ss:RIS}
RISs' role is ever increasing in the communication paradigm where their main goal is to increase the data rate via achieving higher SNR. However, when it comes to localization approaches, RISs are utlized on the one hand to augment the information gathered by the system by introducing curvature of arrival (COA), while on the other hand they provide geometrical diversity by acting as a passive anchor. Specifically, the requirement of localization algorithms for numerous base stations (passive anchors) in order to provide an accurate estimate can be further enhance by the introduction of RISs. This notion has been investigated in the recent published literature, which takes advantage of the phenomenon of multipath to to pinpoint the origin of high-frequency signals~\cite{Wen2020}. Although multipath signals can be destructive for communications, researchers have been able to take advantage of them and enable localization and mapping with a single base station (BS)~\cite{Ge2020}. It becomes obvious that it is possible to reduce infrastructure cost by a great deal by exploiting RISs that act as passive reconfigurable BSs with low energy consumption~\cite{Wymeersch2020, Strinati2021}. 

In mmwave/THz bands high accuracy localization necessitates the need to densify the network with a massive number of anchors in order to counteract blockages. This is of course due to the fact that such bands are more susceptible to blockages compared with their sub-6 GHz counterparts. However, densifying the network only with active nodes, such as small cells and active relays that host power amplifiers, will induce an immense energy consumption due to the power amplifier, which is the most power consuming component of active nodes. Hence, RISs can also notably reduce the network energy consumption for high-accuracy localization. Morever, active nodes inevitably need to be connected to the power grid. There are places though where the power grid cannot reach to for reasons of heavy planning/maintenance costs and aesthetics. In such scenarios it's much more preferable to use RISs that may not even need to be connected to the power grid if they are supplied by alternative sources of energy such as solar radiation through solar panels attached to the RISs in outdoor scenarios~\cite{Tishchenko2023}. This would of course be feasible due to the much lower consumption of RISs compared with active nodes. Thus, RISs are expected to enable novel localization approaches that will revolutionize future high-frequency (mmWave, THz, optical wireless) networks.

Despite the aforementioned advances, breakthroughs made in material research focused on mmWave, THz, and optical-operating materials could provide immense advances in RIS-enabled localization~\cite{Khalid2016}. mmWave, THz, and optical-operating metasurfaces that offer improved reconfigurability and sensing precision are essential for both mmWave, THz, and optical localization and communications due to the strong blockage and directionality exhibited in the mmWave, THz, and optical frequency band. Recent research shows that mmWave, THz, and optical-operating hypersurfaces that use a stack of physical and virtual components to provide lens effects and individualized reflections can offer significant performance increase~\cite{Liaskos2018}. In addition, similar small-scale metasurfaces have been proven capable of efficient mmWave, THz, and optical signal steering at a broad variety of angles, polarization conversion, as well as generation of orbital angular momentum~\cite{Fu2020}. On another note, despite their slow clock speeds and capacitance leakage, CMOS-enabled RISs offer improved energy efficiency and seamless integration to existing systems~\cite{Venkatesh2020}. Similarly, graphene-based metasurfaces offer a promising alternative for the construction of RISs, due to their low-complexity biasing circuits and low power requirements~\cite{Nie2019}. Finally, multiple researcher have investigated the plausibility of micro-electro-mechanical (MEM) devices as a candidate echnology for the realization of RISs. However, several issues must be overcome, such as relatively large footprints, control signals, and switching rates, before they can be a viable option.

\subsection{AI}\label{Ss:AI}
As we enter the data-rich 6G era, AI is envisioned to play an increasingly vital role. The design and development intelligent systems and beings that can think, plan, and make optimum decisions based on probabilistic foundations, sometimes in unpredictable settings, is a vast field~\cite{Ali2020}. To train models beyond explicitly coded rules, most modern AI systems rely on ML, which enables data-driven interdisciplinary techniques. Such data-driven algorithms will be essential to 6G systems and beyond, opening up new possibilities for not just wireless communication but also sophisticated localization methods using mmWave, THz, and optical frequencies.

Fingerprinting and the use of classification and regression techniques were the primary focus of ML approaches to localization~\cite{Zafari2019,Hsieh2019}. We anticipate that ML will be used more frequently in data-rich and complex localization applications; especially, for the GNSS poor indoor and urban outdoor channel conditions, as traditional signal processing and mathematical techniques are not sufficient to address complex issues where we have a large number of noisy and multi-modal observations, as well as the non-linear characteristics of the signal. Instead, we may mimic the behavior of the system—sensor noise and all—using AI techniques. In addition, in many circumstances, pattern recognition and predictive models based on ML approaches are needed to achieve high-level sensing and localization from perhaps high-dimensional low-level raw observations, such as CSI in massive MIMO systems. Consequently, mapping systems and localization will rely increasingly heavily on the application of statistical AI approaches to model complicated radio signal properties and fuse multiple complementary but noisy sensors. These are expected to be supplemented by hybrid signal propagation models that combine conventional physics-based models with sequential Bayesian models and data-driven learning methodologies. Furthermore, it will be hard to manually construct the mathematical model, due to the enormous complexity of the predictive function of mapping low-level observations to high-level goal ideas. In order to build approaches with unprecedented precision and adaptation capabilities for 6G localization, ML offers an alternative framework that is based on cutting-edge DL and probabilistic methods and learns from data by optimizing or inferring previously unknown parameters~\cite{Ciftler2020}.

A number of strategies to improve conventional localization are currently available in the current AI toolkit, based on methods, like probabilistic learning and reasoning, as well as DNNs. However, a lot of these tried-and-true methods have their limits, because of how data-hungry they are and how much processing power and labeled training data they need. Novel inference approaches and hierarchical models need to be created and cleverly coupled in order to learn from restricted, arbitrarily structured, and noisy data. Several current methods might be improved in order to reduce the expense of gathering labeled training data. In order to improve a supervised solution, semi-supervised learning blends a large number of unlabeled with a limited number of labeled ones. Additionally, next-generation wireless systems' localization features are expected to be more autonomous and time-evolving, necessitating the use of adaptive ML algorithms. For example, cooperative localization based on RL~\cite{Peng2019} and crowdsourcing based on FL~\cite{Ciftler2020} might assist in overcoming the difficulties posed by adaptable settings and limited data. These issues must be resolved in order to properly use AI and ML approaches as an enabler for the highly dynamic, large-scale localization of the future.

\subsection{Channel charting}\label{Ss:Charting}
Applying traditional unsupervised ML-based dimensionality reduction techniques to the field of CSI is what channel charting is all about~\cite{DeLima2021}. Users may be discovered and tracked on a chart that is generated automatically based on a big collection of CSI samples collected in a certain setting. The lack of supervision of channel charting makes it useful in cases, when there is insufficient data to construct an accurate geometric representation of the user's position and surroundings~\cite{Studer2018}. Although it is not possible to directly correlate a position on the chart with a user's physical location, it provides a pseudo-location within the cell that is stable over time and users. Therefore, the need for specific measurements of the labeled CSI datasets that are necessary for fingerprinting methods are eliminated. Several network features, including mmWave beam tracking and association, grouping in device-to-device (D2D) situations, resource management, predictive rate and cell-to-cell handoffs may benefit from keeping tabs on the pseudo-location.

Pseudo-locations have certain advantages over an actual position, but they cannot completely substitute it. Initially, channel charting's lack of supervision means that there is no need for prior information (i.e., geometrical/geographical) in order to achieve self-configuration, which aids in deploying emergency and temporary networks. Another aspect of pseudo-locations is privacy, due to the fact that it enables contact tracking without the knowledge of the user's precise location.

To begin with, channel mapping was created for scenarios with intense scattering phenomena and where major RF propagation characteristics are anticipated to remain stable (such as massive MIMO). In order to use this method LOS application found at high frequencies, it will likely need to be expanded to jointly handle signals from numerous transmission locations. Directly related to the ML foundations of channel charting are other unresolved concerns, such as the feasibility of implementing lifelong learning and the appropriate feature design for CSI signals.

\subsection{Radars}\label{Ss:Radars}
Extensive research and development into techniques for device localization has been performed in recent years~\cite{Kocur2018}. It has been shown that sub 5 GHz functioning ultra-wideband (UWB) radars are useful in such scenarios. Most non-metallic materials exhibit minimal attenuation of electromagnetic radiation in this frequency region, making it possible to detect devices that otherwise would be obstructed. Moreover, UWB radars, by taking advantage of the ultra-wide bandwidth, may provide great precision in target localization thanks to their superior range resolution~\cite{Sachs2013}. These features make UWB radars useful for a wide variety of applications, including but not limited to: beamforming in cellular communications; vehicle localization; monitoring critical infrastructure; indoor patient monitoring; search and rescue operations; etc.

Beamforming improvements of the narrow beams are essential for terahertz (THz) and  millimeter wave (mmWave) communications systems to generate enough receive signal strength. High beam training overhead, however, makes it difficult for these systems to serve highly mobile applications like extended reality (XR), drone, or vehicle communications~\cite{Alkhateeb2018}. Beam selection is notoriously sensitive to changes in transmitter and receiver locations as well as environmental geometry and features. This suggests that learning more about the environment and the positions of the transmitters and receivers can assist solve the beam selection problem. Using inexpensive radar sensors, such those used for automotive applications, is a quick and easy approach to get this understanding~\cite{Ginsburg2018}. This paper's objective is to solve the beam selection issue using radar sensory data, and to do so, it presents the first AI-based demonstration of radar-based beam prediction under realistic vehicular communication scenarios.

Frequency-modulated continuous wave (FMCW) radars, which can function in a wide range of environmental conditions, have recently seen widespread use in self-driving automobiles and autonomous robotics. Therefore, it is an intriguing open topic whether or not these radars can be employed for reliable simultaneous localization and mapping (SLAM) in large-scale areas under harsh weather conditions. Moreover, due to these advances in radar technologies other breakthroughs are anticipated in radar odometry~\cite{Aldera2019,Park2020}, estimating mobility~\cite{Ramesh2021}, landmarks extraction~\cite{Cen2018,Cen2019}, and more~\cite{Suaftescu2020,Tang2020}.

\subsection{Sensors}\label{Ss:Sensors}
A continuously increasing number of IoT devices with a wide variety of sensors are being deployed to cover a vast array of use cases, ranging between smart cities~\cite{Kanev2017} to industrial applications~\cite{Liu2021} and decentralized initiatives to protect the planet~\cite{Sommer2016,Tonisson2021}. Information on the specific location of the measurements taken is crucial for making sense of the results. Due to the fact that many applications depend on mobile sensors, the network needs to be able to dynamically detect their positions.

When working on a wireless sensor network (WSN) protocol or application, localization is crucial. For instance, in environmental monitoring applications, it is essential to know the precise location from where readings are taken in order to conduct proper statistical and scientific analysis; without this location information, the acquired data are essentially worthless. Node locations are also crucial to the development of such methods. In target identification and geographical routing, node locations are required~\cite{Tomic2016,Xing2014}. Techniques like mobility management~\cite{Xiao2018}, mobile sensor deployment~\cite{Liao2015}, and topology mapping may also benefit from their utilization~\cite{Gunathillake2018}. There has been a recent uptick in the usage of sensor networks for indoor localization~\cite{Ebner2016,Ebner2017}.

In order to take use of location-based services like navigation, tracking, and monitoring, users may determine their location anywhere in the world using the Global Navigation Satellite System (GNSS)~\cite{Saeed2019}. The GNSS is a collection of navigational satellite systems from various countries, including the Europe's Global Navigation Satellite System (GALILEO), United States' Global Positioning System (GPS), China's BeiDou, and Russia's Global Navigation Satellite System (GLONASS). To locate a user inside or in a hostile environment, GPS and GNSS fall short of expectations. Indoor environments are more difficult and intricate than their outside counterparts for a number of reasons. Multi-path propagation error is caused by interference from several sources, including but not limited to ceilings, walls, equipment, and humans. As a result of obstacles unique to the inside environment, designing accurate indoor positioning systems for next-generation wireless networks is a formidable task.

Utilizing a satellite-based navigation system is the standard method used for this. However, there are a few problems that come along with using GPS~\cite{Bochem2022}. For example, they are only effective in areas with good reception for satellite signals, which limits their use in certain outdoor settings and prevents them from being used inside. Limiting the number of nodes that have GPS sensors may help with the first two issues. These nodes then serve as anchor or seed nodes, guiding subsequent nodes to their correct locations. Static anchors with predetermined positions are often used in place of movable anchor nodes outfitted with GPS receivers.

\subsection{VLP}\label{Ss:visible_light_positioning}
A lot of work has been done to assess the efficacy of various tracking approaches suited for mobility control in indoor settings from the perspective of RF technology~\cite{Anastou2021}. None of the aforementioned improvements may be used in applications, where privacy is paramount, such hospitals and petrol stations, since those establishments must instead use isolated VLC systems. As a further drawback, their precision is limited by electromagnetic interference and the significant multipath phenomena brought on by many reflections in barriers, like walls, furniture, and moving persons~\cite{Luo2017}.

New light-based ways for indoor location without the restrictions of the RF-based traditional techniques have been published as a reaction to the recent developments in semiconductor-based lighting systems that have made LEDs the dominating choice for illumination~\cite{Chen2021}. Since the optical energy is concentrated on the LOS connection, visible light positioning (VLP) devices are more resistant to the multipath propagation effect. Consequently, the accuracy of the AoA improves. However, in scenarios with LOS blockage, novel reflecting surfaces, such as optical RISs and mirrors, can be used to reinstate high quality communications~\cite{Abdelhady2020}. Furthermore, ceiling-mounted LEDs have a very high density since their main job is to provide illumination. Therefore, it is anticipated that VLP systems would increase accuracy and fulfill the tougher standards of indoor settings, while RF-based systems are restricted in sensitive regions~\cite{Zhuang2018}.

The main VLP techniques utilize multiple image and light sensors, as well as positioning-optimization methods, for instance filtering, spring and normalization. The proximity approach used in~\cite{Luo2017} predicts the relative position based on a known base station or access point and is proven to have a huge variable in proximity, hence it does not match localization needs in many interior environments. Other efforts used multiple VLP triangulation methodologies, such as received signal strength (RSS), time difference of arrival (TDoA), time of arrival (ToA), and AoA, to boost position estimate accuracy without added hardware. Although these procedures are accurate, they are computationally intensive.

Most of the aforementioned contributions agree that numerous access mechanisms have been suggested to separate and analyze transmitted mixed signals, a necessary operation for a VLP system. Time-division multiplexing (TDM) and frequency-division multiplexing (FDM) are two common methods that have been studied in the literature~\cite{Maheepala2020}. Demand for tighter coordination between individual LEDs, however, increases this complexity cost. When the location data is based on light of varying wavelengths, precision is increased since the signals are more easily distinguished. In order to limit the amount of LEDs utilized for localization, some traditional VLP systems use square waveforms whose spectral content include odd-integer harmonic frequencies. However, a sinusoidal waveform has significantly less complicated spectrum information~\cite{Costanzo2019}. This allows for a greater number of LEDs to be used, which in turn greatly improves both localization precision and breadth.

\section{System models}\label{S:System_model}
This section focuses on the system and channel models that lay the foundation of the discussed mmWave/THz and optical localization systems. In more detail, Section~\ref{Ss:THz} discusses the proposed mmWave/THz system model, which includes both LOS and RIS-aided NLOS scenarios, while Section~\ref{Ss:VLC} delves into the VLC-based localization techniques that can be split into LOS and optical RIS-aided NLOS system models.
\begin{figure}
    \centering\includegraphics[width=0.9\columnwidth]{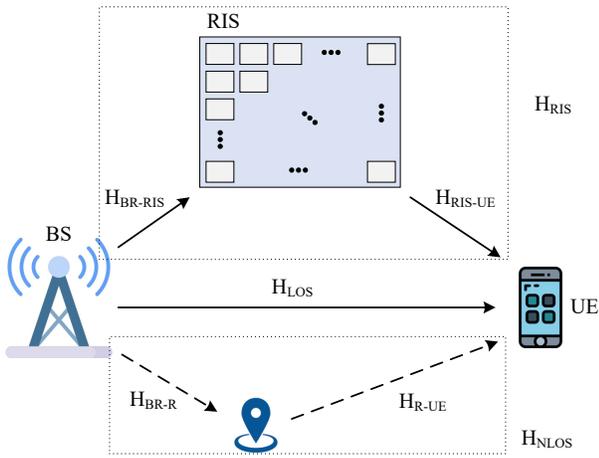}
    \caption{mmWave/THz localization system architecture.}
    \label{fig:THz}
\end{figure}

\subsection{mmWave/THz}\label{Ss:THz}
THz and mmWave techniques have been researched extensively during the past couple of years aiming to reduce the ``THz gap''. From a hardware perspective, small-scale graphene-based THz transceivers have been demonstrated with groundbreaking sensitivity, precision and power. Specifically, SotA photonic- and electronic-enabled transceivers have accomplished unprecedented signal transmission and modulation~\cite{Sengupta2018,Sengupta2019,Kenneth2019}.

In order to properly configure and analyze the performance of a system, a channel model with robust accuracy is required. Channel modelling may make use of either stochastic, deterministic, or a combination of such approaches. Initially, the simplest model, i.e., deterministic, is analyzed for the far-field MIMO channel in the localization system, which assumes a planar waves in contrast to the near-field models that assume spherical waves. As depicted in Fig.~\ref{fig:THz}, in the uplink, we can split the channel matrix, $\textbf{H}$, of the channel into two distinct submatrices as
\begin{align}
    \textbf{H} = \textbf{H}_\mathrm{LOS} + \textbf{H}_{\mathrm{NLOS}} ,
\end{align}
with $\textbf{H}_\mathrm{NLOS}$ and $\textbf{H}_\mathrm{LOS}$ denoting the NLOS and LOS channels, respectively.

\subsubsection{LOS}\label{Sss:THz_LOS}
The LOS channel coefficient is valid in scenarios where the BS and the user equipment (UE) share an unobstructed communication path. In such cases, the channel matrix can be written as in~\cite{Heath2016}
\begin{align}
    \textbf{H}_\mathrm{LOS}= \rho e^{-j 2 \pi\left(f \tau-\nu t\right)} G_{\mathrm{BS}} G_{\mathrm{UE}} \mathbf{a}_{\mathrm{BS}} \mathbf{a}_{\mathrm{UE}}^T ,
\end{align}
where $\mathbf{a_{BS}}$, $\mathbf{a_{UE}}$, $G_{BS}$, and $G_{UE}$ represent the steering vectors and the gains of the BS
and the UE, respectively. Moreover, $\tau$, $\nu$, $f$, and $\rho$ denote the signal delay, Doppler shift, frequency, and path gain, respectively. 

In more detail, the path gain can be expressed as in~\cite{Shahmansoori2017}
\begin{align}
    \rho = \frac{c K_\alpha}{4 \pi f d} ,
\end{align}
with $d$ being the propagation distance and $K_\alpha$ denoting the attenuation coefficient, which is a function of the distance and transmission frequency. It is important to highlight that, in THz systems $K_\alpha$ can be evaluated based on high-resolution transmission molecular absorption caused by water vapor and other gases in the atmosphere, while in mmWave ones it denotes the atmospheric attenuation.

Furthermore, by assuming negligible Doppler effect and identical $K_\alpha$ between the subcarriers, the channel becomes frequency-flat and can be obtained as
\begin{align}
    \textbf{H}_{\mathrm{LOS}}& = c_k \rho e^{-j \xi} e^{-j 2 \pi \Delta f_k \tau} \mathbf{a}_{\mathrm{BS}} \mathbf{a}_{\mathrm{UE}}^T , 
\end{align}
where
\begin{align}
    &\rho = \frac{\lambda_c {K_\alpha} G_{\mathrm{BS}} G_{\mathrm{UE}}}{4 \pi d_{BU}} , \\
    &\xi = 2 \pi f_c \tau,
\end{align}
and 
\begin{align}
    &\tau = \frac{d_{BS-UE}}{c} + B.
\end{align}
Likewise, $B$ and $c$ are the clock synchronization offset and the speed of light, respectively, while $f_k$ and $c_k$ denote the frequency of the $k$-th subcarrier and the ratio between the central and the subcarrier frequency, respectively. Finally, $d_{BU}$ represents the distance between the BS and the UE.

\subsubsection{NLOS}\label{Sss:THz_NLOS}
\underline{Without RIS:} When all LOS paths are obstructed, the transmitted mmWave/THz signal may arrive at the user's device after being reflected on random (i.e., obstacles, walls, metallic structures, etc) or carefully placed (i.e., RISs) objects. Specifically, THz NLOS communication scenarios are characterized by increased losses and sparsity, while scattering and diffraction phenomena are introduced from rough surfaces~\cite{Ma2019b}. Thus, such channels are commonly modelled with ray-tracing and stochastic techniques. On the other hand, mmWave NLOS can be characterized by modeling the stochastic reflection loss and geometrical modeling~\cite{Fascista2019}.  

If we assume M reflectors between the BS and the UE, the channel matrix for each reflector can be expressed as
\begin{align}
    \textbf{H}^{(m)}_{\mathrm{NLOS}}& = c_k \rho^{(m)} e^{-j \xi^{(m)}} e^{-j 2 \pi \Delta f_k \tau^{(m)}} \mathbf{a}_{\mathrm{R}} \mathbf{a}_{\mathrm{UE}}^T ,
    \label{Eq:H_NLOS}
\end{align}
where
\begin{align}
    &\rho^{(m)} = \frac{\lambda_c {K_\alpha}^{(m)}_{(R-UE)} G^{(m)}_{\mathrm{BS}} G^{(m)}_{\mathrm{UE}}}{4 \pi d^{(m)}_{R-UE}} , \\
    &\xi^{(m)} = 2 \pi f_c \tau^{(m)}
\end{align}
and
\begin{align}
    &\tau^{(m)} = \frac{d^{(m)}_{BS-R} + d^{(m)}_{R-UE}}{c} + B.
\end{align}
In~\eqref{Eq:H_NLOS}, $d^{(m)}_{BS-R}$ and $d^{(m)}_{R-UE}$ denote the distances between the BS and the $m$-th reflector as well as between the $m$-th reflector and the UE, respectively. Also, $\mathbf{a}_{\mathrm{R}}$ represents the steering vector of the reflector, while ${K_\alpha}^{(m)}_{(R-UE)}$ is the attenuation coefficient for the RIS-UE channel.

\underline{With RIS:} In the case of RIS-aided systems, the previous analysis is augmented with RIS-reflected signal coefficients. The reflection of the signal on the RIS modifies its phase and amplitude with the channel matrix given by~\cite{Tang2020b}
\begin{align}
    \textbf{H}_{\mathrm{RIS}}& = c_k^2 \rho e^{-j \xi} e^{-j 2 \pi \Delta f_k \tau} \mathbf{a}_{\mathrm{BS}} \mathbf{a}_{\mathrm{RIS}}^T \mathbf{\Omega} \mathbf{a}_{\mathrm{RIS}} \mathbf{a}_{\mathrm{UE}}^T ,
    \label{Eq:HRIS}
\end{align}
where
\begin{align}
    &\rho = \frac{\lambda_c^2 {K_\alpha}_{(BS-RIS)} {K_\alpha}_{(RIS-UE)}}{16 \pi^2 d_{BS-RIS} d_{RIS-UE}} , \\
    &\xi = 2 \pi f_c \tau
\end{align}
and
\begin{align}
    &\tau = \frac{d_{BS-RIS} + d_{RIS-UE}}{c} + B ,
\end{align}
In~\eqref{Eq:HRIS}, ${K_\alpha}_{(BS-RIS)}$ is the attenuation coefficient for the BS-RIS channel, while ${K_\alpha}_{(RIS-UE)}$ is the attenuation coefficient for the RIS-UE channel. In addition, $d_{BS-RIS}$ and $d_{RIS-UE}$ denote the distances between the BS and the RIS as well as between the RIS and the UE, respectively. Finally, $\mathbf{a}_{\mathrm{RIS}}$ represents the steering vector of the RIS.

Finally, the complete NLOS channel matrix can be obtained by summing all the individual reflection components, both random and RIS ones, as
\begin{align}
    \textbf{H}_{\mathrm{NLOS}} = \sum^M_{m=1} \textbf{H}^{(m)}_{\mathrm{NLOS}} + \textbf{H}_{\mathrm{RIS}} .
\end{align}

\subsubsection{Localization geometry}\label{Ss:THz_geometry}
From a localization perspective, the system consists of a base station, a RIS and the end user's device. In order to determine the position of the latter in space, we need to define a global coordinate system defined as 
\begin{align}\mathbf{p} = [x,\,y,\,z]^T.\end{align} 
As discussed in~Section~\ref{Ss:RIS}, the channel contains LOS as well as NLOS paths generated by RIS, reflector and/or diffractor elements.
\begin{figure}
    \centering\includegraphics[width=1\columnwidth]{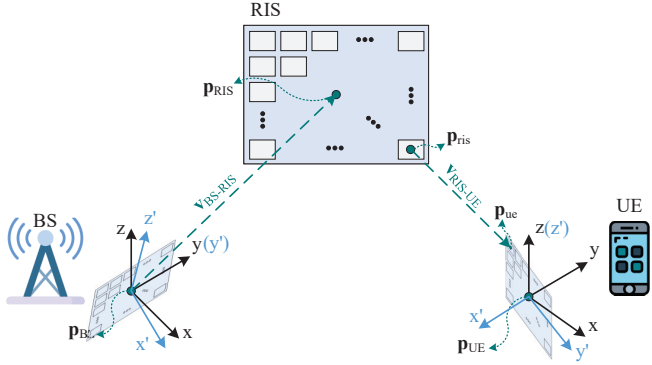}
    \caption{Global vs local coordinate system.}
    \label{fig:Coordinates}
\end{figure}

To simplify the analysis, a local coordinate system needs to be used, which can be translated to the global one by using an Euler angle vector and a xyz rotation sequence as
\begin{align}
    \mathbf{p}_\mathrm{i} = \mathbf{R} \mathbf{\tilde{p}}_\mathrm{i} + \mathbf{p}
\end{align}
with $\mathbf{p}_i$ and $\mathbf{\tilde{p}}_i$ representing the position of the $i$-th element with $i=\mathrm{BS}, \mathrm{UE}, \mathrm{RIS}$ in the global and local coordinate system, respectively, while $\mathbf{R}$ denotes the rotation matrix and $\mathbf{p}$ is the center of the array. The global and local coordinate systems are presented in~Fig.~\ref{fig:Coordinates}.

Furthermore, in the case that the user transmits a signal towards the BS, the direction vector in the global coordinate system can be written as
\begin{align} \label{Eq:v_BS_UE}
    \mathbf{v}_\mathrm{BS-UE} = \frac{\mathbf{p}_\mathrm{UE}-\mathbf{p}_\mathrm{BS}}{d_\mathrm{BS-UE}} ,
\end{align}
where 
\begin{align}
d_\mathrm{BS-UE}=||\mathbf{p}_\mathrm{U}-\mathbf{p}_\mathrm{BS}||
\end{align} 
denotes the distance between the UE and the BS. \eqref{Eq:v_BS_UE} can be translated in the local coordinate system as
\begin{align}
    \mathbf{\tilde{v}}_\mathrm{BS-UE} = \mathbf{R}^{-1} \mathbf{v}_\mathrm{BS-UE} ,
\end{align}

The AoA/AoD pairs can be derived based on the definitions provided for the global and local coordinate systems by using an azimuth, $\phi$, and an elevation, $\theta$, angles. The former is defined as the angle between the y axis and the projection of the direction vector on the xy plane, while the latter is the angle between the direction vector and the xy plane. Due to the fact that the AoA/AoD angles can only be measured in the local coordinate system, the direction vector needs to be rewritten in the local coordinate system as 
\begin{align} \label{Eq:AoA/AoD}
    \mathbf{\tilde{v}}_\mathrm{BU} = f(\mathbf{\tilde{\varphi}}_\mathrm{BU}) = 
    \begin{bmatrix} 
        \mathrm{cos}(\tilde{\phi}_\mathrm{BU}) \mathrm{cos}(\tilde{\theta}_\mathrm{BU}) \\
        \mathrm{sin}(\tilde{\phi}_\mathrm{BU}) \mathrm{cos}(\tilde{\theta}_\mathrm{BU}) \\
        \mathrm{sin}(\tilde{\theta}_\mathrm{BU})
    \end{bmatrix} ,
\end{align}
where $\mathbf{\varphi}$ and $\mathbf{\tilde{\varphi}}$ denote the AoA/AoD angles in the global and local coordinate system, respectively, while $f(\mathbf{\tilde{\varphi}}_\mathrm{BU})$ represents the mapping between the direction vector and the AoA/AoD angles. On the contrary, \eqref{Eq:AoA/AoD} can be expressed in terms of the direction vector as 
\begin{align}
    f(\mathbf{\tilde{\varphi}}_\mathrm{BU}) = 
    \begin{bmatrix} 
        \tilde{\phi}_\mathrm{BU} \\
        \tilde{\theta}_\mathrm{BU}
    \end{bmatrix} = 
    \begin{bmatrix} 
        \mathrm{arctan2}(\mathbf{\tilde{v}}_\mathrm{BU,x},\mathbf{\tilde{v}}_\mathrm{BU,y}) \\
        \mathrm{arcsin}(\mathbf{\tilde{v}}_\mathrm{BU,z})
    \end{bmatrix} ,
\end{align}
with $\mathrm{arctan2}\left(\cdot\right)$ denoting the four-quadrant inverse tangent. This procedure can be used to calculate the global AoA/AoD values by substituting $\mathbf{\tilde{v}}_\mathrm{BU}$ with $\mathbf{v}_\mathrm{BU}$, while the same methodology can be applied to any of the aforementioned channels, either LOS or NLOS.

\subsection{VLP}\label{Ss:VLC}
\begin{figure}
    \centering\includegraphics[width=1\columnwidth]{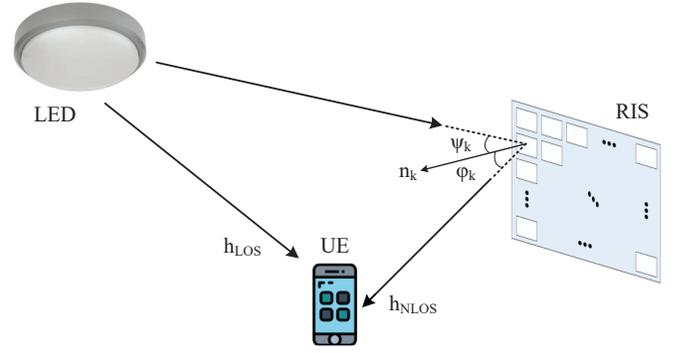}
    \caption{VLC localization system architecture.}
    \label{fig:VLP}
\end{figure}

Visible-light based positioning (VLP) systems are distinct from their radio-based counterparts in a number of ways. As a result, it is important to develop appropriately designed VLP system that incorporate the optical transmission particularities and are both precise and reliable. As depicted in~Fig.~\ref{fig:VLP}, signals in VLP systems travel from an light emmiting diode (LED) source to a photodiode (PD) receiver via a wireless channel. The LED is often categorised as a Lambertian source~\cite{Kahn1997}, meaning it disseminates signals in accordance with Lambert's emission law (i.e., when observing a perfect diffuse reflecting surface, the radiant intensity is proportional to the cosine of the angle between the direction of incoming light and the surface). 
The following analysis, assumes that the channel between the receiver and the transmitter may be obstructed, in which case it enlists an optical RIS to extend the VLP system coverage.

\subsubsection{LOS}\label{Sss:VLC_LOS}
The LOS process may be represented by a Lambertian model, which means that after being reflected off of an object like a wall, the light's intensity drops to such a low level that it may be safely disregarded. Changing the reflection route of the optical connection and achieving beamforming through a multi-mirror reflection link are both possible thanks to RIS based on mirror arrays, as described in~\cite{Qian2021}. In the LOS channel the system model can be expressed as in~\cite{Zhuang2018}
\begin{align}
    y = h \, x + n ,
\end{align}
where $x$ and $y$ represent the transmission and received signal, respectively. Also, $n$ denotes the additive white Gaussian noise (AWGN), while $h$ is the channel gain. In more detail, the LOS channel gain can be written as
\begin{align}
        h^{\mathrm{LOS}} &=\frac{A(m+1)}{2 \pi d^2} \cos ^m\left(\phi\right) \cos \left(\psi\right) T_o G\left(\psi\right),
\end{align}
where 
\begin{align}
        m &= - \frac{1}{2} \log_2 (\cos(\Phi_{1/2})),  
\end{align}        
and
\begin{align}
        G\left(\psi\right) &= \frac{\alpha^2}{\sin^2(\Psi)}, \text{ for } 0\leq \psi \leq \Psi. 
\end{align}
Moreover, $\phi$ and $\psi$ denote the irradiance and incidence angles, respectively. In addition, $G\left(\psi\right)$ is the optical concentrator gain, while $T_o$ represents the optical filter gain. Finally, $m$ is the Lambertian model index, $\Phi_{1/2}$ is the LED's half-power angle, $\Psi$ is the field of view (FOV), $\alpha$ is the PD's refractive index and $A$ is the PD's area.

\subsubsection{NLOS}\label{Sss:VLC_NLOS}
In the NLOS scenario, a RIS that is made up of $\mathrm{N_x} \times \mathrm{N_y}$ identical rectangular mirrors, each of which is $d$ in size. Each mirror is mounted such that its optical centre is in a plane perpendicular to the $y$ axis. To determine the roll $\omega$ and the yaw $\gamma$ angles, which spin clockwise in the positive direction of the $x$ axis and counterclockwise in the positive direction of the $z$ axis, respectively, we need two mirrors placed at right angles to one another.

The RIS-enabled NLOS pathways can be split into two categories: (i) the RIS component is a RX that handles the signals from LED; and (ii) to re-transmit the optical signal collected in the prior stage, the RIS component is assumed to be a point source and its magnitude is multiplied by the reflection coefficient. As higher-order reflections have an unaccounted-for effect on VLC systems, we restrict the following analysis to first-order reflections~\cite{Tang2020c}. Moreover, we assume that the LED's incident light falls on each of the RIS elements that are called meta-atoms (MAs) at its geometric centre as well as that there are $K$ MAs of area $D$. As a result, the channel gain of the $k$-th RIS element can be expressed as in~\cite{Abdelhady2020}
\begin{align}
    \begin{split}
        h_k^{\mathrm{NLOS}}=&\rho_k^\mathrm{RIS} \frac{(m+1) A D}{2 \pi^2 (d_k)^2 (d_k^r)^2} \cos ^m\left(\phi_k\right) \\
        & \times \cos \left(\psi_k^r\right) \cos \left(\psi_k\right) \cos \left(\phi_k^r\right) T_o G\left(\psi_k^r\right) ,
    \end{split}
\end{align}
where $d_k$ and $d_k^r$ represent the distances of the $k$-th RIS element from the LED and the PD, respectively, while, $\rho_k^\mathrm{RIS}$ denotes the reflection coefficient of the $k$-th element. Moreover, if we assume that the receiver, the transmitter and the $k$-th RIS element are located in $(x_r,y_r,z_r)$, $(x,y,z)$, and $x_k,y_k,z_k$, correspondingly, $\phi_k^r$ and $\psi_k$, which are based on the roll and yaw angles of the $k$-th RIS element, can be evaluated~as
\begin{align}
    \begin{split}
        \cos \left(\phi_k^r\right)=&\frac{\left(x_r-x_k\right)}{d_k^r} \sin (\gamma_k) \cos (\omega_k) \\ 
        &+ \frac{\left(y_r-y_k\right)}{d_k^r} \cos (\gamma_k) \cos (\omega_k) \\ 
        &+ \frac{(z_k-z_r)}{d_k^r} \sin (\omega_k) ,
    \end{split}
\end{align}
and 
\begin{align}
    \begin{split}
        \cos \left(\psi_k^i\right)=&\frac{\left(x-x_k\right)}{d_k} \sin \left(\gamma_k\right) \cos \left(\omega_k\right) \\ 
        &+ \frac{\left(y-y_k\right)}{d_k} \cos \left(\gamma_k\right) \cos \left(\omega_k\right) \\ 
        &+\frac{\left(z_k-z\right)}{d_k} \sin \left(\omega_k\right) .
    \end{split}
\end{align}
Finally, the $k$-th element's norm vector can be written as
\begin{align}
    n_k = \left(\sin(\psi_k) \cos(\phi_k^r), cos(\psi_k)cos(\phi_k^r),-sin(\phi_k^r)\right)^T .
\end{align}
As a result, the complete channel gain of the VLP system can be expressed as
\begin{align}
    h = h^{\mathrm{LOS}} + \sum_{k=1}^K h_k^{\mathrm{NLOS}} (\omega_k,\gamma_k) .
\end{align}

\subsubsection{Localization geometry}\label{Ss:VLC_geometry}
To dig a little deeper into the VLP approaches, the most common geometrical approaches can be categorized in to RSS-based, ToA/TDoA-based and AoA-based~\cite{Zhuang2018}. Specifically, the most well known localization RSS-based technique for VLP is trilateration, where, after the RX measures the strength of the transmitted signals, the distance between the RX and each TX can be calculated, allowing for the drawing of concentric circles whose radii correspond to the measured distances. Existing measurement flaws often cause the recorded distance to be just slightly off from the real ones. Thus, almost always some overlap exists between the areas where the circles are drawn. In this case, the trilateration problem transforms into the least-squares estimate~\cite{Zhang2014}, the solution to which is provided as a solution to the system 
\begin{align}
    \left\{
        \begin{array}{l} 
            \left(x-x_1\right)^2+\left(y-y_1\right)^2=d_1^2 \\
            \left(x-x_2\right)^2+\left(y-y_2\right)^2=d_2^2 \\
            \left(x-x_3\right)^2+\left(y-y_3\right)^2=d_3^2
        \end{array}
    \right. .
\end{align}

As far as the ToA/TDoA-based methods are concerned, they also leverage trilateration algorithms since, as light travels at a fixed speed over a given distance, it generates phase delay in the RX. Such systems need at least three TX, i.e., at least three phase differences, with each one having its own unique frequency id. The power of each TX can be expressed as
\begin{align}
    P_i(t) = P_c + P_s cos(2\pi f_i t+ \phi_0) , 
\end{align}
where $P_s$, $P_s$, and $\phi_0$ denote modulated signal power, the continuous signal power, and the initial phase of the signal, respectively. Also, $f_i$ is the modulation frequency of each transmitter ($i=1,2,3$). As a result, the power recieved by the PD can be written as
\begin{align}
    E(t)=K \left| R \sum_{i=1}^3 P_i(t) h_i(t)\right| ,
\end{align}
with $R$, $K$, and $h_i(t)$ represent the PD's responsivity, the constant proportionality, and the channel impulse response, accordingly. The received signal is processed by band pass filters specialized on each modulated frequency and, afterwards, it is passed by a frequency down converter comprised by a mixer and a band pass filter to unify the frequency. Thus, the system is capable to use Hilbert transform to calculate the two phase differences (from three transmitters). However, a third phase difference is required for accurately obtaining the position of the user, which can be obtained by altering the frequency of two out of the three transmitters~\cite{Jung2011}. As a result, the phase differences can be obtained by
\begin{align}
    \left\{
        \begin{array}{l}
            \Delta \varphi_{12}=2 \pi f_1 \frac{d_1-3 d_2}{c_{12}}=\arctan \left(I_{12} / Q_{12}\right) \\
            \Delta \varphi_{13}=2 \pi f_1 \frac{d_1-5 d_3}{c_1}=\arctan \left(I_{13} / Q_{13}\right) \\
            \Delta \varphi_{21}=2 \pi f_1 \frac{d_2-3 d_1}{c}=\arctan \left(I_{21} / Q_{21}\right)
        \end{array}
    \right. ,
\end{align}
with $f_1$ denoting the reference frequency, and
\begin{align}
    \left\{
        \begin{array}{l}
            I_{ab} = E_a(t) \mathrm{Hilb}[E_b(t)] - \mathrm{Hilb}[E_a(t)] E_b(t) \\
            Q_{ab} = E_a(t) E_b(t) + \mathrm{Hilb}[E_a(t)] \mathrm{Hilb}[E_b(t)]
        \end{array}
    \right. ,
\end{align}
where $E$ represents the RSS after the band pass filter and the down conversion, while $\mathrm{Hilb}[\cdot]$ denotes the Hilbert transform. Finally, the user's position can be calculated based on the following distances
\begin{align}
    \left\{
        \begin{array}{l}
            d_1=-\frac{1}{8} \frac{c}{2 \pi f_1}\left[\tan^{-1}\!\left(I_{12} / Q_{12}\right)\!+\!3 \tan^{-1}\!\left(I_{21} / Q_{21}\right)\right] \\
            d_2=\frac{1}{3}\left[d_1-\tan^{-1}\!\left(I_{12} / Q_{12}\right) \frac{c}{2 \pi f_1}\right] \\
            d_3=\frac{1}{5}\left[d_1-\tan^{-1}\!\left(I_{13} / Q_{13}\right) \frac{c}{2 \pi f_1}\right]
        \end{array}
    \right. .
\end{align}

\section{Localization KPIs}\label{S:KPIs}
Naturally, better location and orientation accuracy and precision would be at the top of the list, when developing a localization system. But other goals, including complexity, coverage and mobility, are crucial to the system's success as a whole. This section provides the fundamental KPIs shared between SotA localization systems.

\subsection{Accuracy}\label{Ss:Accuracy}
The accuracy with which a system can estimate its location and orientation is one of the most important KPIs of its localization capability. In more detail, the accuracy of a system is often evaluated by calculating the distance between the real location and the predicted one. It can be often measured as the mean squared error (MSE) or the CDF of measurements bound by a predefined threshold, which can be expressed as
\begin{align}
    \begin{split} \label{Eq:e_MSE}
        e_{\mathrm{MSE}}(\hat{\mathbf{w}})=&\mathbb{E}\left[\|\hat{\mathbf{w}}-\mathbf{w}\|^2\right] \\
        =&\operatorname{tr}(\mathbf{C}(\hat{\mathbf{w}}, \mathbf{w}))+\|\mathbb{E}[\hat{\mathbf{w}}]-\mathbf{w}\|^2 ,
    \end{split}
\end{align}
with $w$, $\mathrm{C}(\hat{\mathrm{w}}, \mathrm{w})$ and $\operatorname{tr}(\cdot)$ denoting the location, the covariance matrix, and the matrix trace, respectively. Also, $\mathbb{E}[\cdot]$ is the expected value. In more detail, $\|\mathbb{E}[\hat{\mathbf{w}}]-\mathbf{w}\|^2$ represents the bias, while  $\operatorname{tr}(\mathbf{C}(\hat{\mathbf{w}}, \mathbf{w}))$ is the variance of the location estimate. Moreover, it is possible to get a worst case estimate of the accuracy by assuming that no bias exists and using the Cramér-Rao bound, which is illustrated in the following equation, provided the noise and signal models are known. 
\begin{align} \label{Eq:C_w}
    \mathbf{C}(\hat{\mathbf{w}}, \mathbf{w}) \succeq \mathbf{I}^{-1}(\mathbf{w}) ,
\end{align}
where $\mathbf{I}^{-1}(\mathbf{w})$ is the inverse of the Fisher matrix~\cite{Guvenc2009}. Finally, after performing the substitution between \eqref{Eq:e_MSE} and~\eqref{Eq:C_w}, based on the available/selected measurement such as ToA, AoA, RSS, and more, the localization lower bound can be extracted~as
\begin{align}
    \begin{split}
        e_{\operatorname{MSE}}(\hat{\mathbf{w}}) &=\mathbb{E}\left[\|\hat{\mathbf{w}}-\mathbf{w}\|^2\right] \\
        & \geq \operatorname{tr}(\mathbf{C}(\hat{\mathbf{w}}, \mathbf{w})) \\&\geq \operatorname{tr}\left(\mathbf{I}^{-1}(\mathbf{w})\right).
    \end{split}
\end{align}
It should be highlighted that various published works utilize the root MSE (RMSE) instead of the MSE, which results in the following lower bound:
\begin{align}
    \begin{split}
        e_{\operatorname{RMSE}}(\hat{\mathbf{w}}) &=\sqrt{\mathbb{E}\left[\|\hat{\mathbf{w}}-\mathbf{w}\|^2\right]} \\
        & \geq \sqrt{\operatorname{tr}(\mathbf{C}(\hat{\mathbf{w}}, \mathbf{w}))} \\ & \geq \sqrt{\operatorname{tr}\left(\mathbf{I}^{-1}(\mathbf{w})\right)}.
    \end{split}
\end{align}
However, the performance of the system may fall short of the Cramér-Rao bound since many realistic estimators are skewed due to signal NLoS propagation and other causes. To this end, tighter constraints have been introduced, such as the Zik-Zakai, Weiss-Weinstein, and Bayesian Cramer Rao bounds\cite{Xiao2022}.

\subsection{Precision}\label{Ss:Precision}
In terms of localization precision, it reveals the variance of the estimated location. The KPI of localization precision was created to statistically characterize the accuracy as it fluctuates throughout numerous localization attempts. One example of precision KPIs is the geometrical dilution of precision (GDoP) that measures the variation in localization errors and can be expressed as 
\begin{align}
    \mathrm{GDoP}=\frac{e_{\mathrm{RMSE}}(\hat{\mathbf{w}})}{e_{\mathrm{RMSE}}(\hat{\mathbf{d}})} ,
\end{align}
with the estimated range and location RMSEs serving as numerator and denominator, correspondingly. Minimizing the GDoP can provide the optimal fixed node selection and placement as it model the precision of the location estimation with regard to the distribution of the fixed nodes in space.

Another KPI of the localization precision is the localization error, which can be measured by two metrics, namely its CDF and outage. The former represents the likelihood that location estimates will be precise to a specified degree and is given by
\begin{align}
    \operatorname{F_e}(e_{\mathrm{th}})=1-\operatorname{P_{out}}(e_{\mathrm{th}}) ,
\end{align}
while the latter expresses the likelihood of the localization error to overcome a predefined value and can be expressed as
\begin{align}
    \operatorname{P_{out}}\left(e_{\mathrm{th}}\right)=\operatorname{Pr}\left\{\|\hat{\mathrm{w}}-\mathrm{w}\| \geq e_{\mathrm{th}}\right\} .
\end{align}
In actual situations, the outage and/or CDF reflects the likelihood of confidence in the estimated location. In the case when two localization methods have equivalent accuracy, the approach that produces the smaller outage and/or larger CDF values is more precise and thus preferred.

\subsection{Latency}\label{Ss:Latency}
\begin{figure}
    \centering\includegraphics[width=0.4\columnwidth]{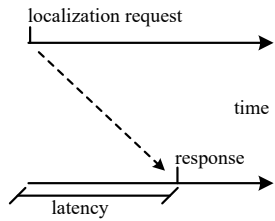}
    \caption{Localization latency.}
    \label{fig:Latency}
\end{figure}
In a localization system, the term latency refers to the delay that occurs between a device making a location request and receiving the response, as shown in Fig.~\ref{fig:Latency}. This is determined by the processing speed of the chosen localization method and the length of the position reference signals utilized. However, latency only describes part of the time-related performance. The pace at which a localization indicator is updated is known as its update rate. This KPI is latency-dependent, in the sense that its maximum value cannot exceed (latency)$^{-1}$, and use-case-dependent.

\subsection{Coverage}\label{Ss:Coverage}
As the distance between each fixed and the mobile network node grows, localization performance often decreases. The system's coverage is the largest geographic region across which reliable localization services may be provided with specified levels of latency, precision, accuracy, and other KPIs. Depending on the available localization infrastructures, the coverage may be broadly broken down into three categories: indoor, local, and global. The probability of coverage, which represents the probability that a specific user achieves an SINR higher than a certain threshold, is the most common KPI of coverage and can be expressed as
\begin{align}
    P_c = P[\gamma \geq \gamma_{\mathrm{th}}] ,
\end{align}
with $\gamma$ and $\gamma_{\mathrm{th}}$ denoting the SINR of the wireless system and the threshold that must be satisfied in order to consider the UE inside the coverage area, respectively.

However, the coverage KPI disregards the system's adaptability when expanding the scope of localization. There is a risk of wireless channel congestion as localization coverage expands, and the localization system will need to execute increased computations and measurements. To this end, the KPI of scalability has been introduced, which quantifies the capacity of a system to accommodate a growing number of end devices. If we consider a wireless networks with $N$ nodes, there is a point after which the residual capacity becomes negative. This point is the maximum number of nodes, $k$, that can be supported by the system. Thus, the scalibility of the wireless network is defined as the number of nodes, after which the residual capacity decreases for every new node added in the system. By setting the residual capacity at $k$ equal to $0$, the scalability can be obtained as in~\cite{Ramanathan2012}
\begin{align}
        C_R(N=k) &= 0,
\end{align}
which leads to
\begin{align}
    \eta\, C_a &= \sum_{i}(1+\Gamma_i)\, L_i\, (1+Y_i),
\end{align}
where $C_a$ is the available capacity, $\eta$ denotes the efficiency, while $L_i$, $\Gamma_i$, and $Y_i$ represent the average load, the contention factor, and the transit factor, respectively.

\subsection{Complexity}\label{Ss:Complexity}
Various aspects of the localization approach, such as the hardware, the signal detection method, and the algorithm's computation requirements, contribute to the overall system's complexity. Depending on the specifics of the intended use case, all or a subset of these types of complexity may be relevant. With regard to hardware, the complexity of hardware realization and deployment is directly related to the complexity of communication and localization algorithms. While, as far as software complexity is concerned, it can be challenging to analytically determine the complexity formula of various localization methods. Thus, it is often considered equal to the computational complexity of the location estimators. Moreover, in the context of localization, complexity is intertwined with accuracy and precision. In addition, there is a crucial relation between complexity and the time elapsed between location updates for a specific mobile node, which is measured by two KPIs, namely update rate and latency.

\subsection{Stability}\label{Ss:Stabililty}
In high-frequency systems with small beamwidths, where pointing errors might cause an outage, loss of tracking or deafness can be a critical issue. Thus, deafness, which is defined as the power leakage brought on by the estimate error, is often used as a metric in the assessment of localization algorithms. Deafness is represented as a percentage and normalised to the half power beamwidth (HPBW) as in~\cite{Stratidakis2019}
\begin{align}
    \mathcal{D} = \frac{\|x-\hat{x}\|}{\mathrm{HPBW}} ,
\end{align}
and, thus, the stability can be written as 
\begin{align}
    \mathcal{S} = 1 - \mathcal{D}.
\end{align}
If it reaches 100\%, the localization process must restart since the estimate failed because the UE is outside the beam. Therefore, a system's stability may be thought of as the variation in localization accuracy over time, particularly in a mobile setting.

\begin{table*}
\centering
    \caption{SoTA Localization Methods}
    \label{tab:localization_methods}
    \begin{tabular}{p{0.001\linewidth}p{0.1\linewidth}p{0.22\linewidth}p{0.22\linewidth}p{0.23\linewidth}p{0.08\linewidth}}
        \toprule
        & \textbf{Algorithm} & \textbf{Application scenarios} & \textbf{Use cases} & \textbf{Technology enablers} & \textbf{Metrics}\\ 
        \midrule
        \parbox[t]{2mm}{\multirow{30}{*}{\rotatebox[origin=c]{90}{Conventional}}} 
        
        & Triangulation 
        & Local trust zones \cite{Hanssens2016, Ottoy2016} 
        \par Mapping \cite{Kristalina2016, Landstrom2016, Prateek2021} 
        \par Sustainable develop. \cite{Tazawa2016, Steendam2018, Lee2020} 
        \par Security \cite{Yan2018} 
        \par Massive twinning \cite{Soltanaghaei2018, Adama2021} 
        & Outdoor \cite{Prateek2021} 
        \par Indoor \cite{Hanssens2016, Ottoy2016, Landstrom2016, Tazawa2016, Steendam2018, Yan2018, Soltanaghaei2018, Lee2020, Adama2021} 
        & mmWave/THz \cite{Landstrom2016} 
        \par VLP \cite{Steendam2018} 
        \par Sensors \cite{Yan2018, Adama2021, Prateek2021} 
        & AoA, ToA, AoD, ToF, RSS, RMSE \\ \cline{3-6}
        
        & Kalman \par Filters 
        & Local trust zones \cite{Werner2015, Yousefi2015, Rastorgueva2018, Kanhere2021, Ammous2022} 
        \par Massive twinning \cite{Koivisto2016, Koivisto2017} 
        \par Robots \cite{Bai2019} 
        & Outdoor \cite{Werner2015, Koivisto2016, Koivisto2017, Rastorgueva2018, Kanhere2021, Ammous2022} 
        \par Indoor \cite{Yousefi2015, Bai2019} 
        & mmWave/THz \cite{Werner2015, Koivisto2016, Rastorgueva2018, Koivisto2017, Bai2019, Kanhere2021, Ammous2022} 
        \par RIS \cite{Ammous2022} 
        \par Sensors \cite{Yousefi2015} 
        & ToA, DoD, RSS, RMSE \\ \cline{3-6}
        
        & Compressive \par sensing 
        & Local trust zones \cite{Cao2017, Abdelreheem2018, Salari2018, Harlakin2021, Jiang2021, Jiang2021b} 
        \par Mapping \cite{Ramadan2017} 
        \par Massive twinning \cite{Gligoric2018} 
        & Indoor \cite{Ramadan2017} 
        \par Outdoor \cite{Harlakin2021, Jiang2021, Jiang2021b} 
        & mmWave/THz \cite{Abdelreheem2018, Harlakin2021, Jiang2021} 
        \par Sensors \cite{Cao2017, Ramadan2017, Salari2018, Jiang2021, Jiang2021b} 
        \par Beamforming \cite{Abdelreheem2018} 
        \par Radar \cite{Harlakin2021} \ ,\ \ VLP \cite{Gligoric2018}
        & RSS, TDoA, DoA, RMSE \\ \cline{3-6}
        
        & Multidimensional \par Scaling 
        & Robots \cite{Oliveira2014} 
        \par Local trust zones \cite{Franco2015, Saeed2015, Chaurasiya2014, Jiang2016, Kumar2016b, Morral2016, Stojkoska2017, Fan2020b, Fan2021} 
        \par Massive twinning \cite{Cui2016} 
        & Indoor \cite{Oliveira2014, Cui2016, Stojkoska2017} 
        \par Outdoor \cite{Saeed2015, Chaurasiya2014, Jiang2016, Morral2016, Fan2021}
        \par Non-terrestrial \cite{Kumar2016b, Fan2020b} 
        & Sensors \cite{Oliveira2014, Franco2015, Saeed2015, Chaurasiya2014, Jiang2016, Kumar2016b, Morral2016, Stojkoska2017, Fan2020b, Fan2021, Cui2016}
        & ToF, TDoA, RSS, RMSE \\ \cline{3-6}
        
        & Direct \par Localization 
        & Local trust zones \cite{Weiss2004, Garcia2017, Yin2017, Chen2018, Han2019, Ma2019, Zhao2020, Xia2021, Yu2021} 
        \par Sustainable develop. \cite{Hossain2019b, Bai2022} 
        & Outdoor \cite{Weiss2004, Garcia2017, Yin2017, Chen2018, Han2019, Ma2019, Zhao2020, Xia2021, Yu2021, Hossain2019b, Bai2022} 
        & mmWave/THz \cite{Garcia2017, Han2019, Hossain2019b} 
        \par Sensors \cite{Weiss2004, Yin2017, Chen2018, Ma2019, Zhao2020, Xia2021, Yu2021, Bai2022} 
        \par Radar \cite{Yu2021} 
        & ToA, TDoA, AoA, RMSE \\ \cline{3-6}
        
        & Swarm \par Intelligence 
        & Robots \& Mapping \cite{Nedjah2020} 
        \par Local trust zones \cite{Daely2016, Arsic2016, Kulkarni2016, Monica2016, Tuba2018, Strumberger2018, Strumberger2018b, Phoemphon2018, Kulkarni2019, Strumberger2019, Arafat2019, Akram2021} 
        \par Sustainable develop. \cite{Singh2018} 
        & Indoor \cite{Nedjah2020, Monica2016, Kulkarni2016, Monica2016} 
        \par Outdoor \cite{Nedjah2020, Daely2016, Arsic2016, Tuba2018, Strumberger2018, Strumberger2018b} 
        & Sensors \cite{Nedjah2020, Daely2016, Arsic2016, Kulkarni2016, Monica2016, Tuba2018, Strumberger2018, Strumberger2018b, Phoemphon2018, Kulkarni2019, Strumberger2019, Arafat2019, Akram2021, Singh2018} 
        & AoA, ToA, RMSE, RSS \\ \cline{3-6}
        
        & Fingerprinting 
        & Sustainable develop. \cite{Zhou2015, Zhou2015b, He2016, Souza2016, Li2016, Chen2016b, Zhang2017c, Zhou2017d, Pecoraro2018, Duan2019, Moghtadaiee2019, Zhang2019fing, Gao2021} 
        \par Mapping \cite{Li2019c} 
        \par Local trust zones \cite{Shen2021} 
        & Indoor \cite{Zhou2015, Zhou2015b, He2016, Souza2016, Li2016, Chen2016b, Zhang2017c, Zhou2017d, Pecoraro2018, Duan2019, Moghtadaiee2019, Li2019c, Zhang2019fing, Gao2021} 
        \par Outdoor \cite{Shen2021} 
        & Sensors \cite{Zhou2015, Zhou2015b, He2016, Souza2016, Li2016, Chen2016b, Zhang2017c, Zhou2017d, Pecoraro2018, Duan2019, Moghtadaiee2019, Li2019c, Zhang2019fing, Gao2021} 
        \par mmWave/THz \cite{Shen2021} 
        & RSS, RMSE\\ \cline{3-6}
        
        & SLAM 
        & Mapping \cite{Schlegel2018, Holder2019, Cui2020, Rodriguez2018, Lin2019, Chen2020b, Wang2021b, Chen2021b} 
        & Outdoor \cite{Schlegel2018, Rodriguez2018, Lin2019, Chen2020b, Wang2021b} 
        \par Non-terrestial \cite{Holder2019} 
        \par Underwater \cite{Cui2020} 
        \par Indoor \cite{Chen2021b} 
        & Sensors \cite{Schlegel2018, Cui2020, Lin2019, Chen2020b, Chen2021b} 
        \par Radar \cite{Holder2019, Wang2021b} 
        \par & RMSE, RSS \\ 
        
        \midrule
        \parbox[t]{2mm}{\multirow{43}{*}{\rotatebox[origin=c]{90}{Learning-based}}}
        
        & K-nearest \par neighbor 
        & Sustainable develop. \cite{Alhajri2019, Peng2016b} 
        \par Massive twinning \cite{Wang2019b, Guo2018}
        \par Local trust zones \cite{Sun2018, Wang2019c} 
        \par Mapping \cite{Boudani2020, Chen2016} 
        & Indoor \cite{Alhajri2019, Peng2016b, Wang2019b, Guo2018, Boudani2020, Chen2016} 
        \par Outdoor \cite{Sun2018, Wang2019c} 
        & mmWave/THz \cite{Sun2018, Boudani2020, Wang2019c} 
        \par VLP \cite{Chen2016} 
        \par Sensors \cite{Alhajri2019, Peng2016b, Wang2019b, Guo2018} 
        & RMSE, RSS \\ \cline{3-6}
        
        & Support vector \par machine 
        & Sustainable develop. \cite{Zhang2018b, Ray2016, Cheng2016} 
        \par Mapping \cite{Li2017} 
        \par Massive twinning \cite{Kram2019, Yang2018, Zhang2019d} 
        \par Local trust zones \cite{Wu2018, Wu2019d, Sanam2018, Mahfouz2015, Burghal2020, Petric2019, Zhang2017b, Rezgui2017, Khalajmehrabadi2017, Wang2016c, Guo2019, Yan2017}
        & Indoor \cite{Zhang2018b, Kram2019, Wu2018, Wu2019d, Sanam2018, Mahfouz2015, Petric2019, Zhang2017b, Rezgui2017, Khalajmehrabadi2017, Wang2016c, Guo2019, Yan2017, Cheng2016} 
        \par Outdoor \cite{Li2017, Ray2016, Yang2018} 
        \par Non-terrestrial \cite{Li2017} 
        & Sensors \cite{Zhang2018b, Kram2019, Wu2018, Wu2019d, Sanam2018, Mahfouz2015, Petric2019, Zhang2017b, Rezgui2017, Khalajmehrabadi2017, Wang2016c} 
        \par Radars \cite{Li2017, Yan2017, Yang2018, Cheng2016} 
        \par mmWave/THz \cite{Ray2016} 
        & ToA, TDoA, RSS, RSME \\ \cline{3-6}
        
        & Decision \par trees 
        & Local trust zones \cite{Ahmadi2015, Akram2018, Guo2018b, Malmstrom2019, Rashdan2020} 
        \par Massive twinning \cite{Musa2019, De2019}
        \par Sustainable develop. \cite{Zhang2019e, Chen2019c, Apostolo2019}
        & Indoor \cite{Ahmadi2015, Musa2019, Chen2019c, Apostolo2019, Zhang2019e, Akram2018, Guo2018b}
        \par Outdoor \cite{De2019, Malmstrom2019, Rashdan2020}
        & Radars \cite{Ahmadi2015, Musa2019, De2019}
        \par Sensors \cite{Chen2019c, Apostolo2019, Zhang2019e, Akram2018, Guo2018b}
        \par mmWave/THz \cite{Malmstrom2019, Rashdan2020} 
        & AoA, RSS, RMSE \\ \cline{3-6}
        
        & Gaussian \par Processes 
        & Local trust zones \cite{Savic2015, Prasad2017, Prasad2018b, Kumar2016, Yu2018, Wang2021, Prasad2018} 
        \par Sustainable develop. \cite{Homayounvala2019, Sun2018b}
        \par Mapping \cite{Zhao2018b, Yiu2017}
        & Indoor \cite{Homayounvala2019, Yiu2017, Kumar2016, Sun2018b, Yu2018}
        \par Outdoor \cite{Savic2015, Prasad2017, Prasad2018b, Wang2021, Prasad2018}
        & Sensors \cite{Homayounvala2019, Yiu2017, Kumar2016, Sun2018b} 
        \par Radars \cite{Savic2015, Zhao2018b, Yu2018}
        \par mmWave/THz \cite{Prasad2017, Prasad2018b, Wang2021, Prasad2018} 
        & ToA, RSS, RMSE \\ \cline{3-6}
        
        & Neural \par networks 
        & Sustainable develop. \cite{Fang2019, Wu2019c, Wang2016b, Wang2015, Belmonte2019, Feng2019}
        \par Local trust zones \cite{Yang2019, Lee2018, Ge2019, Jaafar2018, Sobehy2019, Decurninge2018, Orabi2021, Gante2019, Comiter2017}
        \par Massive twinning \cite{Zhang2019c, Boudani2020}
        & Indoor \cite{Fang2019, Jaafar2018, Wang2016b, Wang2015, Zhang2019c, Sobehy2019, Belmonte2019, Feng2019}
        \par Outdoor \cite{Yang2019, Lee2018, Ge2019, Decurninge2018, Orabi2021, Gante2019, Comiter2017}
        & Sensors \cite{Fang2019, Wu2019c, Ge2019, Jaafar2018, Wang2016b, Wang2015, Zhang2019c, Belmonte2019, Feng2019, Sobehy2019}
        \par mmWave/THz \cite{Yang2019, Lee2018, Decurninge2018, Orabi2021, Gante2019, Comiter2017}
        & RSS, ToA, RMSE\\ \cline{3-6}
        
        & Autoencoders 
        & Sustainable develop. \cite{Khatab2017, Kim2018, Song2019, Belmannoubi2019, Zhang2017, Yazdanian2018, Abbas2019} 
        \par Local trust zones \cite{Wang2017}
        & Indoor \cite{Khatab2017, Kim2018, Song2019, Belmannoubi2019, Zhang2017, Wang2017, Yazdanian2018, Abbas2019}
        & Sensors \cite{Khatab2017, Kim2018, Song2019, Belmannoubi2019, Zhang2017, Yazdanian2018, Abbas2019}
        \par mmWave/THz \cite{Wang2017}
        & RSS, RMSE\\ \cline{3-6}
        
        & Convollutional \par NNs 
        & Local trust zones \cite{Jang2018b, Ebuchi2019, Ibrahim2018, Soro2019, Chen2017, Wang2018, Jing2019, Li2019b, Wu2021, Vieira2017} 
        \par Sustainable develop. \cite{Liu2019b, Hsieh2019, Huang2018, Bregar2018} 
        \par Massive twinning \cite{Mittal2018, Niitsoo2019}
        & Indoor \cite{Jang2018b, Mittal2018, Liu2019b, Ibrahim2018, Soro2019, Chen2017, Wang2018, Jing2019, Li2019b, Hsieh2019, Huang2018, Bregar2018}
        \par Outdoor \cite{Ebuchi2019, Niitsoo2019, Wu2021, Vieira2017}
        & Sensors \cite{Jang2018b, Mittal2018, Liu2019b, Ibrahim2018, Soro2019, Chen2017, Wang2018, Jing2019, Li2019b, Hsieh2019, Huang2018, Bregar2018, Niitsoo2019}
        \par mmWave/THz \cite{Ebuchi2019, Wu2021, Vieira2017} 
        & ToA, RSS \\ \cline{3-6}
        
        & Recurrent \par NNs 
        & Local trust zones \cite{Turabieh2019, Zhang2019b, Wu2019b, Yu2020, Tarekegn2019}
        \par Sustainable develop. \cite{Canton2017, Rizk2019c, Rizk2019b, Chen2019b, Qian2019, Adege2019}
        \par 
        & Indoor \cite{Turabieh2019, Canton2017, Rizk2019c, Rizk2019b, Chen2019b, Qian2019, Yu2020} 
        \par Outdoor \cite{Zhang2019b, Wu2019b, Adege2019, Tarekegn2019}
        & Sensors \cite{Turabieh2019, Canton2017, Rizk2019c, Rizk2019b, Chen2019b, Qian2019, Zhang2019b, Adege2019}
        \par Radar \cite{Tarekegn2019, Yu2020} 
        \par mmWave/THz \cite{Wu2019b}
        & RSS, RSME \\ \cline{3-6}
        
        & Unsupervised \par learning 
        & Local trust zones \cite{Le2018, Zafari2019, Sikeridis2018, Guo2019, Saeed2016, Saeed2019} 
        \par Massive twinning \cite{Choi2019, Choi2019b, Jang2018, Ye2018}
        \par Sustainable develop. \cite{Gao2017}
        & Indoor \cite{Le2018, Choi2019, Choi2019b, Zafari2019, Sikeridis2018, Guo2019, Saeed2016, Saeed2019}
        \par Outdoor \cite{Sikeridis2018}
        & Sensors \cite{Le2018, Zafari2019, Guo2019, Saeed2019, Choi2019, Choi2019b, Jang2018, Gao2017}
        \par mmWave/THz \cite{Sikeridis2018, Ye2018, Saeed2016}
        & ToA, RSS, RMSE\\ \cline{3-6}
        
        & Federated \par learning 
        & Local trust zones \cite{Niknam2020, Chen2017b, Ciftler2020, Li2020, Nguyen2021, Liu2019c, Yin2020} 
        & Indoor \cite{Niknam2020, Chen2017b, Ciftler2020, Li2020, Nguyen2021, Liu2019c, Yin2020}
        \par Outdoor \cite{Niknam2020, Chen2017b, Ciftler2020, Nguyen2021}
        & Sensors \cite{Chen2017b, Li2020, Nguyen2021, Liu2019c, Yin2020}
        \par mmWave/THz \cite{Niknam2020}
        & ToA, TDoA, AoA, RMSE\\ \cline{3-6}
        
        & Reinforcement \par Learning 
        & Local trust zones \cite{Shao2018, Li2019, Dou2018, Peng2019} 
        \par Mapping \cite{Mirama2021, Fahad2018}
        & Indoor \cite{Shao2018, Li2019, Dou2018, Peng2019}
        \par Outdoor \cite{Mirama2021, Li2019, Peng2019, Fahad2018}
        & Sensors \cite{Shao2018, Li2019, Dou2018, Peng2019}
        \par mmWave/THz \cite{Mirama2021, Fahad2018}
        & RMSE, RSS\\ \cline{3-6}
        
        & Transfer \par learning 
        & Local trust zones \cite{De2020, Li2021, Zou2017, Zhou2017, Behera2020, Yoo2019, Lei2019} 
        \par Sustainable develop. \cite{Xiao2018b, Liu2017, Zhang2018}
        \par Mapping \cite{Xu2020, Zhou2017b}
        & Indoor \cite{Li2021, Xiao2018b, Liu2017, Zou2017, Zhang2018, Xu2020, Zhou2017b, Behera2020, Yoo2019}
        \par Outdoor \cite{De2020, Lei2019}
        & Sensors \cite{Xiao2018b, Liu2017, Zou2017, Zhang2018, Xu2020, Zhou2017b, Behera2020, Yoo2019, Lei2019} 
        \par Radar \cite{Li2021} 
        \par mmWave/THz \cite{De2020}
        & AoD, AoA, RSS \\ 
        \bottomrule
    \end{tabular}
\end{table*}

\section{Localization algorithms}\label{S:Methodologies}
This section delves deeper into the various localization methodologies and algorithms that exist. Differentiating between the conventional and learning-based methods is selected for the following analysis. Learning-based methods refers to methods that use ML frameworks, hence NNs would fall under this umbrella. Keep in mind that distinguishing between the two may be challenging; thus, some of the publications described here might be categorized in multiple ways. Not only that, but fingerprint matching makes use of probabilistic approaches. As a result, due to the ambiguous position between ML and analytical solutions, they are categorized under the conventional methods umbrella.

\subsection{Conventional}\label{Ss:Classical}
This section provides an overview of conventional algorithms for deriving the location of a mobile node in the considered area. As also summarized in Table \ref{tab:localization_methods}, these methods include triangulation, Kalman filters, compressive sensing, multidimensional scaling, direct localization, swarm intelligence algorithms, and fingerprinting.

\subsubsection{Triangulation}\label{Sss:Triangulation}
\begin{figure}
    \centering\includegraphics[width=0.8\columnwidth]{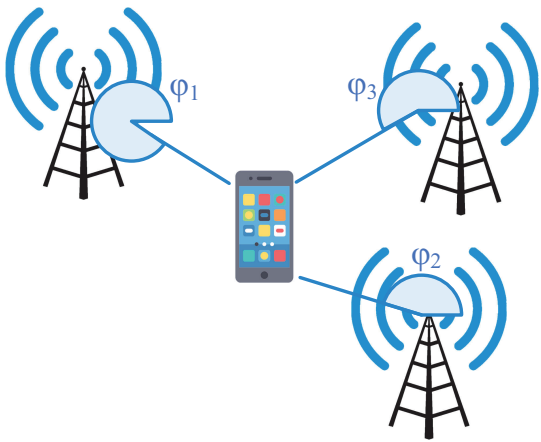}
    \caption{Triangulation architecture.}
    \label{fig:triangulation}
\end{figure}
Triangulation uses the AoA or AoD measurements of radio signals exchanged between the mobile node and the access points to estimate the location of a node at the intersection of lines through simple geometric relationships \cite{Laoudias2018}. A minimum of two access points is required to estimate a node's 2D coordinates. An indicative example of triangulation is shown in Fig. \ref{fig:triangulation}. The mobile node estimates the AoA measurements from three different access points, namely $\phi_1$, $\phi_2$, and $\phi_3$. Using these measurements and the known locations of the access points, the location coordinates of the mobile node can be obtained.

The authors of \cite{Hanssens2016} presented a UWB triangulation scheme that leverages the geometrical properties of the propagation path based on the AoD, AoA, and ToA measurements. Ottoy and Strycker, in \cite{Ottoy2016}, developed a triangulation algorithm that autonomously selects the appropriate AoA measurements out of a set of possible values. In \cite{Kristalina2016}, the authors presented a geometric triangulation approach that leverages the AoA measurement between the mobile node and the anchor nodes. Moreover, Landstrom and Beek in \cite{Landstrom2016} proposed an approach for transmitter localization suitable for multipath mmWave 5G scenarios. The proposed approach combines stochastic ray-shooting with the triangulation concept, to provide accurate localization based on the path relations between the receivers and the TX. Tazawa et al., in \cite{Tazawa2016}, investigated the impact of extending the number of antenna elements on localization accuracy. In this respect, the triangulation technique is applied to the AoD and RSS measurements originating from multiple receivers. Steendam in \cite{Steendam2018} presented a 3D triangulation algorithm based on the maximum likelihood principle. Using the proposed iterative algorithm, each AoA estimate is updated based on the previous position estimate. A method that detects unreliable AoA measurements is presented in \cite{Yan2018}. The proposed method uses the characteristics of estimated locations of various nodes to detect unreliable nodes and mitigate localization error.

Moreover, the authors of \cite{Soltanaghaei2018} presented a localization solution for WiFi wireless networks, which leverages multipath reflection for estimating a node's location with respect to the receiver. The proposed technique offers orientation information and decimeter-level localization based on the AoA, AoD, and ToF measurements. In \cite{Lee2020}, the authors presented a RSS-based indoor triangulation approach that employs the incoherent reception of the transmission from a node of unknown location by several identical indoor nodes. Also, the nodes are able to communicate with each other in order to construct an indoor propagation model. Adama and Asutkar, in \cite{Adama2021}, combined the weighted prediction and grey prediction algorithms in order to develop a triangulation approach for reducing the estimation error of unknown nodes and improving the positioning accuracy. The authors of \cite{Prateek2021} proposed a method consisting of a range-free approach for detecting symmetric triangulations, which, combined with semidefinite programming, can enhance localization accuracy.

\subsubsection{Kalman filters}\label{Sss:KF}
\begin{figure}
    \centering\includegraphics[width=0.9\columnwidth]{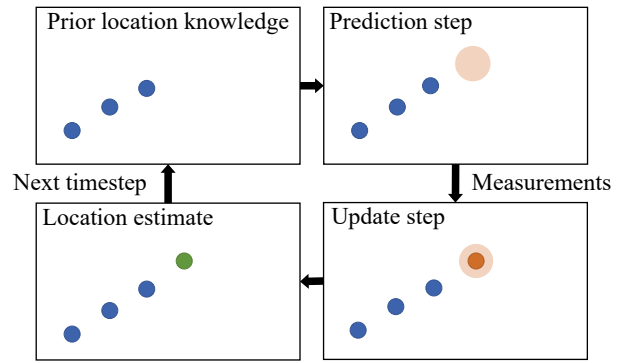}
    \caption{Kalman filters architecture.}
    \label{Fig:Kalman}
\end{figure}
KF is a fundamental technique for the study of noisy or inaccurate measurements, offering a clear knowledge of how a signal evolves over time. The KF eliminates random noise before estimating the state of the monitored process over time. KFs are widely used in navigation, radar applications, and movement control. KF involves two steps: a) the prediction step, which projects the current state of the model and associated uncertainties into the subsequent time step; b) the update step, where the projection is adjusted by calculating the weighted average of the projected state and the measurements. KF-based methods estimate and fuse multiple approximations of an unknown value in order to generate an accurate approximation. In this respect, KFs are able to exploit the estimates from multiple anchor nodes in order to accurately obtain the location of a mobile node.

The authors of \cite{Werner2015} presented an extended KF (EKF) method that tracks the location of a node based on the fusion of time of arrival (ToA) and direction of arrival (DoA) estimates. The simulation results showed that the joint Doa and ToA estimation outperforms the DoA-only estimation. In \cite{Yousefi2015}, the authors reported the use of an unscented KF (UKF) for the ToA-based localization of a mobile node in non-line-of-sight scenarios. The evaluation results revealed a high localization accuracy in presence of low-variance noise, making it suitable for high-resolution UWB localization. Moreover, Koivisto \textit{et al.} in \cite{Koivisto2016} documented two approaches based on EKF and UKF for 3D localization scenarios. The proposed approaches leverage a number of densely deployed anchor nodes that enable the fusion of DoA and ToA estimates. Based on the simulations, both methods can achieve sub-meter scale accuracy, with the EKF-based method slightly outperforming the UFK-based one. Similarly, a two-stage cascaded EKF method for mmWave localization scenarios is adopted in \cite{Rastorgueva2018}. In the first stage, the direction of departure (DoD) is estimated at each anchor node, whereas in the second stage, all DoD estimates are fused into the final 3D location estimates. In \cite{Koivisto2017}, the authors presented a cascaded EKF-based solution to track the DoA and ToA of various devices. In more detail, the first EKF is employed to generate the individual DoA and ToA estimates from the anchor nodes based on the reference signals, whereas the second EKF is employed to fuse the DoA and ToA estimates from multiple anchor nodes. 

Furthermore, Bai \textit{et al.} in~\cite{Bai2019} develop an UKF-based algorithm for obtaining the 2D coordinates of a mobile node that exploits the packet loss errors to estimate and improve the localization accuracy. The simulation results indicate an average localization error of 0.39m when the packet loss rate is lower than 90\%. Moreover, Kanhere and Rappaport in \cite{Kanhere2021} leverage an EKF to combine the DOA and time of flight (ToF) from multipath components in LoS and NLoS environments. The respective experimental results report an average error of 24.8 cm at 140 GHz. Finally, the authors of \cite{Ammous2022} consider a mmWave localization scenario involving a base station and a RIS and design an EKF-based approach to obtain the location of a node using the time-difference of arrival (TDOA) and round trip time (RTT) measurements.

\subsubsection{Compressive Sensing}\label{Sss:CS}
\begin{figure}
    \centering\includegraphics[width=1\columnwidth]{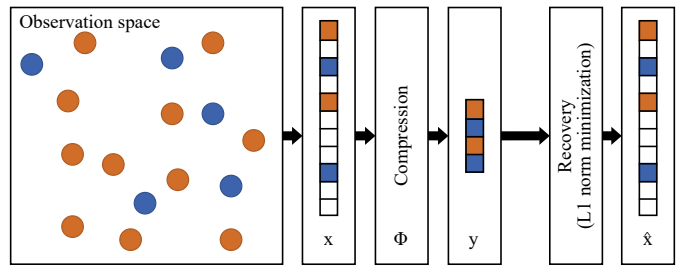}
    \caption{Compressive sensing architecture.}
    \label{Fig:Compressive_sensing}
\end{figure}
Compressive sensing (CS) theory, as illustrated in Fig.~\ref{Fig:Compressive_sensing}, provides a novel framework for recovering sparse signals under a certain basis, with a significantly lower number of samples compared to conventional methods (e.g., the Nyquist theorem) \cite{Donoho2006, Candes2008}. Nonetheless, the acquisition of the original signal requires a time-consuming or expensive process, which appears futile as, during compression, the vast majority of the recovered information is discarded. CS provides and alternative procedure, which by acquiring only the required linear and non-adaptive measurements it is possible to obtain a compressed form of the signal directly. As a result, even with fast recovery techniques, CS predicts the reconstructed signal from severely under-sampled non-adaptive data adequately and can be employed to recover various estimates, such as TDOA, RSS, AoA, from the signals using fewer signal samples.

An approach for estimating the TDOA using CS theory was presented in \cite{Cao2017}. Specifically, the approach utilizes the discrete Fourier transform and a maximum likelihood estimator for obtaining accurate TDOA estimates. In \cite{Ramadan2017}, the authors presented an indoor localization approach based on the received signal strength (RSS) obtained CS processing.  Additionally, the authors of \cite{Abdelreheem2018} designed a multi-level beamforming technique for LoS mmWave links, which leverages CS theory to estimate the signal AoD. An approach for deriving the 2D location coordinates in a VLC environment was presented in \cite{Gligoric2018}. In particular, CS was employed to obtain the location estimates from multiple overlapping light beams. Furthermore, Salari \textit{et al.}, in \cite{Salari2018}, documented a CS-based approach for obtaining TDOA estimates, which requires a limited number of signal samples, leading to a significantly lower computational complexity. A CS-aided MIMO scheme for aerial scenarios was reported in~\cite{Harlakin2021}. In more detail, the authors combined MIMO techniques for 2D antenna arrays and CS-based DoA estimation for 3D target tracking. An aerial scenario leveraging CS theory was also investigated in \cite{Jiang2021} and \cite{Jiang2021b}. The Bayesian CS approach was applied to estimate ground node locations using RSS samples collected by the UAV sensors. Particularly, in \cite{Jiang2021}, a trajectory optimization method was employed to minimize the errors of CS-based location estimation, whereas, in \cite{Jiang2021b}, the Bayesian CS-based approach was enhanced using an a priori knowledge-aided algorithm.

\subsubsection{Multidimensional Scaling}\label{Sss:MDS}	
Multidimensional scaling (MDS) is a mathematical approach for classification and visualization of high-dimensional data. MDS is applied on a matrix formed by the distances between nodes and, for each node, it generates a vector of coordinates in the Euclidean space \cite{Seco2018}. Consequently, MDS is widely utilized in localization problems, in which several nodes have to estimate their relative locations from the measured distances between themselves.

Assuming a network consisting of a number of anchor nodes and $N$ mobile nodes, let $\mathbf{D}$ denote the $N \times N$ distance matrix
\begin{equation}
    \mathbf{D}= 
    \begin{bmatrix}
        0 & d_{1,2}^2 & d_{1,3}^2 & ...& d_{1,N}^2\\
        d_{2,1}^2 &  0 & d_{2,3}^2& ... & d_{1,N}^2\\
        d_{3,1}^2 & d_{3,1}^2 &  0 &  ... & d_{3,N}^2\\
        \multicolumn{5}{c}{$\vdots$}     \\
        d_{N,1}^2 & d_{N,2}^2 &  d_{N,3}^2 &  ... & 0\\
    \end{bmatrix},
\end{equation}
where $d_{i,j}$ stands for the shortest distance between the $i$-th and $j$-th mobile nodes. Classical all-pairs shortest-path algorithms, such as the Floyd–Warshall or Dijkstra algorithms, can be used for obtaining the distances. 

The rank of $\mathbf{D}$ can be reduced by double centering as
\begin{equation}
    \mathbf{C}=-\frac{1}{2}\mathbf{J}\,\mathbf{D}\,\mathbf{J},
\end{equation}
where 
\begin{equation}
    \mathbf{J}=\mathbf{I}-\frac{1}{N}\mathbf{e}\mathbf{e}^\intercal,
\end{equation}
$\mathbf{I}$ is the $N \times N$ identity matrix, $\mathbf{e}$ is a column vector of length $N$ with all elements equal to 1, and $\intercal$ is the matrix transpose operator. Additionally,
\begin{align}
    \mathbf{C}\simeq\mathbf{X}\,\mathbf{X}^\intercal,
\end{align}
where $\mathbf{X}$ is a $N \times 2$ matrix with each nodes' coordinates translated so as the centroid of the network is at $(0,0)$.

Therefore, the matrix containing the relative node locations can be estimated by minimizing the following formula:
\begin{equation}\label{eq:mds_func}
    \hat{\mathbf{X}}_{\text{MDS}}=\argmin_x \|\mathbf{C}-\mathbf{X}\mathbf{X}^\intercal\|^2.
\end{equation}

Eigenvalue decomposition can be applied for solving \eqref{eq:mds_func} as
\begin{equation}
    \mathbf{C}=\mathbf{V}\mathbf{\Lambda} \mathbf{V}^\intercal,
\end{equation}
where $\mathbf{V}$ stands for the eigenvector matrix and 
\begin{align}
    \mathbf{\Lambda}=\rm{diag}(\lambda_1, \lambda_2, ..., \lambda_N) ,
\end{align}
stands for the respective eigenvalue matrix. Since $\mathbf{C}$ is positive-definite and symmetric, and 
\begin{align}
    \rm{rank}(\mathbf{C})\leq 2 ,
\end{align}
it can be represented by its two largest eigenvalues. Consequently,
\begin{align}
    \mathbf{C}\simeq\mathbf{V}_s\mathbf{\Lambda}_s\mathbf{V}_s ,
\end{align}
and the solution to \eqref{eq:mds_func} can be obtained as
\begin{equation}
\hat{\mathbf{X}}^{\rm{signal}}_{MDS}=\mathbf{V}_s\mathbf{\Lambda}_s^{1/2},
\end{equation}
with $\mathbf{\Lambda}_s^{1/2}$ is a diagonal matrix with the two largest eigenvalues. Of note,  the obtained position estimate is not referred to any particular reference system, and will be, in general, rotated, translated, and reflected from the  physical node arrangement. Therefore, the actual mobile node locations can be derived  by the known positions of the anchor nodes.

MDS-based localization offers two significant advantages; the approach is robust to individual error ranges, provided that there exists data redundancy in the square distance matrix. Also, it provides a closed-form localization estimate, regardless of network size. The authors of \cite{Oliveira2014} designed a relative localization algorithm for obtaining the inverse of the distance between any pair of communicating nodes, based on the RSS metric using MDS, and deriving the network topology. Aiming to solve the geometrical uncertainties imposed by the conventional MDS algorithm, an enhanced MDS algorithm was presented in \cite{Franco2015}. In \cite{Saeed2015}, Saeed and Nam combined MDS and Procrustes analysis to develop a two-phase localization algorithm for cognitive networks. Specifically, an approximated distance-based approach is employed to maximize the accuracy based on the nodes' proximity information available in the network. The authors of \cite{Chaurasiya2014} designed a MDS-based 3D localization algorithm that aims to reduce communication and computational overheads towards minimizing energy consumption. In \cite{Cui2016}, Cui \textit{et al.} introduces a 3D localization algorithm that is based on the combination of polynomial data fitting and MDS. The algorithm is able to achieve high localization accuracy without requiring information about the channel noise and the node movement model. 

Moreover, Jiang \textit{et al.} in \cite{Jiang2016} presented a solution for finding a node's 2D coordinates through the TDOA measurements by minimizing the MDS-based cost function. A similar approach was presented in \cite{Kumar2016b}, in which the authors introduce an approach that integrates the relative velocity information, available via Doppler shift measurements between mobile nodes, into the MDS cost function. Furthermore, a low-complexity majorization algorithm was leveraged to minimize the MDS cost function. Morral and Bianchi in \cite{Morral2016} reported a distributed algorithm for node self-localization in WSNs based on sporadic measurements of the RSS metric. The nodes' positions can be recovered from the principal components of the similarity matrix which is constructed from the squared inter-nodes distances. The authors of \cite{Stojkoska2017} designed an indoor UAV localization algorithm that is based on distance measurements between the access points and the UAV. The distance measurements are derived through the RSS samples, while two different localization techniques are utilized, namely the MDS and the weighted centroid localization. In \cite{Fan2020b}, Fan \textit{et al.} a combination of fast-clustering and MDS is presented for 2D localization in mobile networks. In more detail, in the inner-cluster relative localization stage, the advantages of combining classical MDS and iterative MDS are leveraged, while in the inter-cluster coordinate registration stage, the least squares method is used to reduce the registration error. In~\cite{Fan2021}, the authors presented a cooperative 3D localization algorithm based on filtering and MDS techniques, which are employed to mitigate any abnormal measurements due to errors.

\subsubsection{Direct Localization}\label{Sss:DirectLocalization}		
In the aforementioned approaches, the location of a node is estimated by various parameters such as the AoA, AoD, TDOA, and ToA. An alternative approach to addressing localization challenges is direct localization \cite{Weiss2004}. In direct localization, the source's location is estimated straight from the data, with no need to estimate any intermediate parameters, like the AoAs or others. Specifically, the echo signals from all node pairs are accumulated and the target location is extracted directly. To achieve this, signals, or some function thereof, must be sent to a fusion centre that makes location estimations. The fusion operation is carried out on the signal level, resulting in less computational overhead and, therefore, improved performance. Such a topology is often simpler to implement in scenarios, where reduced distances are involved. However, for cellular networks, cloud radio access networks (C-RAN) is expected to provide the necessary backbone as BSs forward incoming signals to a centralised unit that performs the baseband processing~\cite{Hossain2019b}.

In~\cite{Garcia2017}, the authors presented a direct source localization approach based on AoA and ToA estimates, which determines the source location by jointly processing data snapshots acquired at each BS. Moreover, the authors of~\cite{Yin2017} developed a single-step direct localization algorithm for obtaining the location of multiple stationary nodes by combining AoA and Doppler information. The numerical results indicate that the proposed direct localization algorithm can attain the corresponding Cramer-Rao constraint. A direct passive localization method that jointly exploits the AoA and TDOA information was presented in~\cite{Chen2018}. By exploiting spatio-temporal processing, the proposed method did not require prior knowledge about the number of source nodes. Han \textit{et al.} in~\cite{Han2019} reported a joint near-field and far-field direct localization approach for massive MIMO environments. Specifically, the approach uses the AoA measurements and divides the BSs into far-field and near-field ones to mitigate localization errors. The authors of \cite{Ma2019} developed a TDOA-based direct localization approach under the assumption that multiple nodes work with independent clocks. The approach leverages an expectation-maximization (EM) algorithm and a Gauss-Newton algorithm for coarse and refined parameter estimation, respectively. 

Moreover, in~\cite{Zhao2020}, the authors presented a framework for direct localization comprised of beamspace design and position determination. In addition, they derived the respective beamspace direct localization performance bound and present a localization algorithm with low computational complexity and communication overhead. A direct localization approach that leverages random spatial spectrum was presented in~\cite{Xia2021}. According to the approach of~\cite{Xia2021}, the co-channel signals were transformed into the spatial spectrum, enabling the suppression of multipath components. Moreover, a spatial sparse clustering algorithm was utilized to distinguish the nodes in the spatial domain. In~\cite{Yu2021}, the authors developed a factor graph-based approach for direct localization in distributed MIMO radar environments. A graph representation was carefully designed for the direct localization problem, which was solved efficiently through message passing based on expectation propagation and belief propagation. Finally, Bai \textit{et al.} in \cite{Bai2022} presented a direct localization algorithm for multipath orthogonal frequency division multiplexing (OFDM) environments. The algorithm utilizes the data observed by multiple nodes and estimatest the source node in one step by exploiting the orthogonality between the noise subspace and the array response vector.

\subsubsection{Swarm Intelligence}\label{Ss:SI}
Swarm intelligence is a subset of algorithms that studies group dynamics, where most approaches take inspiration from the social behaviours of animals, including bees, birds, ants, and others. Complex patterns often form through the coordination of the activities of multiple entities that are relatively unsophisticated in their mode of operation. All swarm intelligence-based approaches use a search strategy based on the dichotomy and interplay of two conventional procedures, namely exploration and exploitation~\cite{Nedjah2020}. The goal of exploration is to increase the fraction of the search space that has been uncovered and, hence, reduced to workable solutions. The term ``exploitation'' refers to the swarm's effort to enhance the quality of previously found solutions by increasing the intensity of its search efforts in the area around those solutions.

In \cite{Daely2016} and \cite{Arsic2016}, the authors respectively utilized the dragonfly and fireworks algorithms for estimating the node locations that were randomly deployed in a specific area. The authors of \cite{Kulkarni2016} presented a multi-stage localization method that leverages the artificial bee colony algorithm. Moreover, they compared the proposed method against the particle swarm optimization in terms of localization accuracy, number of nodes, and computation time. Monica and Ferrari in \cite{Monica2016} investigated the problem of UWB indoor localization in WSNs by employing the particle swarm optimization algorithm. A two-stage WSN localization approach based on the firefly algorithm is proposed in \cite{Tuba2018}. In the first stage, four anchor nodes, that were placed at the area edges, were considered, whereas, in the second stage, the closest anchor nodes are considered. Furthermore, the authors of \cite{Strumberger2018} utilized the monarch butterfly optimization algorithm for addressing the localization problem in WSNs. Also, the same authors in \cite{Strumberger2018b} presented an adaptation of the hybridized moth search algorithm for solving the node localization problem and compared it against alternative state-of-art algorithms. Singh \textit{et al.} in \cite{Singh2018} focused on the 2D localization problem and compared the performance of four swarm intelligence algorithms, namely the biogeography-based optimization, the firefly algorithm, the particle swarm optimization, and the H-best particle swarm optimization algorithms. 

Moreover, in \cite{Phoemphon2018}, the authors introduced a fuzzy logic model for localization in WSNs. Towards increasing the efficiency of the model, the concept of resultant force vectors is applied, while the particle swarm optimization algorithm is employed to minimize the irregular deployment effects. The authors of \cite{Kulkarni2019} utilized the firefly algorithm and the artificial bee colony algorithm for approximating the distance of a mobile node from the anchor nodes. Also, a comparison between the two algorithms in terms of computation time and localization accuracy is presented.	Similarly, in \cite{Strumberger2019}, the authors investigated the performance of two swarm intelligence algorithms, namely the elephant herd optimization and tree growth algorithms, with respect to solving the localization problem in WSN environments. In \cite{Arafat2019}, Arafat and Moh utilized the particle swarm optimization algorithm for the localization of UAV nodes. To increase the convergence speed of the algorithm, the boundary box technique is leveraged, leading to smaller initial search space. Akram \textit{et al.} in \cite{Akram2021} presented an adaptation of the multi-objective particle swarm optimization algorithm for jointly maximizing the number of localized nodes and minimizing the energy consumption and computation time.Finally, in \cite{Nedjah2020}, the authors compared three swarm intelligence algorithms, namely the firefly algorithm, the artificial bee colony algorithm, and the particle swarm optimization algorithm, in SLAM scenarios.

\subsubsection{Fingerprinting}\label{Sss:Fingerprinting}
The fundamental idea behind fingerprinting is to estimate the mobile node location by comparing the received signal fingerprint to a database of known signal locations that have already been recorded \cite{Vo2016}. The database of signal fingerprints has to be generated in advance. Consequently, fingerprinting consists of two phases, namely the populating or training phase and the matching phase. In the first phase, the area of interest is explored to create a signal map, whereas, in the second phase, the node location is estimated by comparing the current fingerprint signal with the fingerprints database.

Various metrics can be used for generating the signal fingerprint database during the training phase. For instance, using the reference RSS, denoted by $\rm{RSS}_{\rm{R}}$, the $(x,y)$ coordinates of a node can be expressed as
\begin{equation}
    \mathbf{F}(x,y)= [\rm{RSS}_{\rm{R}}(1), \rm{RSS}_{\rm{R}}(2), ..., \rm{RSS}_{\rm{R}}(N)],
\end{equation}
where $N$ is the number of access points.

For the matching phase, an Euclidean distance matching method can be leveraged. The method calculates the Euclidean distances between the measured signal fingerprint, denoted by $\rm{RSS}_{\rm{M}}$, and each reference location in the database as follows:
\begin{equation}\label{eq:fingerprint_distance}
    D=\sqrt{\sum_{j=1}^N\left(RSS_{\rm{R}}(j)-RSS_{M}(j)\right)^2}
\end{equation}
Finally, a nearest neighbor algorithm can be used for finding the closest fingerprint based on \eqref{eq:fingerprint_distance}.

Zhou \textit{et al.} in \cite{Zhou2015} and \cite{Zhou2015b} investigated the correlation between the access point deployment and the localization precision based on the RSS and designed a simulated annealing algorithm for optimizing the access point positions. The authors of \cite{He2016} designed an approach that automatically updates the fingerprint database by employing a clustering algorithm to filter out the altered signals to achieve high localization accuracy. Using the observed signals and the calculated location of a node, the fingerprint database is constantly updated without manual intervention. In \cite{Souza2016}, the authors combined RSS-based fingerprinting with context-aware data regarding the node's environment (i.e., building floor plan) and demonstrated that the proposed method can reduce the required access point number and mitigate the effects of wireless interference. The authors of \cite{Li2016} developed a collaboration-based fingerprinting approach, in which several assistant nodes around an unknown node are selected, based on the RSS sequences similarity, and distances between them are used as auxiliary information to improve the positioning accuracy. 

Moreover, Chen \textit{et al.} in \cite{Chen2016b} designed a cooperative fingerprinting approach that takes into consideration the physical constraint of pairwise distances to refine and improve the estimated positions of multiple nodes, thereby increasing the robustness against outdated database entries and distance errors. In \cite{Zhang2017c}, the authors proposed a fingerprint localization method based on the path-loss measurement for the training and matching phase. The two-step method consists of a path-loss-based fingerprinting scheme, which aims to improve precision, and a dual-scanned fingerprinting that guarantees localization robustness. A crowdsourcing method for indoor fingerprinting-based localization is proposed in \cite{Zhou2017d}. The method leverages the RSS data collected from multiple smartphones and generates the database of fingerprints. A similar method is presented in \cite{Pecoraro2018}. Specifically, the method aims to improve localization accuracy by generating the fingerprint database using the channel frequency response, which can be obtained from the CSI. 

In \cite{Duan2019}, the authors designed a fingerprinting scheme based on the achieved data rates, which can be directly obtained by the access points. To mitigate data rate fluctuations, the scheme employs a time-window mechanism that takes into account multiple access points and various transmission power levels. In \cite{Moghtadaiee2019}, the authors developed a fingerprinting method that integrates information about the signal propagation effects and architecture of an indoor area to generate fingerprints using a  low number of RSS measurements. Li \textit{et al.} \cite{Li2019c} proposed a crowdsourcing fingerprinting approach that integrates wireless, inertial. Also, they defined the fingerprinting accuracy indicators and assessed their performance in predicting location estimation errors and outliers. In \cite{Zhang2019fing}, the authors introduced a phase decomposition method to calculate the multipath phase and apply principal component analysis to derive the fingerprint based on the CSI. The authors of \cite{Gao2021} developed a fingerprinting prototype that eavesdrops on smartphone signals in order to acquire the CSI. Furthermore, it employs a joint outlier detection and clustering approach to detect signal changes. Finally, Shen \textit{et al.} in \cite{Shen2021} proposed a fingerprint training scheme that leverages information entropy theory and maximum likelihood estimation. The scheme uses uplink channels to extract the link states and the AoA measurements and applies a weighted mean square error algorithm to estimate the node location.

\subsubsection{Simultaneous localization and mapping}\label{Sss:SLAM}
The challenge that simultaneous localization and mapping (SLAM) is tasked with solving is building a map of a moving agent's surroundings, alongside estimating its trajectory. Due to its pivotal role in the development of autonomous robots, SLAM has attracted a lot of attention from the research community during the last couple of decades. In simultaneous localization and mapping (SLAM), the map, which can be found in various formats, such as occupancy grid, point cloud, and more, serves as the model of the physical world, while its estimate is entangled with the inference of the robot's trajectory. Building a consistent representation of the world requires collecting data from a wide range of sources, regardless of the shape that representation takes. For the purpose of analysing the connections between the available data, they are split into data frames, each of which can be anything from a collection of laser range scan to a series of photographs. The collection of such data frames presents an additional difficulty in that the obtained measurements are always realtive to the ``uncertain'' location of the mobile entity that captures them.

Schlegel \textit{et al.} in \cite{Schlegel2018} presented an open source SLAM system that uses localization and image processing techniques to generate a virtual 3D map. Four processes were employed for map generation, namely image-based triangulation, motion estimation, pose-graph-based map management, and relocalization. A radar-based SLAM method was presented in \cite{Holder2019}. The method constructs a map from radar measurements combining the pose graph and the iterative closes point algorithm for matching the scans. Also, GNSS information is leveraged to improve localization accuracy. The authors of \cite{Cui2020} reported a SLAM method for autonomous underwater vehicles. The method utilizes simulated annealing and applies an EKF on the depth, compass, and acoustic sensor measurements. Rodriguez \textit{et al.} in \cite{Rodriguez2018} emphasized the criticality of the accurate representation and quantification of uncertainties to correctly report the associated confidence of the robot's location estimate in every step of the SLAM process. In \cite{Lin2019}, the authors developed an intelligent filter-based SLAM approach for enhancing the localization performance of mobile robots. The proposed approach can be also applied to underwater and aerial scenarios using the appropriate radar-based measurements. In \cite{Chen2020b}, the authors investigated the relations among the graphical structure of pose-graph SLAM, Fisher information matrix (FIM), Cramer Rao lower bound, and the optimal design metrics. A SLAM framework that employs light detection and ranging (LIDAR) and image processing techniques for localization was presented in \cite{Wang2021b}. The framework considers both light intensity information and geometry information to improve localization accuracy. Chen \textit{et al.} in \cite{Chen2021b} presented an active SLAM algorithm for mobile robots able to derive a collision-free trajectory. The algorithm is based on a predictive control framework that leverages graph topology that approximates the uncertainty minimization problem as a constrained non-linear least squares problem. After applying convex relaxation to the original problem, a convex optimization algorithm and a rounding process based on singular value decomposition are employed for solving the problem.

\subsection{Supervised learning}\label{Ss:Supervised}
This section examines contributions that use supervised learning based ML methodologies for tackling the localization problem. In more detail, these techniques discussed entail both traditional ML approaches, such as kNN, SVM, decision trees, and Gaussian processes, and DL based ones, like NNs, autoencoders, CNNs, and RNNs. 

\subsubsection{K-nearest neighbors}\label{Sss:kNN}
\begin{figure}
    \centering\includegraphics[width=1\columnwidth]{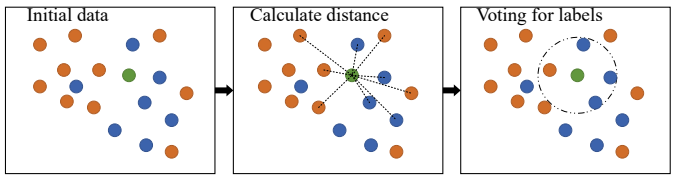}
    \caption{kNN architecture.}
    \label{Fig:kNN}
\end{figure}
With k-nearest neighbors (kNN), the average position of the k-nearest fixed network nodes from the fingerprint database is used to estimate the target's coordinate. In the localization process as presented in~Fig.~\ref{Fig:kNN}, the selected region is where the majority of the k-neighbors are placed. The label for an observed point in an $M$-class classification issue can be provided by
\begin{align}
    \hat{y}_o=\max _c \sum_{i \in \mathcal{K}_o} \mathbb{1}\left(y_i=c\right) ,
\end{align}
where $\mathbb{1}\left(y_i=c\right)$ is equal to unity in the case of the $i$-th neighbor's label is equal to $c$, while $\mathcal{K}_o$ denotes the neighbors of the observed data point. Using the available data and background information, we can determine an appropriate value for the design parameter $k$ and specify the set of individuals who qualify as ``neighbors''. As a result the kNNs can be selected based on a variety of criteria, including the Euclidean distance between locations, which can be expressed~as 
\begin{align}
    d\left(\boldsymbol{x}_i, \boldsymbol{x}_0\right)=\sqrt{\sum_{k=1}^N\left|x_{i, k}-x_{0, k}\right|^2} ,
\end{align}
with $x_{i, k}$ denoting the measurement between the $i$-th mobile and the $k$-th fixed network node, while $N$ represents the total number of mobile nodes. It is important to remember that kNN is a non-parametric model. The new labels are derived by comparing the fresh observations to the training examples, hence it is in the same category as instance-based learning.

Several cutting-edge ML systems use weighted kNN as a last-ditch effort to forecast the location, with varying proposals on how to define the ``distance'' between fingerprints and weights. Weighted kNN uses kNN as a fusing method, but assigns different weights to the positions of the k-nearest neighbors before averaging them. Moreover, kernel functions and euclidean distance are examples of methods that may be used to compute the distances and weights, respectively. In addition, kNN has been taken into consideration in various localization methods, notably fingerprint based localization, because of its low computational cost and straightforward structure. Methods for building fingerprint databases, for adaptively choosing k, and for generating meaningful similarity metrics have all been the subject of studies pertaining to kNN. Finally, various features have also been employed in kNN. The authors of~\cite{Alhajri2019} developed a kNN-based cascading localization system, whereby a kNN is first used to identify the environment, and then a kNN is used to localization using a variety of characteristics. They demonstrate that hybrid features outperform RSS alone in a variety of settings and highlight the relevance of doing so.

The gathered RSS data are then used to build the radio map, an early example of this method. When the localization step takes place online, the online RSS measurement from the three indoor fixed nodes is compared to the previously stored RSS dataset. Typically, the Euclidean distance is used for the comparison. However, considerable localization estimate errors may occur when employing a fixed k value, as is the case with the typical kNN, if the network nodes are moved or the RSS value changes. Several papers have investigated tweaks to kNN as a potential solution to this problem, with a number of techniques being available, including adjusting the similarity metric~\cite{Peng2016b, Wang2019b}, and sub-selecting a subset of the kNN~\cite{Guo2018, Wang2019b}. For instance, the method in~\cite{Wang2019b} clusters the k-nearest fixed network nodes and then uses their mean in the "delegate" cluster to determine the final location. Both the Euclidean distance and the cosine similarity are proposed as similarity metrics and used to build the weights for a weighted kNN in the work cited in~\cite{Peng2016b}. Lastly, the crowd-sourced indoor localization technique in~\cite{Chen2016} uses a smartphone's orientation sensor and an optical camera to pinpoint a user's precise location within a building. In this approach, a crude estimate of the position is derived via kNN over RSS from WiFi, speeding up the search space of image-based localization, which may be further limited by information from orientation sensors.

Another feature that has been employed for the kNN algorithm is the AoA. For example, in~\cite{Sun2018}, AoA is used to pick a subset of k fixed network nodes in a massive-MIMO OFDM system. However, the channel observations are numerous and computationally intensive to store and sort through. To this end, a sparse channel representation based on the angle-delay domain that allows for effective compression has been proposed for database development. It employs two tiers of fingerprint clustering and categorization in order to search across multiple fingerprints. A combined angle and delay similarity measure, which is dependent on the amount of overlap between the scatterers, is suggested to quantify the channel observation-fingerprints similarity. Finally, a weighted kNN is used to accomplish localization based on ToA, AoA, and the corresponding weights. In this respect, localization is achieved by minimizing the Euclidean distance between the features, while the representation of the angle-delay domain is created by applying a fast method to retrieve the fixed network nodes and compressing the database~\cite{Wang2019c}.

\subsubsection{Support vector machine}\label{Sss:SVM}
\begin{figure}
    \centering\includegraphics[width=1\columnwidth]{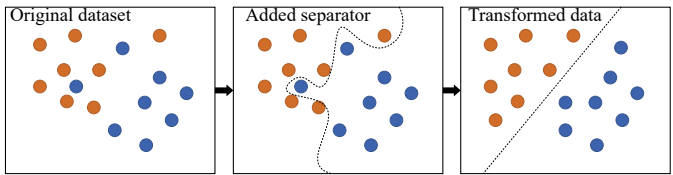}
    \caption{SVM architecture.}
    \label{Fig:SVM}
\end{figure}
The support vector machine (SVM) is a very potent ML kernel based algorithm. Similarity between vectors may be captured by a kernel function, which can be included into instant based learning approaches. Intuitively, the distance between two vectors may be used to represent their similarity to one another. One such kernel is the radial basis function (RBF), which changes as one moves further away and is used in the Gaussian Kernel. Other kernel are sigmoid, polymonial, and linear ones. The kernel function seeks to identify the optimal hyperplane for classifying entities. Specifically, the hyperplane is selected to increase the gap between the groups. The kernels provide an efficient means of locating the product of observed features and training sessions in higher dimensional spaces, which can improve the features' distinguishability. A high-level illustration of the aforementioned SVM method is presented in~Fig.~\ref{Fig:SVM}.

Other characteristics have been applied to SVM with various kernel approaches. Using Gaussian kernel ridge regression, the authors of~\cite{Zhang2018b} presented a localization approach based on RSS and ToA. Of note, in order to avoid being affected by measurement timing and random synchronization mistakes, the method suggested in~\cite{Li2017} employs ToA as fingerprints and is based on the Gaussian Kernel. Moreover, the energy decay time is only one example of characteristics that may be extracted from the channel impulse response and used as part of a UWB localization system~\cite{Kram2019}. The goal is to reduce the size of the impulsive response while still capturing relevant environmental information. For hierarchical area classification, the features are sent to a support vector machine algorithm. Using CSI from multiple sub-carriers, the authors of~\cite{Wu2018} built a visibility graph to capture frequency correlations between neighboring sub-carriers for SVM-based localization. next, in~\cite{Wu2019d}, the authors documented an SVM and kernel regression based method for localizing and recognizing activities based on CSI. In this approach, SVM performs classification of the target into an activity class while localization is accomplished by a regression model. Based on the CSI of a MIMO system,~\cite{Sanam2018} was able to gather CSI and transmit it to a centralized server in a device-free localization system. Finally, in~\cite{Mahfouz2015}, localization using an accelerometer and RSS was employed in WSN target tracking; the RSS based kernel approach produces a coarse position estimate, which is then passed along with the accelerometer measurements to a Kalman filter that performs instantaneous localization.

Localization has seen widespread use of SVM with a great variety of kernel approaches. Applying SVM in WSN localization for a small number of sensors with known positions was one of the first research efforts that use kernel-based localization~\cite{Burghal2020}. The RSS between different sensors is used in SVM trained with a Gaussian kernel to classify the area. At last, more precise coordinates are calculated by averaging the sensors' respective area centers. For indoor mobile phone positioning using RSS,~\cite{Petric2019} employed Gaussian and linear kernels. Another approach combined SVM with a radial basis function (RBF) kernel and PCA to perform localization~\cite{Zhang2017b}, while the authors of~\cite{Rezgui2017} offered a strategy to handle the devices' variety by first ranking the RSSs only from trustworthy fixed nodes and then using SVM to categorize the desired location. To reduce the number of correlated fixed network nodes,~\cite{Khalajmehrabadi2017} presented using real-time node selection. Then, it uses a kernel to assign weights with regard to the similarity between fixed nodes and the corresponding RSS values; based on the assigned weights it estimates the location.

The localization problem is very sensitive to the choice of kernel. To better capture the connection between the coordinates before localization and the RSS space, kernel canonical correlation analysis is presented, where the Matern kernel is utilized for the physical space and the Gaussian kernel is used for the signal space~\cite{Wang2016c}. To account for the training set's spatial organization, a kernel that takes use of any potential connection between the coordinates has been proposed~\cite{Guo2019}. According to~\cite{Yan2017}, a hybrid kernel consists of two separate kernels: a global and a local kernel that together account for the influence of both nearby and far-off fixed network nodes. In~\cite{Yang2018}, Import vector machine (IVM) was employed for NLOS classification for ToA-based ranging in UWB systems; however, the authors noted that an ISM has less complexity and offers a greater classification probability.

Other facets of localization had also made use of kernel approaches, such as a SVM for estimating range errors with regard to CSI in an UWB system and an SVM for recognizing poses, which is subsequently utilized to find a good match with fixed network node~\cite{Zhang2019d}. Prior to utilizing particle filters for location determination,~\cite{Ray2016} classified spaces as indoor or outdoor. Using SVR to recreate the RSS values of the unselected fixed nodes,~\cite{Cheng2016} improves the localization process's resilience against noise by proposing a method for classification and node selection.

\subsubsection{Decision trees}\label{Sss:Trees}
\begin{figure}
    \centering\includegraphics[width=0.9\columnwidth]{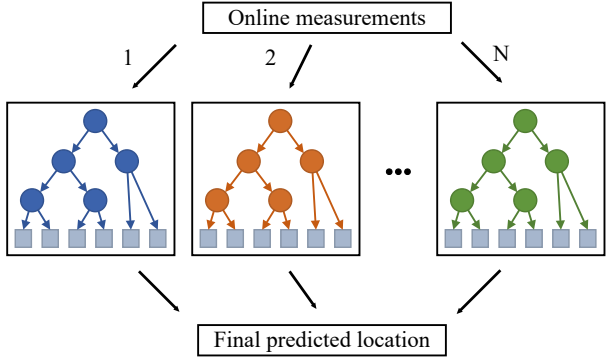}
    \caption{Decision trees architecture.}
    \label{Fig:Trees}
\end{figure}
Rule sets for categorizing data are commonly derived via decision trees, which do so by segmenting the space of possible labels and possible observations. Specifically, Fig.~\ref{Fig:Trees} outlines a our two-stage positioning approach that makes use of various weighted decision trees. During the first stage (training), data is gathered to create a database known as the radio map. For $R$ reference and $A$ access nodes, $N$ location and $\theta$ direction measurements per node, the radio map can be expressed as in~\cite{Sanchez2015}
\begin{align}
    \Psi^\theta=\left(\begin{array}{cccc} 
    \varphi_{1,1}^\theta[\tau] & \varphi_{1,2}^\theta[\tau] & \cdots & \varphi_{1, R}^\theta[\tau] \\
    \varphi_{2,1}^\theta[\tau] & \varphi_{2,2}^\theta[\tau] & \cdots & \varphi_{2, R}^\theta[\tau] \\
    \vdots & \vdots & \ddots & \vdots \\
    \varphi_{A, 1}^\theta[\tau] & \varphi_{A, 2}^\theta[\tau] & \cdots & \varphi_{A, R}^\theta[\tau]
    \end{array}\right) .
\end{align}
This data is comprised of location and direction measurements from fixed nodes. The ensemble model is constructed using this dataset. In the second stage, which is called testing, the mobile nodes make localization requests utilizing online measurements. In more detail, inputs such as online measurements from the fixed nodes are used to estimate the mobile node's location, which can be represented as
\begin{align}
    \Psi_r^\theta=\left(\begin{array}{c} 
    \varphi_{1,r}^\theta[q] \\
    \varphi_{2,r}^\theta[q] \\
    \vdots  \\
    \varphi_{A,r}^\theta[q] 
    \end{array}\right) ,
\end{align}
with $r$ denoting the mobile node's location. 

Both the localization system's accuracy and the time required to construct and assess the ensemble model are affected by the dataset's size. Moreover, a variety of localization issues, such as localization based on fixed node selection and clustering, coordinate prediction, and LOS identification, have benefited from the use of decision trees~\cite{Ahmadi2015, Musa2019}. Many of the articles, however, resort to extensive use of decision trees to solve the difficulties at hand or apply them to relatively straightforward categorization issues. 

The employment of numerous learning solutions to arrive at the final solution has been demonstrated to deliver outstanding performance even when constructed as an ensemble of fundamental ML solutions, such as decision trees. Adaptive and gradient boosting are two ensemble learning approaches that have been extensively applied~\cite{Chen2019c, Apostolo2019, Zhang2019e}. When the RSS distance is not indicative of the actual physical separation between two points,~\cite{Chen2019c} provided a fingerprinting approach that uses gradient boosting to convert raw RSS into features using a learnt non-linear mapping function. The main idea is to first build negative and positive pairs of fixed nodes and afterwards develop a loss function to guarantee that the similarity is maintained between them. Finally, after training the mapping function, weighted kNN was used with the revised mapping function to carry out the localization. On the other hand, adaptive boosting was used for passive localization in~\cite{Zhang2019e}, where phase information from CSI was employed to build a fingerprint map. This method iteratively refines the sample weights of the training sets in order to facilitate classification. The best four anticipated nodes' positions are weighted equally to provide a final location estimate.

Another of the most prominent approaches of ensemble learning is random forest (RaF), which uses a forest of decision trees to make a prediction. For instance,~\cite{De2019} employed classification-based localization with multi-path information and a RaF to obtain the TDoA through volume cross-correlation between CSI values. As a means of improving localization accuracy, the suggested technique combines ray-tracing with empirical data. In order to determine both room predictions and spatial coordinates, the RSS of WiFi was investigated in~\cite{Akram2018}. The offline phase comprises the training of ensemble classifiers and a RaF regressor, as well as fingerprint data preprocessing for location prediction. Room prediction, establishing membership in a soft cluster, and preprocessing are all performed in the online phase. Another example uses a multi-antenna system to construct fingerprints from a variety of parameters, including power spectral density (PSD), RSS, and other statistical data~\cite{Guo2018b}. A RaF is then trained to serve as a classifier for each individual feature. The approach employs a number of samples and classifiers to boost the reliability of position estimations, and an entropy measure is utilized to choose both a stable time instant and a reliable classifier. The location is therefore the median of the location predictions, which is limited to fall inside the union of the projected locations generated by the chosen classifier and the chosen time instant.

\subsubsection{Gaussian processes}\label{Sss:Gaussian}
\begin{figure}
    \centering\includegraphics[width=1\columnwidth]{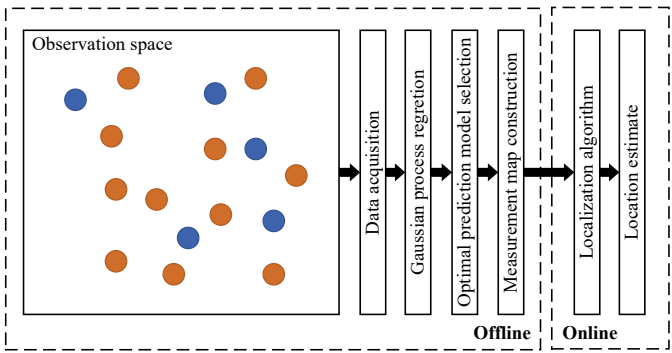}
    \caption{Gaussian processes architecture.}
    \label{Fig:Gaussian}
\end{figure}
Gaussian process (GP) regression models have been successfully used to forecast the spatial distribution of RSS measurements using minimal known labeled fixed nodes, which has been proven to increase the precision of localization schemes. Fig.~\ref{Fig:Gaussian} depicts the overall structure of an indicative localization system. In particular, the radio map used for localization in this example requires very few training measurements. Therefore, it significantly reduces the required time and energy for deployment and radio map construction procedures. Moreover, during the operation phase, RSS measurements are collected and used to pinpoint an exact location of the mobile nodes.

Multiple alternative ML strategies rely on GPs, which can substitute Bayesian methods in the case of posterior distribution estimation of the labeling function~\cite{Savic2015, Prasad2017, Homayounvala2019, Zhao2018b, Prasad2018b}. During the training process, it is important to find optimal values for the GP's covariance and mean, while kernel functions, like Matern and RBF, are often used to capture it. For example, in~\cite{Yiu2017}, GP regression was used for the construction of an RSS-based continuous distribution. The authors used a MLE technique to determine the location of the target based on the RSS data available to them. Researchers examined the quadratic, Matern, and Gaussian kernels in order to determine which one best captures the relationship between geographical points. Moreover, in~\cite{Kumar2016}, a GP was used to represent the probability distribution of the RSS values, and then for a particular RSS observation, the location was estimated using a weighted combination of the fixed network node's positions in accordance with Bayes rule. An approach that applies GP-based dynamic calibration and estimation of the radio map with regard to RSS measurements is described in~\cite{Sun2018b}. The standard deviation of the trained GP model was used to quantify the accuracy of the predicted position, and the final location is calculated using weighted kNN.

In order to achieve a RMSE near the Cramer-Rao constraint in massive-MIMO systems, the authors of~\cite{Prasad2018b} presented numerical approximation GP techniques, which were validated using simulations. Also, several UWB channel characteristics, including RMS delay spread, RSS and ToA, were retrieved from the power delay profile and put to use in a ranging scheme in~\cite{Yu2018}. The initial stage is to employ a kernel PCA to project the chosen channel parameters onto a nonlinear orthogonal high-dimensional space; from there, a subset of these projections is used as an input for GP regression to get a ranging estimate.

\subsubsection{Neural networks}\label{Sss:NN}
\begin{figure}
    \centering\includegraphics[width=0.9\columnwidth]{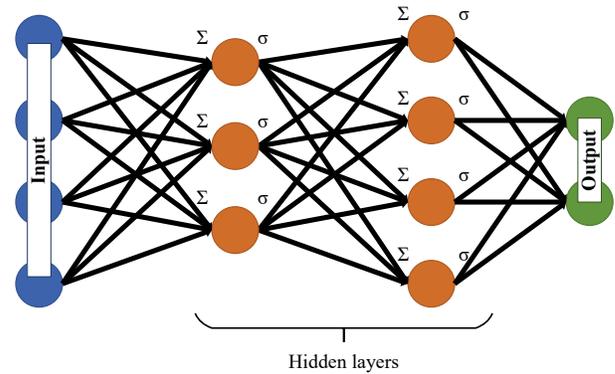}
    \caption{NN architecture.}
    \label{Fig:NN}
\end{figure}
Today, neural networks (NNs) are widely used in ML. This is because they form the basis of many of the most effective learning designs. NNs are robust computational models that aim to mimic human cognitive abilities. Figure~\ref{Fig:NN} presents a visual representation of the structure of a NN, which comprises of many layers, each one with a set of parallel neurons. After calculating a weighted combination of the input based on the summation function, $\Sigma$, the neuron feeds it into the activation function, $\sigma$. Thus, we may think of a neuron as a perceptron or an extension of logistic regression. Each node in one layer is linked to all other nodes of the next layer in the simplest structures, often called fully-connected NNs. With non-linear activation functions and at least one hidden layer, NNs may estimate any function with arbitrary precision, according to the universal approximation theorem. Much like the logistic regression example, we require an optimization technique even to train a single neuron. Efficiency in training is especially crucial when dealing with deep NNs. Backpropagation is a typical technique for training NNs, where the loss function's gradient is back-propagated through the layers using the chain rule and then techniques like gradient descent may be used to adjust the parameters. 

Various NN-based localization approaches have been offered in past literature. Among these reported techniques, NNs were employed for passive localization, where the calibrated phase and amplitude of CSI were used as a hybrid complex input feature to the NN to localize the user~\cite{Fang2019}. Also, in~\cite{Yang2019}, localization was achieved in a mmWave communication system by integrating NNs into a collaborative weighted least square estimator, while, in~\cite{Wu2019c}, the authors employed RSS data as input to a NN, which subsequently calculates an estimate for the ranging error; the modified range values may then be used by a localization technique like least squares. Another instance,~\cite{Lee2018} utilizes the user's position in an cellular network along with the RSRP of the three strongest fixed nodes for the user's localization. Moreover, a three hidden layer NN that leverages RSS from several fixed nodes has been recently proposed with its robustness being increased by data augmentation methods. In addition, two separate networks are developed based on the RSS and ToA measurements of a WiFi system, and the final user's location was calculated as the weighted average of the two NNs' outputs~\cite{Ge2019}. A hierarchical localization technique that utilized a NN trained for cellular and WiFi indoor location has been implemented in~\cite{Jaafar2018}. This technique is based on real-time RSS data to recognize the environment during the online phase before being fed to the matching trained NN. 

As far as MIMO-OFDM systems are concerned,~\cite{Wang2016b} presented a fingerprint-based localization approach that relies on the CSI magnitude from three antennas. This technique represents fingerprints using the weights of a four hidden layer NN. For the NNs' training, a greedy learning approach was used that stacks radial basis functions. Following initial training and supervised tweaking, a NN's output is a faithful recreation of the input data. During the online phase, the observed data likelihood probability is represented by a set of RSS values when the $i$-th location is true, and then the Bayes rule is used to get the posterior probability of location $i$. The destination is determined by averaging all fixed network nodes equally. A linear modification of the phase value as a calibration step increases the stability of the phase values, as proposed in~\cite{Wang2015}. Fingerprints may be created from the weights of a NN with three hidden layer.

With the use of directional antennas in a WiFi network, the authors of~\cite{Zhang2019c} constructed localization fingerprints based on RSS measurements and employed a classifier in the form of a two hidden layer NN to determine which fingerprint is most similar to the observed values. It is noted that, an improvement in localization precision may be achieved by using an approximation of the AoA. For example, the CSI amplitude of a MIMO system can be used as input to several NNs with various hyperparameters, with the final location estimation being a fusion of the output from all NNs~\cite{Sobehy2019}. From the numerous combining strategies that were analyzed, the best result was provided by taking both a weighted average and a median of the location estimations. Furthermore, in~\cite{Belmonte2019}, a framework was presented that integrates characteristics from many communication protocols, including XBee, Bluetooth, and WiFi, to aid in the tracking of targets. Yaw readings and RSS values from various nodes are used as inputs to the proposed NN. After the fingerprint probabilities have been calculated, the output of the Gaussian outliers filtering technique can be aggregated to generate the position estimate, which can subsequently be fed into a particle filter for target~tracking.

Extreme learning machine (XLM) is a NN architecture with a single hidden layer that is presented as an alternative to back-propagation, which is used by the aforementioned designs and requires significantly more time. First-layer weights are often assigned at random, whereas hidden-layer weights are typically determined using a least-squares fit. Several studies, including~\cite{Decurninge2018} and~\cite{Feng2019}, aim to exploit the low-complexity of XLM design in localization. In~\cite{Feng2019}, the data was first grouped offline using k-means clustering, then an XLM was employed to classify the data to one of the clusters, and finally a dedicated XLM was trained for each cluster. During the active online phase, the XLM of the corresponding cluster was employed once the RSS values have been first categorised. As part of the offline process in~\cite{Decurninge2018}, an XLM was trained using data collected from a database of WiFi RSS fingerprints. Fingerprints were still gathered at certain previously identified hotspots throughout the online phase, and the solution was revised to account for the additional data. The newest XLM was then utilized for coordinate prediction.

\subsubsection{Autoencoders}\label{Sss:Autoencoders}
\begin{figure}
    \centering\includegraphics[width=1\columnwidth]{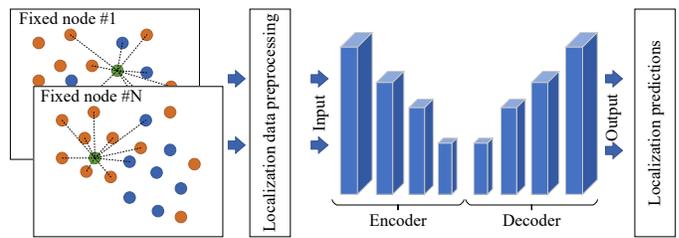}
    \caption{Autoencoder architecture.}
    \label{Fig:Autoencoder}
\end{figure}
The localization literature makes use of autoencoders with a number of pre-existing supervised solutions, despite the fact that autoencoders are typically applied in unsupervised learning. Figure~\ref{Fig:Autoencoder} presents a high-level representation of an device-free localization autoencoder, where the area is partitioned into $N$ grids, each of which has a fixed node. The goal is to predict the position of the target inside one of these N grids. Pretrained features are extracted from data using an autoencoder as the foundation of the proposed network. The localization readings from the stationary nodes are used to build the input topology. Interestingly, the autoencoder's output is a ``copy'' of its input, which delegates autoencoders into the unsupervised learning architectures. The autoencoder may be broken down into three distinct parts. Firstly, the encoder maps the input to the second part, namely the hidden state, $h$, and afterwards, the decoder maps $h$ to the output, which can be expressed as
\begin{align}
    \boldsymbol{y} = f(g(\boldsymbol{x})) ,
\end{align}
where $\boldsymbol{x}$ and $\boldsymbol{y}$ denote the input and output, while $f(\cdot)$ and $g(\cdot)$ represent the encoder and decoder functions. 

The output of the autoencoder is often sent into a localization module, making this method a common choice for extracting strong feature representations. The autoencoder is utilized to extract high-level features, and the encoder's output substitutes the random projection of an XLM~\cite{Khatab2017}. Different regularization algorithms that rely on the encoder output may be easily included in the training procedure, which involves a hierarchical tuning procedure where the classifier parameter is adjusted before moving on to the encoder parameters. On a different approach, a multi-output network that makes use of the encoded information to predict the coordinates, floor, and building was presented in~\cite{Kim2018}. The building structure along with the target's coordinates were used to generate the location using stacked autoencoders~\cite{Song2019}. Another application of stacked autoencoders combined with two fully-connected layers of a deep network applies multi-label classification prediction in an indoor localization scenario~\cite{Belmannoubi2019}, while, in~\cite{Zhang2017}, an autoencoder was used to reduce the dimensionality of the magnetic field signals and RSS. 

Autoencoders have also been used to evaluate the degree of similarity between the gathered fingerprints and observed characteristics in fingerprint-based localization. Bi-modal DL localization (BiLoc) is a fingerprint localization solution that provides the average amplitude across two antennas and AoA estimations as input to autoencoders and utilizes the output weights as fingerprints~\cite{Wang2017}. In the online phase, the degree of similarity determines the likelihood that the target is located at a given time and place based on the observed characteristics recnostructed by the autoencoders. The final position of the target is determined by averaging the locations of the fingerprints according to their respective weights. In~\cite{Yazdanian2018}, the closest static point was used to determine the location by reconstructing the observed signal based on the autoencoder's latent variable and position. Moreover, stacked autoencoders matching to each fingerprint are constructed to provide a probabilistic position estimate in real-time. To achieve this, this method first attempts to rebuild the observed characteristics and then compares the similarity with the radial basis function~\cite{Abbas2019}.

Autoencoders have also been used in device-free localization~\cite{Wang2016, Zhao2019, Zhao2018}. The encoder's output is sent into a classifier in convolutional autoencoder, which then makes a prediction about the predicted location of the target~\cite{Zhao2019}. In this example, the convolutional autoencoder takes as input a picture built from the disparity between the target and real-time RSS values. Based on the detected signals from the RSS, the authors of~\cite{Wang2016} estimated the gesture, activity, and position of the user by denoising the signal through wavelet decomposition, using a sparse autoencoder for dimentionality reduction, and finally incorporating the learnt features into a regression model and softmax classification. Data transformation is another possible use of autoencoders. For instance, the device heterogeneity issue may be addressed by an autoencoder that transforms the characteristics seen by a test device into features that match to the device used for obtaining the database~\cite{Rizk2019}.

\subsubsection{Convolution neural networks}\label{Sss:CNN}
\begin{figure}
    \centering\includegraphics[width=0.9\columnwidth]{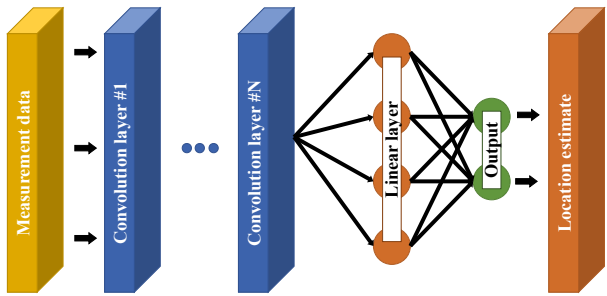}
    \caption{CNN architecture.}
    \label{Fig:CNN}
\end{figure} 
CNNs are efficient NNs designs that have demonstrated remarkable performance in computer vision applications. Through the use of parameter sharing, CNNs are able to construct more complex networks with a reduced set of parameters, compared to a fully-connected NN, like the one presented in~Fig.~\ref{Fig:CNN}. The convolution technique as well as the features maps and kernels play a very important role in the performance of CNNs. Within the convolution operation, the model's parameters are multiplied by the input data. Therefore, its data requirements and the computational complexity are reduced. At this point, it is important to point out two noteworthy characteristics of CNNs. On the one hand, connection sparsity is common in CNNs, in which low dimension filters can be used to generate each output value, while, on the other hand, CNN's sharing of parameters where each pixel uses the same weights. The translation efficiency and invariance of CNNs are appealing qualities made possible by these attributes. The term ``translation invariance'' is used to describe the CNNs' resistance to translation, which allows them to provide consistent results despite changes to the input data. Due to the achievable data and energy efficiency, more complex structures may be developed, allowing for the benefits of deeper networks to be realized. Overfitting is also mitigated by limiting the amount of trainable parameters, allowing the network to be trained on fewer samples of data.

Since CNNs have been proven effective in computer vision, various recent publications have developed CNN-based localization strategies. Several studies restructure the recorded RSS array into 2D or 3D pictures and pass it to a CNN because of the high number of reported RSS values in metropolitan regions. With an applied CNN on the RSS picture, the authors of~\cite{Jang2018b} were able to assign the coordinates to one of numerous floors and buildings. Improvements to the RSS pictures were achieved by adding correlation values to the characteristics in~\cite{Mittal2018}. The augmented image was fed into a hierarchical localization structure, where CNNs first predict the floor number, based on which the corridor number was estimated, and finally the coordinates. In order to quantify the relative importance of the RSS value to the fingerprint, the authors of~\cite{Liu2019b} suggested a set of hybrid RSS characteristics. In a smart parking system, time-series bluetooth RSS measurements were utilized for pedestrian and car localization~\cite{Ebuchi2019}. Another hierarchical localization approach predicts the coordinates, floor, and building by combining RSS values from all fixed nodes over many time occurrences to create an RSS-time 2D representation~\cite{Ibrahim2018}. From a temporal perspective, in~\cite{Soro2019}, continuous wavelet transform was employed to generate a 2D time-frequency picture, which was fed in a CNN that predicts the closest reference points to the target. Then, kNN was used to infer the target's coordinates. 

High dimensional features may be used thanks to CNNs' increased efficiency in NNs. A multi-layer CNN that forms an image using the channel frequency response values at the available sub-carriers at various time instances has been presented in~\cite{Chen2017}. Under this framework, realizations at various antennas can be used at different CNN channels, whose output was the likelihood of the target localization problem. According to~\cite{Wang2018}, by averaging the fixed nodes' expected probability and actual locations, it is possible to estimate the precise position, which is based on the pair-wise phase difference between antennas. The position was predicted by combining the locations of the most likely fixed nodes during the online phase. Next, in~\cite{Jing2019}, the posterior distribution of the position of the end device was approximated by a Markov model using the spatio-temporal information learnt by 3D CNNs. It produces two 3D representations of the phase and calibrated amplitude of the CSI, where the depth, breadth, and height correspond to the phase/amplitude of various TX-RX pairs, sub-carriers, and packets, respectively. Finally, a computationally efficient deep CNN architecture has been used for predicting the most likely fixed nodes in a three-antenna system that generates three-channel images. From these channels, two were utilized for the antenna pairwise phase differences, while the other for the amplitudes of all sub-carrier packet samples~\cite{Li2019b}.

CNNs have been also employed for device-free localization. For instance, in~\cite{Hsieh2019}, an 1D CNN model that takes advantage of either RSS values of packets from multiple antennas has been used or CSI amplitudes from all antennas and sub-carries. Another approach, in~\cite{Huang2018}, generated pictures from a time series of RSS values and associated continuous wavelet transform in order to identify people in a building. From these contributions, it becomes evident that solutions based on CSI outperform others when it comes to localization accuracy. Moreover, in~\cite{Bregar2018}, a CNN was employed to identify NLOS interference and estimate the range error in ultra-wideband systems, while, in~\cite{Niitsoo2019}, the authors recommended the use of time-calibrated complicated channel impulse responses to build pictures in an industrial context, where the the increased multi-path propagation limits the communication and localization range. 

Lastly, CNNs have been successfully applied in massive-MIMO systems. According to~\cite{Wu2021}, a uniform planar array can benefit from 3D pictures with power values in the delay, as well as vertical and horizontal domains. A CNN network is then fed these pictures and trained using the inception module in combination to various kernel sizes, which are imposed by the distinct sparsity of each domain. In popular Deep NNs designs, such as AlexNet, GoogLeNet, and more, the inception module is utilized to combine the results of many kernels for more accurate feature extraction~\cite{Vieira2017}.

\subsubsection{Recurrent neural networks}\label{Sss:RNN}
\begin{figure}
    \centering\includegraphics[width=0.85\columnwidth]{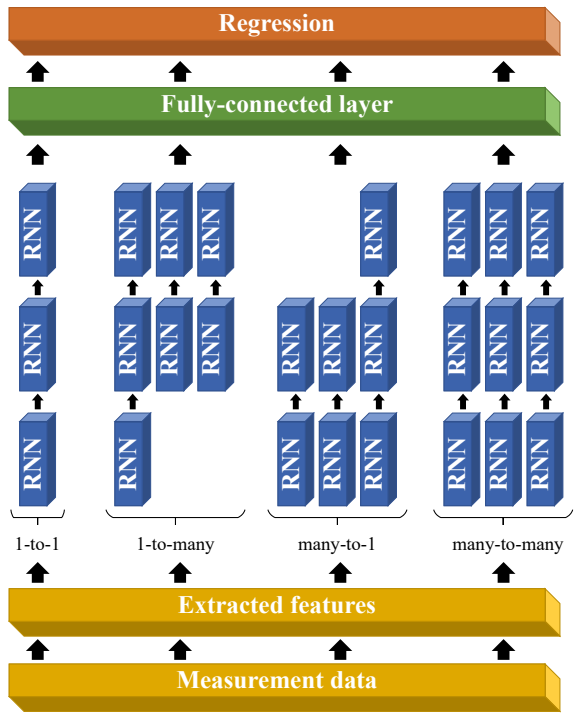}
    \caption{RNN architecture.}
    \label{Fig:RNN}
\end{figure}
Modeling sequential data is a common use of a specific form of NN called a recurrent neural network (RNN). With natural language processing and data processing of time series exhibiting a lot of potential. RNNs have the ability to carry out tasks, in which the outcome is reliant on both the past outcomes and the present input. The results of previous calculations may be viewed as being stored in the "memory" of RNNs. The RNN's hidden state, $\boldsymbol{h}(t)$, may remember the results of previous calculations. The output of the RNN can be written as 
\begin{align}
    y(t)=\sigma_y\left(W_h \boldsymbol{h}(t)+\boldsymbol{b}_h\right) ,
\end{align}
with $W_h$ denoting the weights of the RNN, while its hidden state can be expressed as
\begin{align}
    \boldsymbol{h}(t)=\sigma_h\left(V_x \boldsymbol{x}(t)+V_h \boldsymbol{h}(t-1)+\boldsymbol{b}_h\right) .
\end{align}
In the previous equations, $\boldsymbol{b}_h$, $V_x$, and $V_h$ are the RNN's parameters, while the activation function is denoted by $\sigma$.

In RNNs, the model's parameters are preserved across time. In other words, they are resources that are shared by all links inside a certain layer and are fine-tuned through time via back propagation. However, when a gradient is propagated in time, gradient explosion or vanishing can occur and cause said gradient to increase or decrease, respectively. Another issue is that the model's effect gradually wanes with time; thus, it can not keep in mind too much history. Therefore, plain RNNs are not often employed in the real world. In reality, ``gates'' in RNN cells are used to account for influence loss or deterioration over time, with the most prominent architectures being long short-term memory (LSTM) and gated recurrent units (GRUs). As illustreted in~Fig.~\ref{Fig:RNN}, RNNs come in a wide variety of sizes and shapes, with some even being able to learn in both directions. According to architecture presented in this figure, for each fixed node during a specifide time window, a number of features extracted from the raw measurements are computed. Thus a plethora of data is generated from the initial measurements. These features are then fed into the LSTM RNNs in order to extract higher level features that capture the essential information of the localization data. Finally, the high level features are used as input for the regression, which is realized by using fully-connected layers in order to counterbalance overfitting of the algorithm. All in all, using RNNs to monitor a target's movement over time is a crucial part of localization.

Numerous recent efforts have relied on recurrent neural networks that receive a series of RSS values for localization purposes. For instance, the floor and building can be estimated using a cascaded RNN that receives a succession of RSS values as input~\cite{Turabieh2019}. For real-time systems, RSS measurements accompanied by numerous Bluetooth anchors can be used for localization with high efficiency~\cite{Canton2017}. Another effort monitored the users' whereabouts from the linked cell towers' RSS history and utilized kNN and Markov models to produce synthetic measurements based on the previously observed ones~\cite{Rizk2019c, Rizk2019b}. Moreover, the authors of~\cite{Chen2019b} used a sliding window on a series of RSS data, computing five features for each access point within that window, which were then used as input for an LSTM device as a vector sequence.

In WiFi and cellular networks, RNNs have been used for collecting the RSS fingerprint from several fixed network nodes over different timesteps and feeding it into a single-layer RNN, which returns the coordinates~\cite{Wu2019b}. By comparing the CSI amplitudes at various sub-carriers and antennas, it is possible to achieve localization in a MIMO system, where the SNR, CSI, and correlation matrix are employed as candidate features for various RNN approaches~\cite{Yu2020}. After the CSI has been pre-processed and shifted, polynomial regression is applied and the output data sequentially into the LSTM system, which is proven to performs better than competing systems. Furthermore, predicting the position of UAV BS using WiFi and RSS measurements has been reported in~\cite{Adege2019}. Before being integrated with the other data, the RSS features' dimensions are reduced using principal component analysis. In order to estimate the location, the new feature vector is fed into a RNN. In a similar arrangement,~\cite{Tarekegn2019} utilized linear discriminant analysis to choose the subset of fixed nodes to decrease calculation time, using RSS data from the accessible WiFi fixed nodes to construct a radio-map. It then uses an LSTM system to provide a position prediction based on the RSS sequences.

RNN and CNN hybrids have been mulled upon in the recent literature~\cite{Qian2019, Zhang2019b}. A smooth trajectory of projected locations can be generated from sequential measurements in a cellular system~\cite{Zhang2019b}. At first, the area is split into cells, while afterwards pictures are generated with dimensions that correspond to the grid structure and populated based on the measurements. The photos are then propagated to a CNN, which uses the retrieved spatial characteristics to predict a score for each possible location. Next, the outcomes are then used as input to a multi-layer LSTM, which is responsible for creating the trajectory. With the use of consecutive RSS values from neighboring fixed nodes, localization is achieved. On another note, a 1D-CNN is used to extract features from consecutive RSS values and is followed by a RNN to record the temporal correlation~\cite{Qian2019}. Afterwards, a mixture density network is trained on the RNN's output to discover the conditional probability distributions of the locations.

\subsection{Unsupervised learning}\label{Ss:Unsupervised}
In contrast to supervised learning approaches, unsupervised learning methodologies take advantage of the features' underlying structure and distributions enclosed in unlabeled data. Multiple techniques for pretraining NNs reside in the general category of unsupervised learning and are discussed in this section. Two examples of architectures that have been employed, are deep belief networks and restricted Boltzman machines. Specifically, the former has been used to learn deep features from RSS data, and then feed those features into a different ML solution for location estimation~\cite{Le2018}. Another application of unsupervised learning techniques is to be applied in various localization tasks like device mapping, data filtering, or access point selection in order to provide increased performance.

\subsubsection{Semi-supervised}
On the way towards applying unsupervised techniques of localization, several researcher have developed methods of both semi-supervised and unsupervised nature. One application can estimate the distance of the user from the access points using RSS values based on a variety of cost functions, while, at the same time, evaluating the localization precision in terms of the difference between actual and estimated locations~\cite{Choi2019, Choi2019b}. Another solution is provided by developing a graph-based model, where latent variables are considered with regard to power levels and location~\cite{Zafari2019, Sikeridis2018}. To solve this model, a Gaussian mixture model was employed to calculate the received RSS likelihood under the assumption of independent normally distributed variables. Next, the models parameters are estimated through the expectation maximization algorithm that is initialized with a basic pathloss model and the positions of the fixed network nodes in order to address the identifiability issue.

\subsubsection{Clustering}
On a different note, unsupervised techniques have been successfully applied for constructing the map of a radio-based system. An instance of such a system develops a logical floor plan using accelerometer and RSS measurements from users within the service area~\cite{Jang2018}. This techniques is based on clustering techniques like k-means on RSS stacking difference, while each virtual room is assigned a representative localization fingerprint. Another proposed framework for unsupervised localization utilizes WiFi, gyroscope, compass, gyroscope, and accelerometer values of naturally moving users and tries to identify some fixed structures in the building (i.e., elevators, columns, stairs, and more) that greatly influence their movement. Afterwards, it employs k-means clustering to extract unique sensor signatures that can increase the localization accuracy for users starting from a known spot in the building~\cite{Guo2019}. Localization can be also achieved by the combination of global-local optimization and a Markov model that fits the RSS traces into the structure of the environment based on unlabeled data and considers the solutions that do not violate the signal propagation~\cite{Ye2018}. 

\subsubsection{Dimensionality reduction}
Since many approaches rely on fusion to fulfil the localization procedure, an unsupervised learning approach has been proposed that concatenates the outputs of the best classifiers to develop a joint location-weight estimate and an extended candidate location set within an unsupervised optimization framework~\cite{Guo2018}. Forecasts that are considered more accurate are given more weight, thus the user's actual location should be near the most accurate predictions. Lastly, multidimensional scaling has been widely employed in dimensionality reduction schemes for localization~\cite{Saeed2016, Gao2017, Saeed2019}. Such schemes create a low dimensionality projection of the known distances between wireless nodes that yields the spatial map of the network configuration. Cooperative localization utilizing wireless signals, such as ToA, AoA, RSS and more, has lately been used for cognitive radio, IoT, and RFID to estimate distances between devices~\cite{Saeed2016, Gao2017}.

\subsection{Federated learning}\label{Ss:Federated}
\begin{figure}
    \centering\includegraphics[width=0.8\columnwidth]{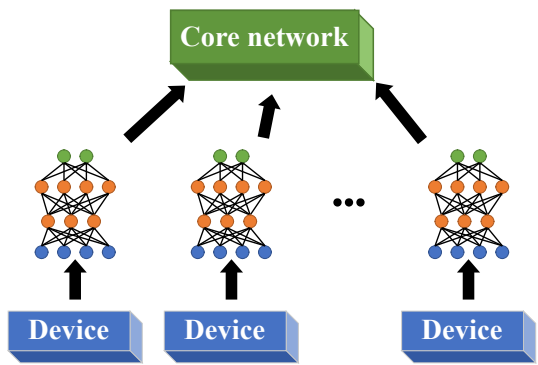}
    \caption{Federated learning architecture.}
    \label{Fig:Federated}
\end{figure}
Federated learning (FL) decouples the capacity to perform ML from the need of storing the data in the cloud, allowing edge nodes to develop a shared prediction model cooperatively while maintaining all the training data locally. This extends the usage of edge device-based prediction models by bringing model training to the edge node itself. Specifically, FL is implemented by deploying a plethora of mobile nodes, each of which is in charge of a certain region that may or may not overlap with its neighbors. In order to perform local training of the global parameters, each mobile node gathers a dataset based on which it performs the minimization of the NN's parameters, $\theta_k$, by solving 
\begin{align}
    \theta_k=\arg \min _\theta \sum_{\forall\left\{x_i, y_i\right\} \in \mathcal{D}_k}\left\|y_i-f\left(x_i ; \theta\right)\right\|_2^2 ,
\end{align}
with $\left\|a\right\|_2^2$ denoting the sum of the square of $a$'s components and
\begin{align}
    y_i=f\left(x_i ; \theta\right)+ n_i ,
\end{align}
where $f$ denoting the regression function. After this is carried out by every mobile node working together, the parameters can be defined as the global objective can be expressed in the form of a sum. The mobile terminals encrypt the communications before sending them to the core network, which then uses homomorphic algorithms to decode the messages. Figure~\ref{Fig:Federated} depicts the FL concept for localization.

Advantages of this collaborative framework include less time spent on training and, more crucially, the protection of individual privacy by removing the need to provide sensitive information to the network~\cite{Niknam2020}. Typically, in FL, it is assumed that a centralized entity would combine the models trained at the edge devices and broadcast back an updated version of the model. Due to the unpredictable nature of mobile surroundings and the privacy leakage limitations introduced by the centralized nature of the data processing architecture employed in object localization, traditional ML localization approaches may be surprisingly fragile~\cite{Chen2017b}. FL has the potential to provide intriguing solutions for allowing accurate and private location services. 

There is an abundance of literature exploring the usage of FL in localization contexts; especially, since location data is one of the most fundamentally private bits of information. For example, a FL technique that uses NNs to anticipate user coordinates in order to increase the reliability and robustness of RSS fingerprint-based localization without compromising participant privacy has been recently presented~\cite{Ciftler2020}. When the users have finished training the model, the central node adjusts the NN weights based on the amount of samples they were exposed to and sends the resulting model back to the users. In addition,~\cite{Li2020} confirmed FL's promise in indoor localization services by addressing the problems of privacy leakage threats and task learning associated with using a centralized AI server. Using the computing power of mobile devices, the suggested architecture combines centralized indoor localization with FL to lighten the burden of fingerprint collecting and save network computational costs while maintaining users' privacy. In order to construct a global statistical localization model, a DL model is run on each node based on unlabeled crowdsourcing and labeled fingerprint data, and then shares the calculated updates with a central server. When compared to other methods of disseminating fingerprint data, FL definitely outperforms them in terms of stability, privacy protection, and transmission cost~\cite{Nguyen2021}.

Similarly, a federated localization strategy for WiFi networks is constructed in~\cite{Liu2019c} using FL. WiFi signals representing known landmarks may be used by mobile devices to create local fingerprints, which can then be fed into a DNN model and subjected to a deep autoencoder to filter out background noise. A centralized server then compiles all of the regional weights into one global one. Local updates are encrypted using a homomorphic encryption method to safeguard the communication channel during the offloading phase. High precision in localization estimates and safety are shown in a laboratory corridor experiment. As another example, in~\cite{Yin2020} a federated localization architecture was suggested for precise IoT localization. FL reduces the privacy risks associated with location estimation by having numerous users work together to develop a model using local fingerprint data. Emerging localization services, such as wireless traffic prediction using BSs, and mobile indoor GPS localization, mobile target navigation and tracking using inertial sensors, may all benefit from FL's utility.

\subsection{Reinforcement learning}\label{Ss:Reinforcement}
\begin{figure}
    \centering\includegraphics[width=0.9\columnwidth]{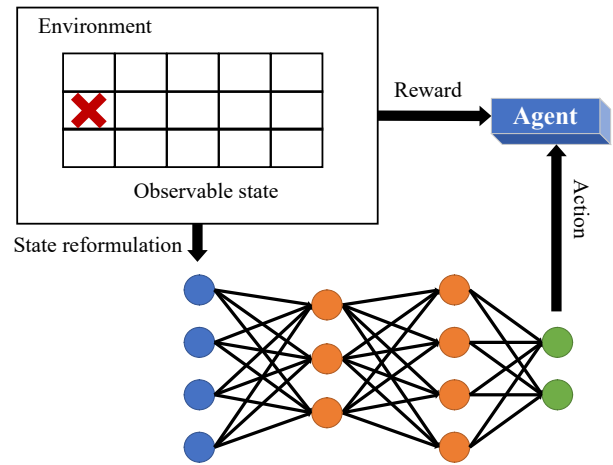}
    \caption{Reinforcement learning architecture.}
    \label{Fig:Reinforcement}
\end{figure}
Reinforcement learning (RL) is among ML's burgeoning research fields and it involves teaching an algorithm how to make judgments by letting it make mistakes, while interacting with the world around it. To optimize their rewards over time, agents conduct a series of behaviors in a predefined setting. A conceptual outline of RL-based localization is provided in~Fig.~\ref{Fig:Reinforcement}. The agent's abstracted state of the world, or environment, from which it draws its action choices at time $t$ is comprised of both the agent's location and the RSS reading make up the state. Moreover, the agent evaluates the state and uses that information to choose what courses of action to conduct. The ``good'' activities explored in localization schemes include traversing the grid while remaining on it. In accordance with the reward function, the agent will get a favorable response after taking the desired course of action. Theoretically, incentives may be determined by the distance between the target position and the agent. However, it is impossible to utilize this approach to handle unsupervised data. A technique, which build on the premise of isolating landmarks with accurate position labels and powerful RSS characteristics for establishing rewards, is proposed as a solution to this problem~\cite{Shao2018}. In this respect, the agent receives a reward if and only if the measured RSS matches the known RSS characteristic at the landmark location.

\subsubsection{Markov decision process}
For dynamic systems, like resource allocation, wireless communication and localization, it has attracted a lot of interest. For instance, a bluetooth based system that uses past locations information and RSS measurements to determine whether or not the reward is achieved, when the solution approaches known RSS values or an RP~\cite{Li2019}. In order to determine a progressive localization algorithm, the WiFi localization problem in a WiFi system is viewed as a Markov decision process with a model-free approach~\cite{Dou2018}. In this example, states may contain the RSS, action history, and center coordinate values. There are five possible motion directions in the action space, each of which may be used to shift the predicted location window by a different set of radius values. Another localization problem that can be tackled with RL methods is scheduling the exchange of signals during cooperative localization~\cite{Peng2019}. As a result, the solution treats the measurement choices as operational tasks and the linkages as autonomous agents. The observations may include the nodes that failed to meet a localization quality criteria, covariance values, and distance.

\subsubsection{Deep RL}
Another application of RL is to enable self-learning localization scenarios in which new information is learned from the outcomes of previous choices. Compared to unsupervised multilateration localization, the accuracy of Q-network in a deep RL structure increases by 37\% when used to predict the position of a pedestrian device~\cite{Li2019}. To better understand human traffic, predicting pedestrian paths can play a very important role. This issue has been tackled by designing a deep RL method that aims at maximizing entropy~\cite{Fahad2018} and has achieved performance similar to LSTM localization. Another approach, used semi-supervised deep RL for estimating the pedestrian distance in localization applications and exhibited a 23\% increase in precision with regard to supervised learning approaches. 

\subsection{Transfer learning}\label{Ss:Transfer}
\begin{figure}
    \centering\includegraphics[width=1\columnwidth]{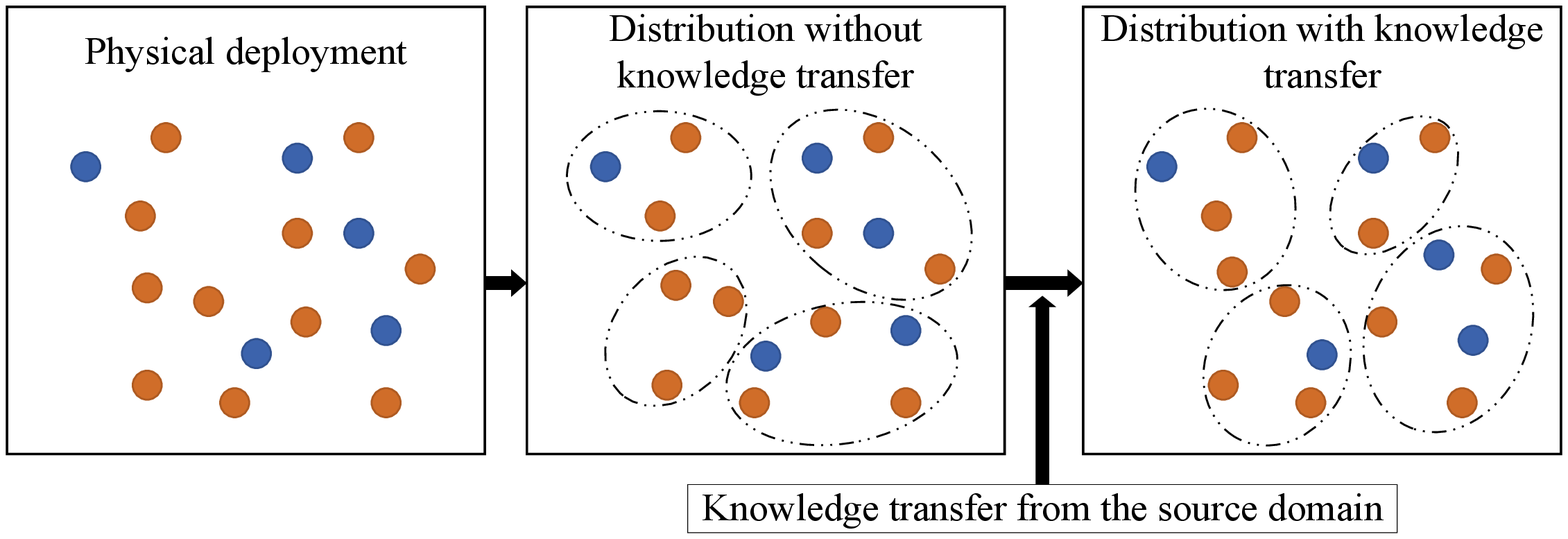}
    \caption{Transfer learning architecture.}
    \label{Fig:Transfer}
\end{figure}
Transfer learning (TL) is utilized to take advantage of the knowledge learned in one domain to a new domain. This is especially useful when it is challenging to acquire data in the target domain, but an abundance of data exists in the source domain. Figure~\ref{Fig:Transfer} depicts the two stages that make up the TL-based localization system, namely offline training and online localization. To be more precise, a number of fixed and mobile nodes are installed in advance, taking into account the physical characteristics of the area and the needs of the individual application. RSS readings from the mobile device may be sent to the surrounding fixed nodes by positioning the mobile ones. Each fixed node then uploads its data to the server, which uses the signal characteristics acquired in order to create a fingerprint database.

Another interesting case of TL is dynamic systems, such as localization applications, that require configuration and retraining on each distinct application. In this scenario, the overhead of retraining the model when the data is accessible in both domains is challenging and can be avoided with TL approaches. Based on these, the challenges of TL enabled localization span across many dimensions, including time, space, and hardware, where the data distribution may vary depending on the dimension. Specifically, TL is presented as a pair of optimization problems: the first involves extracting the underlying semantic manifold of the signal to serve as constraints for the second, which involves labeling the unlabeled data in the target domain. The challenge of TL across devices might be stated as one of multitask learning. For instance, a manifold regularization-based solution for a temporal TL can be applied to localization approaches where data distributions vary between time instances, but are often consistent with one another in low dimensions.

\subsubsection{NNs}\label{Sss:Transfer_NN}
One of the more recent applications of TL involves the utlization of SotA NN techniques. For instance, deep CNNs employ TL between many antenna configurations based on the time-domain CSI. After the network is first trained, it is possible to retrain the network's lower layers with less data points to accommodate the altered antenna configuration~\cite{De2020}. Another example, the transfer of knowledge across settings is explored in order to address the issue of localization based on RSS~\cite{Xiao2018b}. The NN model is pre-trained in one domain before being transferred to another domain, with the two domains being characterized by distinct propagation phenomena. Specifically, the domains are two floors in the same building, both of which have the same access point placement and architectural design. It is proven that with only 30\% of the data in the second floor the TL enabled system can achieve the same level of performance. Finally, deep CNNs for pedestrian localization can be combined with TL approaches with great success (i.e., 45\% increase of the training data and cut down on training time by half)~\cite{Li2021}.

\subsubsection{Conventional TL}\label{Sss:Transfer_classical}
Conventional TL methods augment the real-time measurements gathered from the target domain with labeled fingerprints from the source domain. For instance, a domain-invariant kernel, suitable for use with SVM in TL enabled localization, is learned using data from both domains~\cite{Zou2017}. Metric transfer and metric learning are the two components of the TL-based methodology that operate together to lessen the burden of offline training in the new setting~\cite{Zhang2018}. The metric learning component maximizes the statistical dependency between signal label and feature statistics to learn the distance metrics from source domains. Using a method that minimizes the difference in data between the source and target domains, the metric transfer component determines which metric is optimal for the target domain. In certain cases, the TL result may not take into consideration all aspects of the surrounding environment in order to recreate the radio map. To solve this issue, fuzzy C-means clustering has been used to reduce the impact of external factors~\cite{Liu2017}.

\subsubsection{Manifold learning}\label{Sss:Manifold}
\begin{figure}
    \centering\includegraphics[width=0.9\columnwidth]{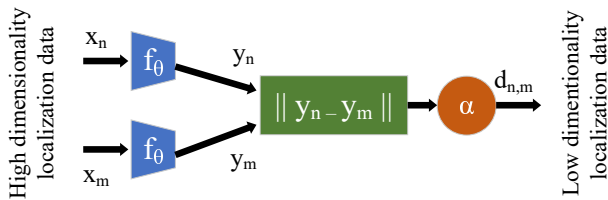}
    \caption{Manifold learning architecture.}
    \label{Fig:Manifold}
\end{figure}
The premise that the observed data sit on a low-dimensional manifold inside a higher-dimensional space is what distinguishes the discipline of manifold learning from others. In the ML world, manifold alignment is widely used for sharing models and data across disparate collections, assuming a common manifold. For instance, the Siamese network design presented in~Fig.~\ref{Fig:Manifold} employs two identical NNs to translate the high-dimensionality input characteristics, $x_n$ and $x_m$, to lower dimensional representations, $y_n$ and $y_m$, taking into account the fact that location affects the values of large scale parameters. The network's objective is to keep the distance between $x$ and $y$ to a minimum. Finally, in a semi-supervised learning situation, the parameter $\alpha$ is a scaling value for distance scale matching.

Data visualization and dimentionality reduction are only two of the techniques that are applied in manifold learning. For example, graphs is one way to approximate the manifold's point relations, while variable manifold alignments can often result in neighborhood graphs with unique distances between graph nodes~\cite{Xu2020}. Specifically, in the area of localization, manifold alignment has found multiple applications for constructing a radio map based on the observed distances between the nodes. Another example is the Laplacian Eigenmap manifold alignment method that develops a weighted graph connecting the data points and maintaining the local geometry~\cite{Zhou2017}. This technique utilizes the Laplacian eigenvectors of the constructed graphs that contain the physical relations of the geometrical observations (such as RSS). Moreover, manifold learning has been utilized under a semi-supervised approach where both labeled RSS data as well as unlabeled timestamped traces are used to build the graphs~\cite{Zhou2017b}. In addition, to keep the distance estimates based on the wireless propagation model, the goal functions can be modified and Gaussian kernels can be used to determine the graph's weights between the RSS measurements~\cite{Behera2020}. Another approach of manifold learning optimizes a time-series graph Laplacian SVM to generate pseudo-labels, which are then used as part of a learning framework to assist with semi-supervised RSS localization. In order to strike a middle ground between the labeled and pseudo-labeled contributions, it incorporates manifold regularization into a transductive SVM~\cite{Yoo2019}. In WiFi, manifold regularization using Laplacian graphs is taken into consideration in the solution for extreme learning machine parameters. Since Bluetooth and WiFi signals display distinct propagation circumstances, two graph Laplacians are developed to reflect the smoothness in Bluetooth and WiFi, respectively~\cite{Jiang2018}. Finally, a Siamese network design, which consists of two identical NNs used to compare two inputs was used as a semi-supervised or supervised CSI localization solution~\cite{Lei2019}. Because of the impact of geography on the values of global parameters, a pair of feed-forward NNs is used to transform the input features into more manageable location representations.

\section{Optimization frameworks for localization}\label{S:Theoretical_framework}
To assess the fundamental bounds of achievable localization performance, problem formulation, design and optimization are crucial. This section starts by giving the formulation of the optimization problem based on the intended system goals, which is presented in Section~\ref{Ss:Problem_formulation}. The system's high-level design considerations are then covered in Section~\ref{Ss:Design}. Finally, Section~\ref{Ss:Optimization} delves into the optimization of system design issues.

\subsection{Problem formulation}\label{Ss:Problem_formulation}
Precision, throughput, signal quality, and energy efficiency are among the most important KPIs that must be optimized in communication systems. When designing a system with precision as a goal, the orientation and position error bounds are put to use for localization. However, these criteria are still a useful and manageable tool for assessing performance in the asymptotic area, even if it is only applicable when the estimator is efficient~\cite{Chen2022}. Aside from the goals listed in Section~\ref{S:6G_aplications}, there are several situations in which they become crucial. However, there is no one, universal definition of an aim; rather, objectives must be defined in light of the context in which they are used in. Compromises are necessary due to the various formulations of objective functions, particularly in the optimization of simultaneous communication and location systems.

The BSs transmit positioning reference signals in low-frequency localization systems, and the accompanying system architecture are mostly offline. For instance, this is the case for antenna array and BS layout design. The improvement in localization performance is a side effect of the beamforming gain in MIMO systems. However, knowing the location of the receivers is essential for beamforming. Thus, combining matrix design, online precoding, and resource allocation are of paramount relevance. THz systems that rely on array-of-subarrays antenna structure perform precoder/combiner optimization at the subarray level rather than the antenna level. Therefore, the beamforming angles of the subarray must be well-designed together with the data symbols from the RFCs. Also, powerful algorithms are needed for optimizing resource allocation and RIS coefficients within a dense network. It is clear that both online and offline improvements will play a significant role in the success of future communication systems. The optimization issue is then formulated, and the impact of the various factors on the goals of the system is discussed.

Specifically, the performance requirements for localization vary depending on the application scenario, as well as the related objectives and KPIs. These goals often overlap, and it is necessary to make concessions. Increases in coverage and update rates may have unintended consequences, such as diminishing accuracy. In order to maximize its effectiveness, a system may have several goals that it must balance. Such a localization system can be modeled under a generic optimization problem such as
\begin{align}
    \begin{split}
        \mathcal{X} = \underset{x}{\arg\min}\ \ &\mathbf{f}\left(\mathcal{X}\right) \\
    \mathrm{s.\ t.}\ \ &\mathbf{g}\left(\mathcal{X}\right) \leq 0 
    \end{split}\ \ ,
\end{align}
where $\mathbf{f}$ and $\mathbf{g}$ denote the objective and constraint functions, respectively. Moreover, $\mathcal{X}$ represents the variable set that is used to achieve the optimal solution for the optimization problem constructed under each localization scenario. Depending on the context, a parameter may represent either an aim or a limitation. For instance, localization accuracy may be used as either a goal to achieve or a constraint to be satisfied, such as spacing, number of devices, beamforming angles, and so on, in addition to other KPIs, like throughput, precision, etc.

\subsection{System design}\label{Ss:Design}
The localization system's design is entangled to the use cases in order to fulfil their requirements. These use cases inform choices on network architectures, cooperative approaches, as well as algorithm development and optimization.

\subsubsection{Network architectures} \label{Sss:Network_architectures}
Under the umbrella of networks, as discussed in the previous sections, three types of channels have been covered LOS, NLOS, and RIS-enabled ones. However, densification is a key characteristic of future communication networks, thus numerous fixed and mobile nodes should be engaged. There are three distinct kinds of topologies for THz communication systems: centralized, distributed, and clustered. In a centralised system, one or more client nodes are linked directly to a centralised server, which uses client/server architecture. This is the most often utilised form of wireless system, in which the clients submit requests to the server and receive the answer. On the contrary, distributed networks aim to eliminate single points of failure by distributing processing across the network and coordinating their efforts. Distributed systems rely on the communication and synchronization between the nodes that are dispersed throughout a network. These nodes often consist of discrete pieces of hardware, however they might also be software processes. Centralized and distributed architectures are often utilized in macro settings with a wide communication distance. The centralized structures may offer greater overall performance with appropriate scheduling, while the scattered ones safeguard user privacy. Finally, a cluster network comprises of two or more computing devices sharing a single computational task. Such networks make use of the hardware's ability to process data in parallel in order to not only boost processing power, but also provide scalability, high availability, and eliminate network dependency on a single device. Nanonetwork settings, which value low power consumption and short communication distances, are well suited to the clustered design~\cite{Ghafoor2020}.

There are four possible architectures to increase the system's performance in large-scale scenarios: (i) cell-free, (ii) RIS-enabled, (iii) 3D and (iv) heterogeneous networks~\cite{Yastrebova2018}. The main advantages of cell-free networks include ultramassive MIMO beamforming and power savings. However, the THz channel's sparser multipath propagation and low rank restricts performance~\cite{Chaccour2022}. In this scenario, due to the lack of cell borders, end devices may have a high coverage probability, while the localization performance can be augmented by the geometrical diversity of the base stations~\cite{Faisal2020}. Next, The network's coverage area may be expanded and the channel reshaped with the use of RISs. THz-frequency RISs are predicted to have modest footprints due to the utilization of short wavelengths, which can allow a higher versatility in deployment. Finally, in a heterogeneous network, many wired and wireless protocols coexist. This is a possible scenario for future networks. In addition to solving the hearing problem, this kind of multi-band network may also significantly shorten the time it takes to go online.

The offline design of a system includes hardware selection to strike a balance between budget and overall system performance. Antenna polarization, hardware imperfections, and phase-shifter quantization are all taken into account throughout the hardware selection process. The main factors that influence signal quality in THz systems are absorption, blockage, device density, and antenna design. Understanding the optimal antenna polarization and device density for a THz network has been investigated in~\cite{Petrov2017}. As a rule, directional antennas are utilized for localization and transmissions, whereas unidirectional antennas are employed during the service discovery phase~\cite{Han2017}. Additionally, a quantized model should be considered in system design due to the discrete nature of phase control and amplitude signals in RISs~\cite{Wu2019}.

\subsubsection{Cooperative approaches} \label{Sss:Cooperative_approach}
Although energy consumption, as well as computational and time requirements are increased due to frequent contacts between the end devices, the localization coverage and precision are enhanced by cooperative localization~\cite{Xiao2022}. For a fair compromise, it is necessary to specify the related performance KPIs. Collaboration between fixed and mobile nodes, data fusion from several kinds of sensors~\cite{Zhang2019,Lu2020}, UAV-assisted localization~\cite{Wang2019} are all crucial. Furthermore, in many real-world use cases, the existence of NLOS propagation makes it difficult for certain agent nodes to directly connect with adequate fixed nodes for localization purposes, which may reduce localization precision. This phenomenon can be mitigated by using cooperative localization strategies~\cite{Xiao2022}. On the contrary, non-cooperative localization requires all mobile nodes to be in constant communication with fixed ones, necessitating either a densely packed network or a wide coverage area. By facilitating communication between mobile nodes, cooperative localization enhances accuracy and increases localization coverage beyond that of non-cooperative localization. The main problem of cooperative localization, which is a parameters estimation issue, may be addressed with probabilistic or deterministic methods. In the first case, the most prominent solutions involve multidimensional scaling, multilateration, traditional linear scaling, and other techniques that often fail in practice because they presume a Gaussian model for all measurement errors~\cite{Wang2020}. As far as probabilistic approaches are concerned, they not only provide location estimates, but also quantify the degree of uncertainty associated with those predictions. For instance, estimation theory and factor graphs are among the most notable techniques of probabilistic approaches, with recent efforts being focused on belief propagation and equivalent Fisher information~\cite{Gao2019}. 

\subsubsection{Algorithm development} \label{Sss:Algorithm_development}
It is still unclear whether THz systems benefit from the use of multi- or single-carrier modulation schemes. The former appear to be advantageous in applications, where frequency-flat channels exist, while, in spite of the poor power efficiency and high complexity, multi-carrier systems are still favoured in frequency-dependent molecule absorption loss and multipath scenarios. The utilization of discrete-Fourier transform spread OFDM may be employed as a ready-made solution in order to mitigate the PAPR impact~\cite{Tarboush2022}. Spatial~\cite{Sarieddeen2019}, index~\cite{Loukil2020}, hierarchical bandwidth~\cite{Hossain2019}, and orthogonal time-frequency space modulations~\cite{Wei2021} are also taken into account for specific scenarios as additional multi-carrier modulations. Additionally, research on non-orthogonal multiple access at THz is being conducted~\cite{Magbool2022}.

When designing wireless systems, the spectrum efficiency or data rate may be impacted by tweaking signal characteristics including packet length, bandwidth, and carrier frequency. These values are also critical for localization to accomplish targeted goals. For better route separation in the delay domain, more bandwidth is preferable, but the higher data size and sampling rate must be kept within the hardware's capabilities. When possible, the messages should be as lengthy as possible to maximize energy acquisition while being as short as possible to minimize delay, particularly in mobile settings. Finally, the localization system's effectiveness is heavily influenced by the design choices made.

\subsection{Optimization}\label{Ss:Optimization}
When a network architecture with no knowledge regarding the location of the end devices is assumed, it is possible to optimize the layout, antenna, and codebook based on data about the immediate fixed surroundings. In this scenario, the optimal placement of fixed and mobile network nodes may be calculated from the Cramér-Rao bound using a predetermined codebook after the quantity of these elements has been established. The design may be optimized for the greatest localization performance by taking into account environmental data. It is also important to ensure optimal orientation of base station antennas. Furthermore, higher beamforming and angular resolution performance is achieved by increasing the array dimensions of the antenna. However, as the number of antennas increases a more complicated and expensive setup is created. Designing an appropriate size subarray is crucial when selecting the antenna architecture. Beamforming gain and accuracy may be improved by increasing the number of active elements per subarray, although coverage can be reduced and hearing loss can occur with narrow beamwidths.

To get from an unconnected state to an active one, a new device must go through the initial access operation of establishing a physical connection with a fixed node~\cite{Hu2020}. The initial access process may be seen as localization without any input from the end device. Blockage and deafness are two problems that arise in THz systems because to the small beams, making initial access difficult~\cite{Barati2020}. This necessitates well thought-out codebook construction and efficient first-time-in processes. Therefore, the codebook layout is determined by search techniques, which can split into science-aware, iterative and exhaustive search ones. In the first category, the beams may be learnt for each partitioned region to lessen initial access procedure's delay, given the fact that the previous user device position or environmental information is provided~\cite{Chen2019}. Next, for iterative search, in order to determine the optimal angular space, hierarchical codebooks may be created for transmission across progressively narrower beams~\cite{Sun2019}. In the last category, the fixed and mobile nodes utilize beamforming to send and receive messages in various directions~\cite{Alkhateeb2017}. In terms of hardware and coverage practicality, an exhaustive search is the best option for these algorithms~\cite{Barati2016}. However, the discovery latency increases linearly with beamforming strength. While iterative search shortens the time it takes to make a discovery, its narrow focus comes at a cost.

On the contrary to aforementioned optimization approaches, when end device location information are available, the three major concepts are active beamforming, RIS coefficient, and resource allocation. Regarding the first case, if you know the exact position of the receiver ahead of time, you may boost the received signal strength by adjusting the beamforming angles such that they point directly at the receiver. However, this boost in SNR does not automatically result in better localization results. For more realistic approaches the Cramér-Rao bound is applied. For instance, beamforming optimization and location estimation can be iteratively applied by using numerous observations~\cite{Zhou2019}. Moreover, the optimum precoders for tracking the angle of departure and arrival can be determined by solving a stated convex optimization problem, if the uncertainty range of the target directions is available~\cite{Garcia2018}. In order to minimize beam assignments, space resource allocation often makes use of directional antennas and array-of-subarray structures. Accuracy and range are directly impacted by how well the analog beamforming angles are optimized. In multi-end-device environments, the beams must be strategically distributed across the various end devices so that communication and localization KPIs are met. To improve the overall system performance in a network with several fixed nodes, cooperative beamforming optimization is required.

Next, improved signal gain in RIS-assisted systems relies heavily on both adjustment of the RIS coefficients and active beamforming. Multiple broadcasts using randomized beamforming angles of messages may be employed for localization when the end device's location is uncertain. Beamforming angles at mobile and fixed nodes may be simultaneously improved using knowledge of the end device's location and orientation in advance. For improving the localization and communication quality, it is possible to tweak the coefficients of the RIS~\cite{Bjornson2022}. Also, a larger data rate is achievable by optimizing the RIS components to increase the SNR at the receiver, while the higher SNR is not always associated with a low Cramér-Rao bound. One limitation that has yet to be solved is 3D MIMO system optimization methods, which are not currently available.

Finally, allocating resources in a communication network that is servicing numerous end devices or performing different activities is a crucial step, with bandwidth, energy, and time being the three key resources. For bandwidth allocation to be feasible, the THz spectrum is split into a series of spectral windows that change in distance due to the fact that the vapor absorption coefficient varies with frequency. When the connection distance increases, the effective bandwidth window decreases in size~\cite{Gerasimenko2019}. The hierarchical bandwidth modulation, which optimizes device density to maximize capacity, is thought to be possible in THz communications due to the effective bandwidth~\cite{Hossain2019}. For localization purposes, it is necessary to identify and allocate appropriate sub-bands and subcarriers to the end devices at various ranges. As far as time slot allocation is concerned, a compromise between overhead and transmission speed must be found when communicating with a single recipient. It stands to reason that data acquired from multiple transmissions will improve localization accuracy, but at the expense of an increase in overhead and latency. Additionally, channel coherence and end device capabilities should be taken into account by the allocation process. Allocating communication time slots to different users helps ensure that all users within the service region receive the best possible positioning QoS. Finally, with regard to energy consumption, localization precision is often a limiting factor instead of a variable to be adjusted. Specifically, in an indoor positioning scenario, the end user requires to know its position with accuracy in the order of centimetres. Therefore, increasing the transmission power to achieve higher precision is unnecessary. All in all, intelligent energy allocation can meet the performance needs of user devices with minimum wasted resources. All of the aforementioned resource allocation strategies can be written in the form of optimization restriction in a plethora of active and passive beamforming problems.

\section{Insights \& future directions}\label{S:Future_directions}

\begin{figure}
    \centering\includegraphics[width=1\linewidth]{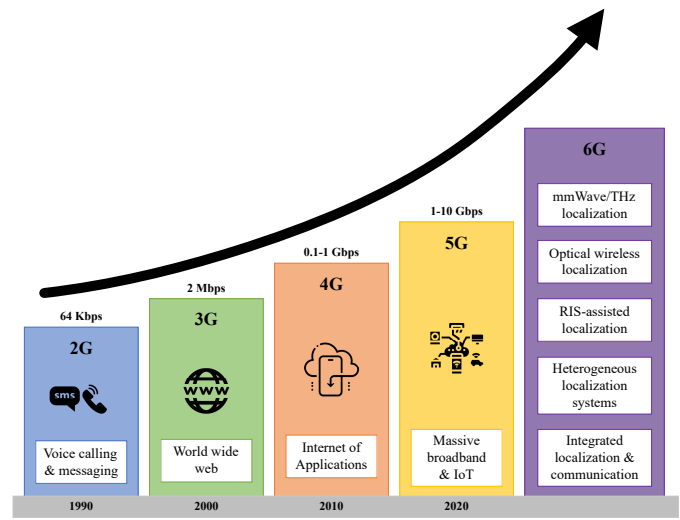}
    \caption{Future directions for 6G localization.}
    \label{fig:future_applications}
\end{figure}

So far, important concerns of future wireless systems' localization have been explored and intriguing findings obtained. In this section, as depicted in~Fig.~\ref{fig:future_applications}, we will address the insights and future directions with regard to mmWave/THz localization, wireless optical localization, RIS-assisted localization, heterogeneous localization systems, integrated localization and communication, as well as low earth orbit (LEO) positioning, navigation and timing (PNT).

\subsection{mmWave and THz localization}
mmWave/THz localization research is still in its infancy, with multiple approaches still to be investigated. In this section, future directions based on fundamental research and intelligent control design are presented.

\subsubsection{Fundamental research}
So far, a deterministic channel model was investigated. However, KPIs of dispersed signals, such as AoA, RSS, and more, are random in practical settings. Modeling the influence of randomly generated signals is necessary, and the impact of reflectors and scatterers on localization performance must be measured. Furthermore, since geometric localization relies on the channel model as its basis, current SotA approaches use an extension of mmWave models by including high-frequency system characteristics as the AOSA structure, the wideband effect, and atmospheric attenuation. However, the channel model may not account for the impact of hardware impairments and other THz-specific factors. Since the localization performance suffers as a result of these model incompatibilities, it is clear that a more precise channel model is necessary. Finally, in most localization tasks, the position and orientation of the BS and RIS are assumed as known anchors and research is focused on the UEs' location and orientation. However, this results in neglecting calibration problems between the reference anchor and additional ones. To overcome this issue, joint UE localization and BS/RIS calibration is of major importance in applications with many~anchors.

\subsubsection{Intelligent protocol design}
While model-based approaches are simple to deconstruct, AI-based approaches are often favoured for learning from or mitigating the impact of unanticipated model mismatches. In this case, it is important to have access to shared databases so that methods will be compared and contrasted. The preservation of user security and privacy, the transfer of a taught model into another domain to save training time, as well as the gathering, sharing, and storage of massive amounts of data, are all pressing concerns that need to be addressed in order to successfully deploy future intelligent 6G wireless networks. For example, optimising the placement of stationary BSs for better coverage or data accuracy is easy when compared to dynamically deployment of the BS by attaching it to UAVs, in temporary heavy traffic settings like stadiums or conference halls. To accommodate the UEs' communication and localization needs, the UAVs' position and path must be optimised. Connectivity in dynamic mmWave/THz UAV networks, however, presents its own unique set of difficulties that must be thoughtfully handled. Moreover, the SLAM method might provide access to a map of the immediate area. The PDF of previous access sites can be employed for scene-aware localization in addition to the map. In order to get around obstructions and make the most of the strong reflectors for localization, the beamforming vectors at the BS/UE and RIS element coefficients must be tuned. The intelligent optimization of the procedure to construct and maintain a map over time is an open question in this area.

\subsection{Optical wireless localization}
The vast majority of VLP techniques use at least three LEDs, with the voltage driving them being sinusoidal with very slight frequency changes. These strategies make use of the generally valid linear response between the LEDs and the PD, as detailed in most manufacturers' data-sheets. By using the RSS parameters and supposing a LOS attenuation model, these methods can determine the PD's static location. In addition, similar techniques have been adapted to carry out position estimate even while the UE is in motion at speeds that are more representative of real-world scenarios. However, by averaging the computed location using a small number of substantially overlapping received signals, the possible accuracy may be greatly improved. The reconstructed route can be smoothed down with the use of AI techniques like Kalman filter-based tracking. With this in mind, a robust modelling framework is needed to examine performance over a wide range of LED frequency, noise level, UE speed, and other design parameters. According to the findings reported in the literature, VLP approaches are effective at accurately localising the UE, which bodes well for their use both inside and outdoors. Finally, in order to go beyond the SotA of VLP wireless systems, it is necessary to pay close attention to positional errors, computing efficiency, as well as localization precision and latency.

\subsection{RIS-assisted localization}
RISs are expected to play a key role in the 6G era by manipulating the propagation condition, creating and/or canceling propagation paths, and practically enabling the ``propagation-as-a-service'' vision. Although a great amount of effort was put on analyzing the benefits and performance of RIS as well as designing RIS-empowered wireless communication protocols, there are only a limited number of contributions that documents the performance envelop of RIS-empowered or assisted localization systems and/or present suitable protocols in this direction. In this section, we briefly present the research gap and provide future directions in RIS-assisted localization~systems.   

\subsubsection{Fundamental research} The initial theoretical studies on RIS-assisted localization system assume: i) deterministic channel model, ii) continuous RIS phase shift capabilities, iii) ideal RIS with no imperfect meta-atom (MA),  iv) analog RIS, v) single-RIS, vi) interference-free systems, and viii) ideal transceivers front-end. Although these assumptions are necessary in order to determine the performance envelop of RIS-assisted localization systems, in several real-world implementations, they are not realistic. This drives a theoretical campaign in this area that would allow the relaxation of the aforementioned assumptions. In more detail, the performance of RIS-assisted localization systems that experience different type of fading, such as Rayleigh, Rice, Nakagami-$m$, Weibul, generalized Gamma, Malaga, Gamma-Gamma, etc. need to be extracted. Notice that several recent works, including~\cite{Papasotiriou2021,Boulogeorgos2022e,Boulogeorgos2022i,Boulogeorgos2019f,Papasotiriouf2022}, have verified these type of fading even in THz systems. In a similar direction, the assumption of continuous phase shift is not valid in practice, since RISs have a discrete number of phases that can use~\cite{Xu2021_d,An2022_d,Xu2021,Shi2022}. As a result, research questions like how many RIS phases we need for localization arises. Likewise, according to~\cite{Taghvaee2020}, the assumption that all the MAs of the RIS are in perfect condition may not be always valid. This motivates an analysis towards the localization approach tolerance to MA failures. Different types of RIS exist, such as analog, hybrid, and digital. Current contributions only deal with analog RISs. The benefits from using hybrid and digital RISs need to be analytically quantified.  Another research question that waits to be answer is concerning the feasibility and benefits of cascaded and parallel multi-RIS localization systems. Moreover, the impact of interference and transceivers hardware imperfections on the performance of RIS-assisted localization systems needs to be analytically assessed. Finally, most published contributions present closed-form or analytical expressions for the Cramer-Rao lower bound or a Fisher information analysis. However, a number of other insightful KPIs exist, such as localization energy efficiency, latency, coverage, etc. that are necessary to be assessed.   

\subsubsection{Intelligent protocol design}
To make the most out of RIS-assisted localization system, while providing adaptability to ever changing conditions, an intelligent protocol that allows the exploitation of the systems resources, namely space, frequency, and time need to be developed. In more detail, the protocol should allow co-design of beam code-words for both the source and the RIS, selects the optimum bandwidth and timeslot for localization. Moreover, it should coordinate the overall localization process allowing changing between beam codewords when needed. Such a protocol should aim at maximizing the localization system energy efficiency without compromising its accuracy and stability. Likewise, the protocol should take into account the nature of the system, i.e. operation frequency band, as well as the type of the RIS, i.e., analog, hybrid, digital. Finally, the protocol should provide guidelines for the optimal RIS placement. To design such protocol, researchers should formulate and solve a number of optimization problems and combine their solution to a reinforcement learning strategy that will ensure the system's adaptability. 

\subsection{Heterogeneous localization system design}
To counterbalance the physical limitations of high frequency wireless localization systems, simultaneous exploitation of microwave and mmWave, THz, and optical bands is required. However, localization systems in different bands have different localization range and accuracy. Moreover, their range and accuracy depends on different phenomena, like atmospheric conditions, size of the scatterers and obstracles, and design parameters, such as antenna gains, beamforming type, etc. Additionally, localization mechanisms have to facilitate the co-existence of several technologies with different coverage that follow different standards. As a result, localization system heterogeneity can be:  i) spectrum, and  ii)  technology~heterogeneity.   

\subsubsection{Spectrum heterogeneity}
refers to scenarios in which the localization agent uses both high  (e.g., mmWave, THz, optical) and low frequencies (e.g., microwaves). The higher the localization frequency, the higher the localization accuracy and the lower the localization range. This observation drives the idea of designing multi-band localizers capable of determining the position of the object of interest in a hierarchical manner. Specifically, such a localizer would use lower frequency to find the location of an object that is outside the high-frequency system range with low accuracy, and high frequencies to increase the localization accuracy of object that are inside their coverage area. A hierarchical localization approach is required for such a localizer that will allow efficient and fast localization of the object of interest and boost its accuracy. Finally, an experimentally verified  theoretical framework that quantify the hierarchical localizer performance and provides design guidelines is needed.      

\subsubsection{Technology heterogeneity}
introduces two scenarios, namely: i) stand-alone and ii) integrated localization systems.  In stand-alone scenarios only one type of localization technology, such as sensors, radards, mmaWave, THz, optical, is used. On the contrary in integrated localization systems, more than one localization technologies are used. In the later, the identification of the technologies that could combined without significantly negatively impacting the localizers energy efficiency needs to be performed. Another parameter that need to be accounted for is the implementation cost. Finally, algorithms for combining localization information from different sources are necessary to be developed. 

\subsection{Integrated localization and communication}
A radio signal can carry both the TX information and localization-related data. As a consequence, a unified framework for integrated communication and localization seems to be a natural next step.  Current cellular networks implement localization functionalities separately from the communication-related functions. The main reason behind this approach has been the different design objectives that these functionalities have. In more detail, the localization system aims at maximizing its accuracy, while  the communication system overall objective is to jointly maximize its reliability, data rate, and transmission range through the wireless channel, which experience fading and noise. Therefore, localization systems used to operate in high frequency bands, where the short wavelength and antenna beamwidth allowed them to pin-point the target with acceptable accuracy. On the other hand, communication have only recently started to exploiting mmWave and THz transmissions. Thus, the time of a localization and communication synergy has come~\cite{Sarieddeen2020,Liu2022a,He2022,Liu2021a}. In order to devise efficient and low-cost integrated localization and communication systems a number of strategically designed steps, starting from suitable KPIs definition and finishing to the design of suitable protocols capable of operating in heterogeneous network environments, need to be taken. Next, we identify and analyze these steps.    

\subsubsection{KPIs definition}
As discussed in~\cite{Ghatak2018,Jeong2015,Destino2017,Destino2018}, there are several trade-offs between localization and communication system goals. In more detail, in~\cite{Ghatak2018}, the authors discussed the localization accuracy and communication data rate trade-off. In~\cite{Jeong2015}, the authors revealed the localization accuracy, power efficiency, data rate trade-off. Similarly, in~\cite{Destino2017}, the authors highlighted the localization accuracy data rate trade-off in single-user mmWave systems, while, in~\cite{Destino2018}, the prior contribution was extended in multi-user scenarios. The aforementioned contributions revealed that there are a number of trade-offs that need to be understand and balanced. To achieve this, the current localization and communication KPIs may be not enough. This motivates rethinking and redefining or creating new KPIs capable of providing insights that in turns optimize the localization and communication synergy. Moreover, new KPIs may need to be defined that quantify the performance of specific applications. For example, QoS may be a better KPI instead of localization accuracy. Finally, for a number of applications in which the reliability of position information is a key design parameter, such as RIS or UAV-empowered wireless systems, the position integrity and availability may be the most important KPIs.   

\subsubsection{Fundamental research}
As we move towards mmWave, THz, and optical bands, the performance of integrated localization and sensing systems are naturally expected to improve. However, new issues and challenges need to be addressed, giving space to fundamental research. Specifically, the research concerning THz channel modeling as well as integrated localization and communication is still at the early stage. Next, we focus in these two topics and provide some possible research~directions. 

Most published contribution in high-frequency integrated localization and communication neglect the impact of multiple scatters and obstacles and employ deterministic channel models. However, in practice, the AOA as well as the amplitude of the scattered signals are random processes. It is of high importance to model the effect of non-LOS signals and of scatters in order to allow the accurate integrated localization and communication system performance evaluation. In this direction, new channel modeling methodologies based on stochastic geometry, random shape theory and electromagnetic tools, like ray tracing or the finite element method, need to be developed. Other phenomena that should be taken into consideration in signal modeling are the wideband effect as well as the impact of atmospheric conditions and/or the transceivers hardware imperfections. Finally, the channel and signal model should also account for the existance of smart materials, such as RISs.  

Building upon the new channel and signal models, a new theoretical framework for the analysis of the performance of integrated localization and communication systems will be developed. This framework should return not only conventional KPIs that quantify the performance either the localization or the communication part of the system, but also the novel KPIs that would be defined and allow the assessment of the integrated localization and communication system. Moreover, different system models that employ different types of localizers with and without RISs, need to be investigated.  The derived expressions are expected to bring insights and allow the optimization of the integrated localization and communication~system.

\subsubsection{Advanced signal processing}
Mobile terminals at various tiers of a multi-tiered network may communicate with one another using distinct frequency bands and independent communication paths. In such scenarios, efficient and dynamic maintenance of wireless networks is a key concern for integrated localization and communication. Maintenance and reconfiguration of wireless connections may be possible via dynamic resource management and the utilization of multiple distinct frequency bands. In addition, cross-layer data exchange depends on efficient signal processing for its operation. In this respect, the use of common hardware for localization and communication services can provide high efficiency for signal processing technologies in integrated communication and localization services, as well as the required reusability of channel estimation units for position extraction. As a result, the most important challenge with respect to advanced signal processing is achieving efficient co-design of signal processing techniques and hardware architecture.

\subsubsection{Waveform design}
Despite the fact that localization and communication waveform designs share a number of common key requirements, including low-latency, low transceiver hardware complexity, and high reliability, in the 4G and the 5G eras, signals and systems for localization and communications were designed separately~\cite{Dammann2016,Raulefs2016}. As we move towards higher frequency wireless systems, the idea of designing a common signal for localization and communication arises. To achieve this we need to determine  optimal signal PSDs that satisfies both the requirements of the localization accuracy and the communication robustness to interference. In more detail, according to~\cite{Dammann2016} and~\cite{Gezici2005}, the Cramer-Rao lower bound of time-based ranging localization systems can be expressed~as
\begin{align}
    \mathrm{CR} = \frac{c^{2}}{8\pi^2\,b^2\,\frac{P}{N_o\,B}},
    \label{Eq:CR}
\end{align}
where $P$ and $B$ are respectively the transmission power and bandwidth, while $N_o$ is the noise PSD. Moreover, $b^2$ is the mean square bandwidth that can be obtained~as
\begin{align}
    b^2 = \frac{\int_{-B}^{B} f^2\,\left|S(f)\right|^2\,\mathrm{d}f}{\int_{-B}^{B}\left|S(f)\right|^2\,\mathrm{d}f},
    \label{Eq:MSB}
\end{align}
with $S(f)$ being the Fourier transform of the transmission~signal. 

From~\eqref{Eq:CR} and~\eqref{Eq:MSB}, for the localization perspective, a signal that has its power concentrated to the edges of the spectrum achieves better performance in terms of localization accuracy, i.e., lower  Cramer-Rao lower bound, in comparison to signals that their power is either uniformly distributed or concentrated at the center of their spectrum. From communications point of view, in order to reduce the inter-symbol interference, signals are designed in a way that their power is concentrated at the center of their spectrum. This observation reveals that there is a trade-off in terms of the PSD requirements for localization and communication. To find the equilibrium, optimization problems for a number of different system models is necessary to be formulated and suitable policies need to be extracted. Finally, adaptability and flexibility in terms of bandwidth, signal power, PSD shape, etc. should be allowed in order to make the most in real-time ever changing wireless~environments.

\subsubsection{Pro-active radio environment mapping and radio resource management}
In order to reap the communication benefits of the proactive radio resource management in areas like beam alignment, channel prediction, cell selection, and more, high-precision radio environment mappings will be necessary. The majority of the radio environment mappings models that exist so far has focused on 2D scenarios. With this in mind, more accurate 3D localization in nLOS, and multi-path applications is required for future 6G networks. In spite of the fact that a number of effective multi-path and NLOS mitigation algorithms have been proposed in the literature, these algorithms are typically quite complex and are only practical for remote localization systems as opposed to self-localization systems where the UE must perform the computation and estimation. However, for the sake of analytical simplicity, these studies often only include first-order reflection, even though higher-order reflections are common in dense multi-path settings and might be crucial to localization in NLOS situations. As a result, further research is required in order to develop more precise and inexpensive localization methods for multi-path scenarios while maintaining the UE's communication performance and allocating radio resources based on its current position. Finally,despite the fact that precise location data improves communication, a better understanding of the underlying connection between varying levels of location precision and transmission rates is required.

\subsubsection{Design of integrated localization and communication protocols for heterogeneous networks}
Future networks will become more diverse by using a wide variety of standards and frequency ranges. For mobile terminals that are constantly moving under unstable radio conditions, the challenge of how to swiftly transition between protocols is of paramount importance in order to maintain reliable localization and communication. Since each network layer can use a unique set of protocols, the creation of logical links between them can be achieved by considering the protocols as gateways. However, the need for cross-layer, -module, and -node information transfer constitutes the integrated protocol design crucial. Finally, a reevaluation of the physical and MAC layers with respect to the protocol design that takes into account localization in addition to the communication metrics is required.

\subsection{Low earth orbit (LEO) localization systems}
In rural/remote areas that lack terrestrial infrastructure or when the signals from the terrestrial infrastructure are blocked due to obstacles, such as in mmWave/THz transmissions, terrestrial- and LEO-based localization can be used in integrated terrestrial non-terrestrial networks to achieve the best from the two options. In future communication, sensing, and localization applications, and, especially, positioning applications with LEO signals, there is currently a dichotomy of approaches. On the one hand, existing systems can be utilized to offer novel services to the end users, while, on the other hand, completely new systems, such as the ongoing efforts towards a LEO-PNT concept, can be designed. In the former, the receiver is responsible for most of the optimization work based on the assumption that the necessary infrastructure is already in place. In the latter, a three-segment optimization is required, which can be achieved either simultaneously or in stages. In order to obtain higher precision for localization with worldwide coverage, future efforts on constructing unique LEO-PNT constellations through optimization are focused on (i) integration of 6G and LEO networks, (ii) intelligent beamforming-based localization, (iii) LEO-based edge computing, and (iv) equip LEO satellites with GNSS receivers.

\section{Conclusion}\label{S:Conclusion}
The potential of the future 6G wireless system for localization was investigated in this article. In more detail, the key localization-based applications and use cases for the next generation of wireless networks were discussed, and the technologies that enable localization services were investigated. Moreover, models of mmWave/THz as well as VLP localization wireless systems were presented , while also taking into account both LOS and NLOS channels that were generated by traditional reflectors as well as RISs. Also, the major localization KPIs were described and mathematical expressions were provided when possible. Furthermore, SotA traditional and learning-based localization techniques were presented alongside their conceptual figures. Finally, the impact of all design aspects of future 6G wireless systems, namely characteristics, assumptions, localization problem formulation, and optimization, were discussed, while lessons learned and future directions were extracted.

% Can use something like this to put references on a page
% by themselves when using endfloat and the captionsoff option.
\ifCLASSOPTIONcaptionsoff
  \newpage
\fi

%\balance
\bibliographystyle{IEEEtran}
\bibliography{IEEEabrv,refs.bib}

\begin{IEEEbiography}
    [{\includegraphics[width=1in,height=1.25in,clip,keepaspectratio]{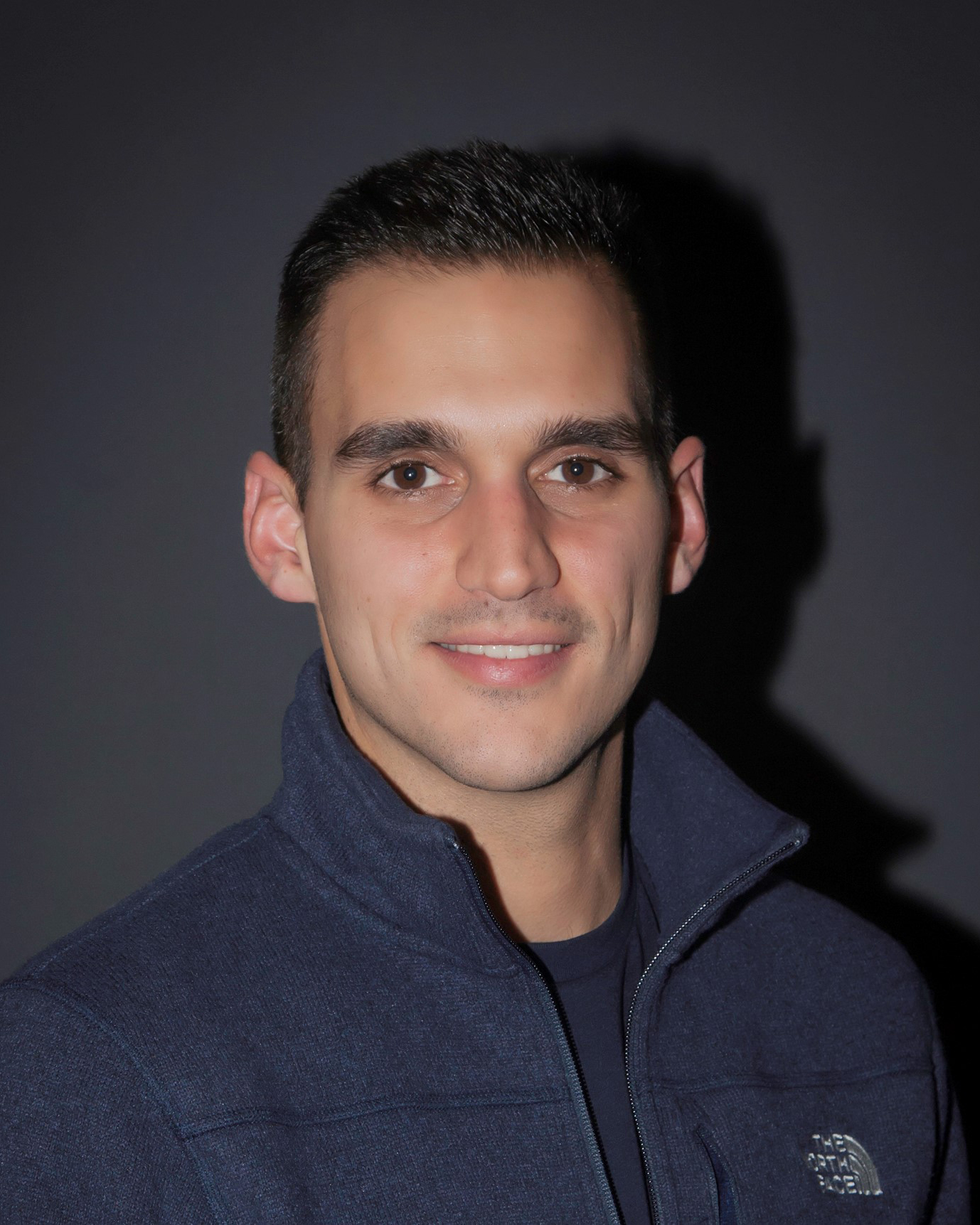}}]
    {Stylianos E. Trevlakis} was born in Thessaloniki, Greece in 1991. He received the Electrical and Computer Engineering (ECE) diploma (5 year) from the Aristotle University of Thessaloniki (AUTh) in 2016. Afterwards, Dr. Trevlakis served in the Hellenic Army in for nine months in the Research Office as well as at the Office of Research and Informatics of the School of Management and Officers. During 2017, he joined the Information Technologies Institute, while from October 2017 until April 2022, he was part of WCIP as a PhD candidate in AUTh. During the same period, he was a teaching assistant at the department of ECE of AUTh. From April 2022 until now, Dr. Trevlakis is working at InnoCube as a postdoctoral researcher with focus on state-of-the-art research in conventional \& AI-enabled Wireless Communication Systems.

    Dr. Trevlakis's research interests lie in the area of Wireless Communications, with emphasis on conventional \& AI-enabled Wireless Communication Systems, as well as Communications \& Signal Processing for Biomedical Engineering.
\end{IEEEbiography}

\begin{IEEEbiography}
    [{\includegraphics[width=1in,height=1.25in,clip,keepaspectratio]{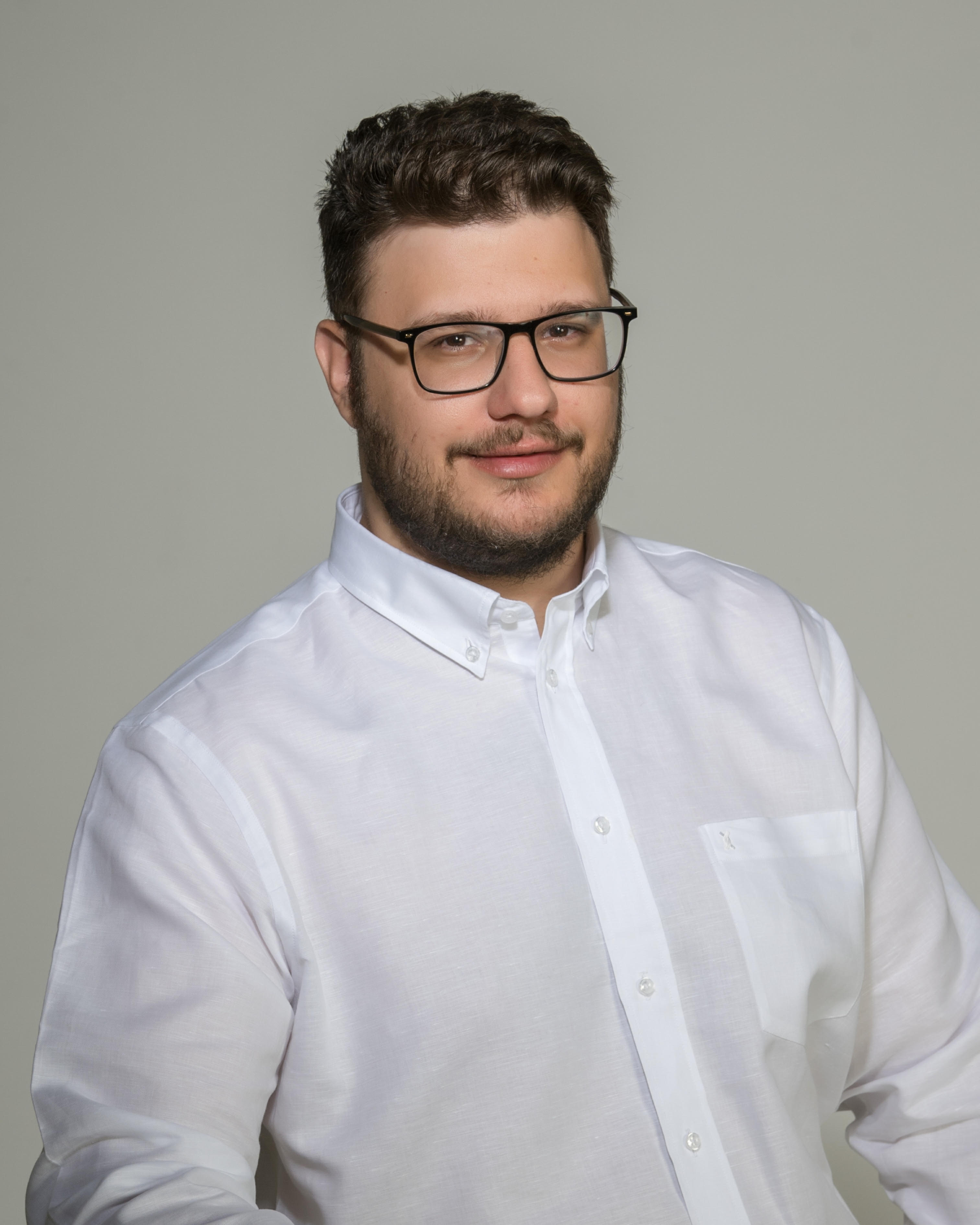}}]
    {Alexandros-Apostolos A. Boulogeorgos} (Senior Member, IEEE) was born in Trikala, Greece, in 1988. He received the Diploma degree in Electrical and Computer Engineering (ECE) and the Ph.D. degree in wireless communications from the Aristotle University of Thessaloniki (AUTh) in 2012 and 2016, respectively. From 2022, he is an Assistant Professor at the Department Electrical and Computer Engineering of the University of Western Macedonia, Greece.  From 2017 to 2022, he served as a Senior Researcher at the Department of Digital Systems, University of Piraeus, where he conducted research in the area of wireless communications. From October 2012 to September 2016, he was a Teaching Assistant with the Department of ECE, AUTh, and from February 2017 to September 2022, he served as an Adjunct Professor with the Department of ECE, University of Western Macedonia, and as an Visiting Lecturer with the Department of Computer Science and Biomedical Informatics, University of Thessaly. He has worked in a number of EU and national projects. 

    Dr Boulogeorgos has (co-)authored more than 120 technical papers, which were published in scientific journals and presented at prestigious international conferences. Furthermore, he is the holder of two (one national and one European) patents.  His current research interests spans in the area of wireless communications and networks with emphasis in high frequency communications, intelligent communication systems, optical wireless communications, and signal processing and communications for biomedical applications.
    
    Dr. Boulogeorgos was awarded the Distinction Scholarship Award from the Research Committee of AUTh in 2014, and was recognized as an Exemplary Reviewer for IEEE COMMUNICATION LETTERS in 2016 (top 3\% of reviewers). Moreover, he was named a Top Peer Reviewer (top 1\% of reviewers) in Cross-Field and Computer Science in the Global Peer Review Awards 2019, which was presented by the Web of Science and Publons. Finally, in 2021, he received the best oral presentation award in the International Conference on Modern Circuits and Systems Technologies (MOCAST) 2021. He has been involved as a member of organizational and technical program committees in several IEEE and non-IEEE conferences and served as a reviewer in various IEEE journals and conferences. He is an IEEE Senior Member and a Member of the Technical Chamber of Greece. He is currently an Editor for IEEE COMMUNICATIONS LETTERS, an Associate Editor for the Frontier in Communications and Networks, and for the MDPI~Telecom. Finally, Dr Boulogeorgos has participated as a guest editor in the organization of a number of special issues in IEEE and non-IEEE journals. 
\end{IEEEbiography}

\begin{IEEEbiography}
    [{\includegraphics[width=1in,height=1.25in,clip,keepaspectratio]{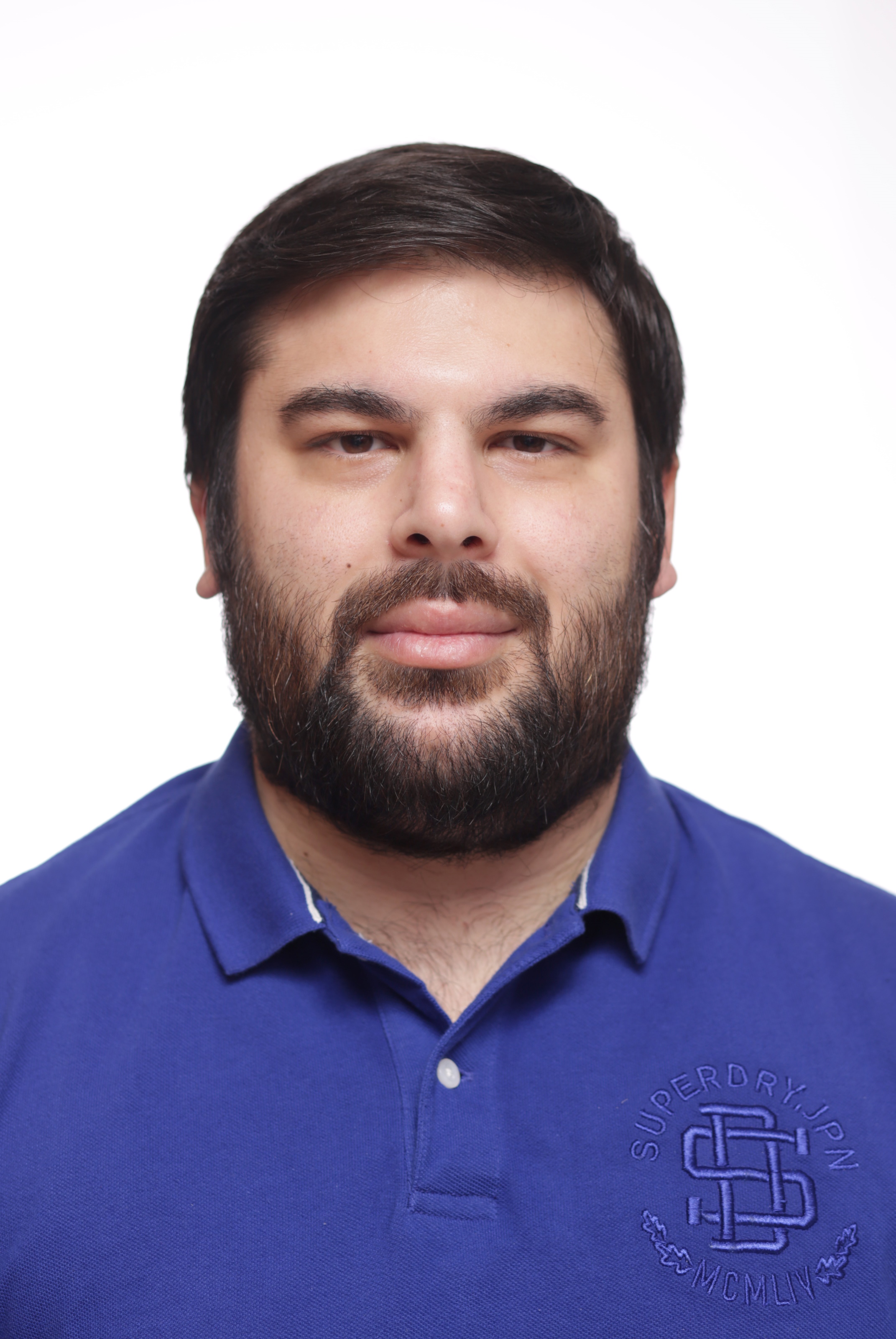}}]
    {Dimitrios Pliatsios} received his diploma degree from the Department of Electrical and Computer Engineering, Aristotle University of Thessaloniki, Greece in 2016 and his PhD from the Department of Electrical and Computer Engineering, University of Western Macedonia, Kozani, Greece in 2022. Currently, he works as a postdoctoral researcher at the ITHACA Lab, Department of Electrical and Computer Engineering, University of Western Macedonia, in EU-funded research projects and participates in drafting research funding proposals. His research interests include resource allocation in wireless communications and edge computing environments, optimization theory, B5G/6G mobile networks, and computer and network security. He is a member of the IEEE and the Technical Chamber of Greece and he has served as a reviewer in several scientific journals (IEEE Internet of Things Journal, IEEE Communication Letters, Elsevier Computer Networks, IEEE Access, MDPI Sensors) and conferences (IEEE GLOBECOM, IEEE ICC, IEEE NetSoft, IEEE CAMAD, IEEE INFOCOM, IEEE PIMRC). His PhD research was funded by the Greek State Scholarship Foundation and he has received the 1st Research Excellence Award by the Research Committee of the University of Western Macedonia.
\end{IEEEbiography}

\begin{IEEEbiography} 
    [{\includegraphics[width=1in,height=1.23in,clip,keepaspectratio]{authors/Ntontin.jpg}}]
    {Konstantinos Ntontin} (Member, IEEE) is currently a research scientist of the SIGCOM Research Group at SnT, University of Luxembourg. In the past, he held research associate positions at the Electronic Engineering and Telecommunications department of the University of Barcelona, the Informatics and Telecommunications department of the University of Athens, and the National Centre of Scientific Research-``Demokritos''. In addition, he held an internship position at Ericsson Eurolab Gmbh, Germany. He received the Diploma in Electrical and Computer Engineering in 2006, the M. Sc. Degree in Wireless Systems in 2009, and the Ph. D. degree in 2015 from the University of Patras, Greece, the Royal Institute of Technology (KTH), Sweden, and the Technical University of Catalonia (UPC), Spain, respectively. His research interests are related to the physical layer of wireless telecommunications with focus on performance analysis in fading channels, MIMO systems, array beamforming, transceiver design, and stochastic modeling of wireless channels.
\end{IEEEbiography}

\begin{IEEEbiography}
    [{\includegraphics[width=1in,height=1.25in,clip,keepaspectratio]{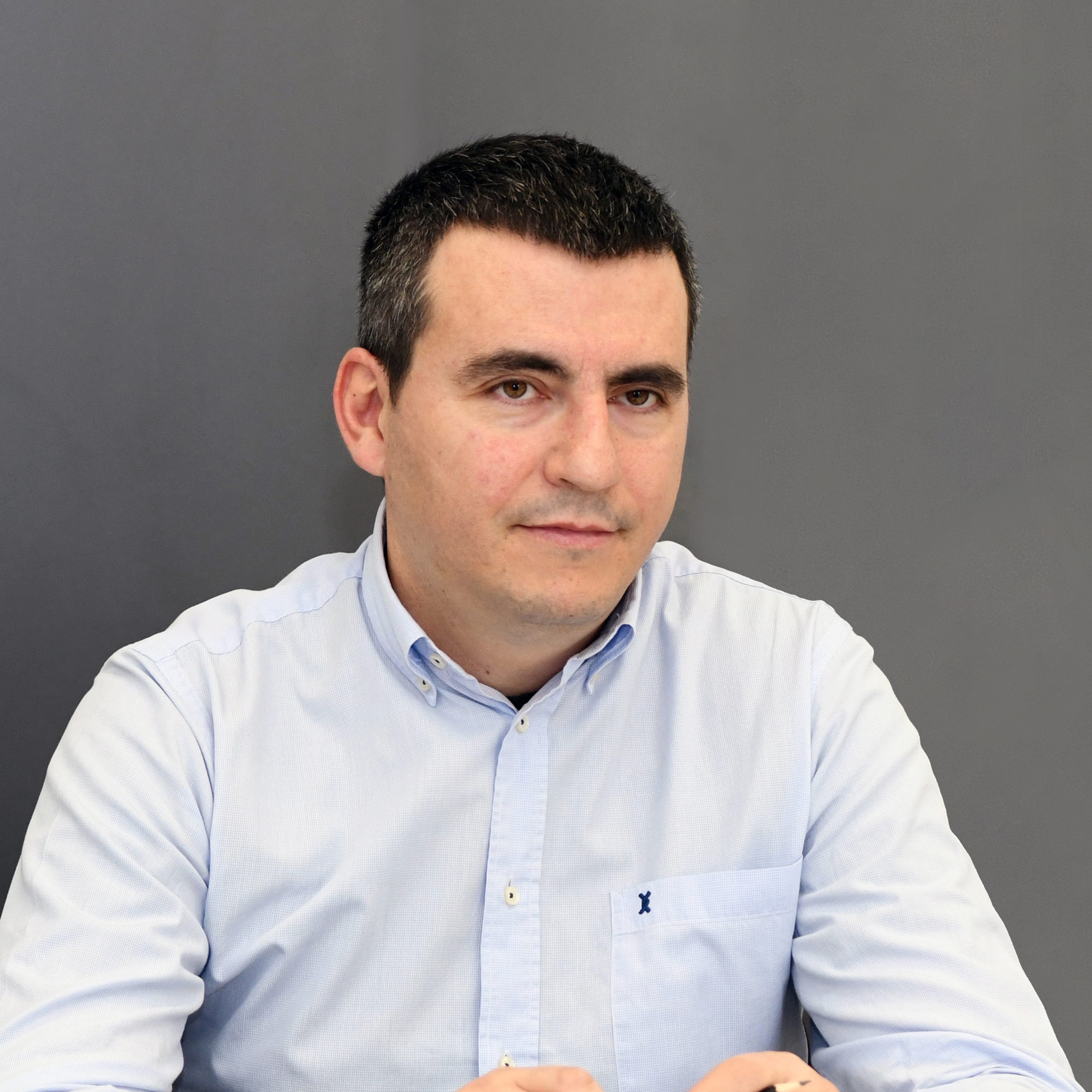}}]
    {Panagiotis Sarigiannidis} is the Director of the ITHACA lab $\text{(https://ithaca.ece.uowm.gr/)}$, co-founder of the 1st spin-off of the University of Western Macedonia: MetaMind Innovations P.C. $\text{(https://metamind.gr)}$, and Associate Professor in the Department of Electrical and Computer Engineering in the University of Western Macedonia, Kozani, Greece. He received the B.Sc. and Ph.D. degrees in computer science from the Aristotle University of Thessaloniki, Thessaloniki, Greece, in 2001 and 2007, respectively. He has published over 260 papers in international journals, conferences and book chapters, including IEEE Communications Surveys and Tutorials, IEEE Transactions on Communications, IEEE Internet of Things, IEEE Transactions on Broadcasting, IEEE Systems Journal, IEEE Wireless Communications Magazine, IEEE Open Journal of the Communications Society, IEEE/OSA Journal of Lightwave Technology, IEEE Transactions on Industrial Informatics, IEEE Access, and Computer Networks. He received 5 best paper awards. He has been involved in several national, European and international projects. He is currently the project coordinator of three H2020 projects, namely a) H2020-DS-SC7-2017 (DS-07-2017), SPEAR: Secure and PrivatE smArt gRid, b) H2020-LC-SC3-EE-2020-1 (LC-SC3-EC-4-2020), EVIDENT: bEhaVioral Insgihts anD Effective eNergy policy acTions, and c) H2020-ICT-2020-1 (ICT-56-2020), TERMINET: nexT gEneRation sMart INterconnectEd ioT, while he coordinates the Operational Program MARS: sMart fArming with dRoneS (Competitiveness, Entrepreneurship, and Innovation) and the Erasmus+ KA2 ARRANGE-ICT: SmartROOT: Smart faRming innOvatiOn Training. He also serves as a principal investigator in the H2020-SU-DS-2018 (SU-DS04-2018), SDN-microSENSE: SDN-microgrid reSilient Electrical eNergy SystEm and in three Erasmus+ KA2: a) ARRANGE-ICT: pArtneRship foR AddressiNG mEgatrends in ICT, b) JAUNTY: Joint undergAduate coUrses for smart eNergy managemenT sYstems, and c) STRONG: advanced firST RespONders traininG (Cooperation for Innovation and the Exchange of Good Practices). His research interests include telecommunication networks, internet of things and network security. He is an IEEE member and participates in the Editorial Boards of various journals.
\end{IEEEbiography}

\begin{IEEEbiography}
    [{\includegraphics[width=1in,height=1.25in,clip,keepaspectratio]{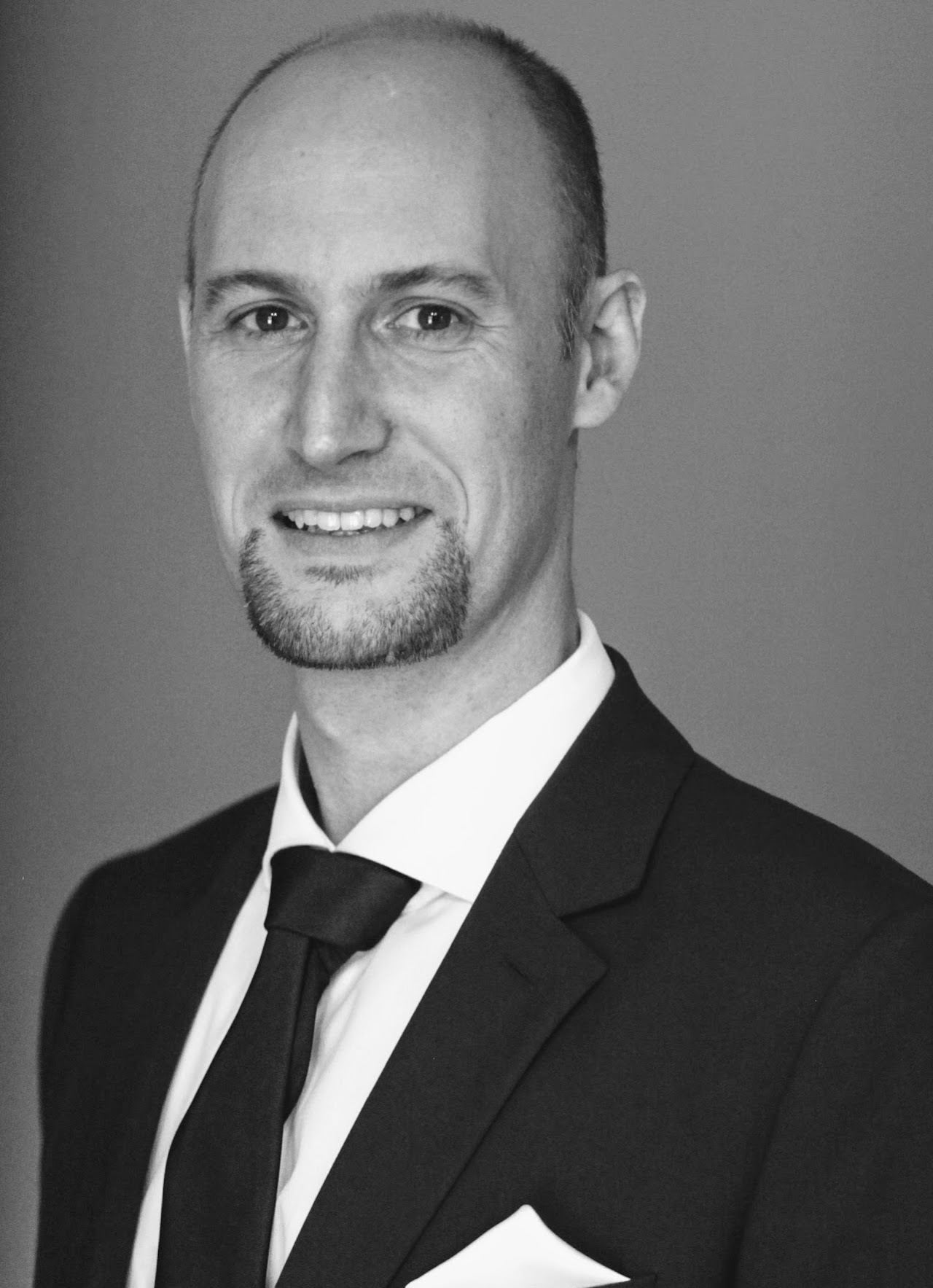}}] 
    {Symeon Chatzinotas} (Fellow, IEEE) is currently Full Professor / Chief Scientist I and Co-Head of the SIGCOM Research Group at SnT, University of Luxembourg. In the past, he has been a Visiting Professor at the University of Parma, Italy and he was involved in numerous Research and Development projects for the National Center for Scientific Research Demokritos, the Center of Research and Technology Hellas and the Center of Communication Systems Research, University of Surrey. He received the M.Eng. degree in telecommunications from the Aristotle University of Thessaloniki, Thessaloniki, Greece, in 2003, and the M.Sc. and Ph.D. degrees in electronic engineering from the University of Surrey, Surrey, U.K., in 2006 and 2009, respectively. He was a co-recipient of the 2014 IEEE Distinguished Contributions to Satellite Communications Award, the CROWNCOM 2015 Best Paper Award and the 2018 EURASIP JWCN Best Paper Award. He has (co-)authored more than 450 technical papers in refereed international journals, conferences and scientific books. He is currently in the editorial board of the IEEE Open Journal of Vehicular Technology and the International Journal of Satellite Communications and Networking.
\end{IEEEbiography}

\begin{IEEEbiography}
    [{\includegraphics[width=1in,height=1.23in,clip,keepaspectratio]{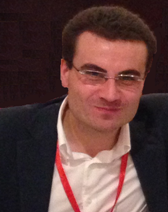}}]
    {Marco Di Renzo (Fellow, IEEE)} received the Laurea (cum laude) and Ph.D. degrees in electrical engineering from the University of L’Aquila, Italy, in 2003 and 2007, respectively, and the Habilitation à Diriger des Recherches (Doctor of Science) degree from University Paris-Sud (now Paris-Saclay University), France, in 2013. Since 2010, he has been with the French National Center for Scientific Research (CNRS), where he is a CNRS Research Director (Professor) with the Laboratory of Signals and Systems (L2S) of Paris-Saclay University – CNRS and CentraleSupelec, Paris, France. In Paris-Saclay University, he serves as the Coordinator of the Communications and Networks Research Area of the Laboratory of Excellence DigiCosme, and as a Member of the Admission and Evaluation Committee of the Ph.D. School on Information and Communication Technologies. He is the Editor-in-Chief of IEEE Communications Letters and a Distinguished Speaker of the IEEE Vehicular Technology Society. In 2017-2020, he was a Distinguished Lecturer of the IEEE Vehicular Technology Society and IEEE Communications Society. He has received several research distinctions, which include the SEE-IEEE Alain Glavieux Award, the IEEE Jack Neubauer Memorial Best Systems Paper Award, the Royal Academy of Engineering Distinguished Visiting Fellowship, the Nokia Foundation Visiting Professorship, the Fulbright Fellowship, and the 2021 EURASIP Journal on Wireless Communications and Networking Best Paper Award. He is a Fellow of the UK Institution of Engineering and Technology (IET), a Fellow of the Asia-Pacific Artificial Intelligence Association (AAIA), an Ordinary Member of the European Academy of Sciences and Arts (EASA), and an Ordinary Member of the Academia Europaea (AE). Also, he is a Highly Cited Researcher.
\end{IEEEbiography}

% that's all folks
\end{document}